\documentclass{aa}

\usepackage[varg]{txfonts}

\usepackage{lineno}
\usepackage{natbib}
\usepackage{amsmath}
\usepackage{graphicx}
\usepackage{xspace}
\usepackage{xcolor}
\usepackage[markup=underlined]{changes}

\usepackage{hyperref}
\hypersetup{
    colorlinks=true,
    linkcolor={red!50!black},
    citecolor={blue!70!black},
    urlcolor={blue!80!black}
}

\graphicspath{{./figs/}{./subs/}}

\usepackage[normalem]{ulem}
\usepackage{placeins}
\usepackage{stfloats}
\usepackage{etoolbox}
\newtoggle{thesis}

\begin{document}

\title{Timing analysis of the black-hole candidate Swift~J1727.8–1613: detection of a dip-like feature in the high-energy cross spectrum}
\titlerunning{Dip feature in Swift~J1727.8–1613}

\author{Pei~Jin\inst{1}\thanks{peijin@astro.rug.nl} \and
Mariano~M\'{e}ndez\inst{1}\thanks{mariano@astro.rug.nl} \and
Federico~Garc\'{\i}a\inst{2} \and
Diego~Altamirano\inst{3} \and
Guobao~Zhang\inst{4,5} \and
Sandeep~K.~Rout\inst{6}}

\institute{Kapteyn Astronomical Institute, University of Groningen, P.O.\ BOX 800, 9700 AV Groningen, The Netherlands
\and Instituto Argentino de Radioastronom\'{\i}a (CCT La Plata, CONICET; CICPBA; UNLP), C.C.5, (1894) Villa Elisa, Buenos Aires, Argentina
\and School of Physics and Astronomy, University of Southampton, Southampton, Hampshire SO17 1BJ, UK
\and Yunnan Observatories, Chinese Academy of Sciences, Kunming 650216, People's Republic of China
\and University of Chinese Academy of Sciences, Beijing 100049, People's Republic of China
\and Center for Astrophysics and Space Science, New York University Abu Dhabi, PO Box 129188 Abu Dhabi, UAE}

\date{Received xxx; accepted xxx}

\abstract{We present a timing analysis of observations with the Hard X-ray Modulation Telescope of the black hole X-ray transient Swift~J1727.8–1613 during its 2023 outburst. We detect, for the first time in a black hole X-ray binary, a prominent dip at $\sim 3$–$15$~Hz in the real part of the cross spectrum between high-energy ($>$25~keV) and low-energy ($<$10~keV) photons in the Low Hard and Hard Intermediate States, during which the QPO frequency rapidly increases and then stabilizes at $\sim 1.0$–$1.5$~Hz. Remarkably, the real part of the cross spectrum reaches negative values at the frequencies around the minimum of the dip, indicative of a phase lag ranging between $\pi/2$ and $\pi$ in this frequency range. 
We fit the power spectra and the real and imaginary parts of the cross spectra simultaneously using a multi-Lorentzian model. Among the lag models, the Gaussian phase-lag model provides a slightly better reduced $\chi^2$ than the constant phase-lag and constant time-lag models, while it  also alleviates the degeneracy associated with those models. From the parameters of the Lorentzian that fits the dip, we estimate the size of the accretion flow, which consistently exceeds 10,000 km as the QPO frequency increases from 0.13 Hz to 2.0 Hz. 
Furthermore, both the energy-dependent phase-lag and fractional-rms spectra of the dip exhibit a change in trend around 15 keV, with the phase lag dropping and rms reaching a local minimum. These spectra closely resemble the shapes predicted by the time-dependent Comptonization model, vKompth, for a low feedback factor, offering a pathway to explain the radiative properties of the corona. Additionally, the coherence function suggests a diversity of variability components, potentially arising from different parts of the corona.}

\keywords{accretion, accretion discs -- stars: individual: Swift~J1727.8–1613 -- stars: black holes -- X-rays: binaries
}

\maketitle 



\section{Introduction}
\label{sec:introduction}

Black hole X-ray binaries (BHXBs) are typically transient systems that stay most of the time in the quiescent state, but show recurrent bright X-ray outbursts lasting for weeks to months~\citep[e.g.][]{1996ARA&A..34..607T, 2006csxs.book..157M}. During an outburst, the spectral and timing properties of these sources change as the source evolves along the hardness intensity diagram (HID). According to \citet{2010LNP...794...53B} and \citet{2011BASI...39..409B}, there are generally four spectral-timing states: Low Hard State (LHS), Hard Intermediate State (HIMS), Soft Intermediate State (SIMS) and High Soft State (HSS).

Short time-scale variability ranging from milliseconds to hundreds of seconds is very common in BHXBs~\citep[see][and references therein]{2011BASI...39..409B, 2019NewAR..8501524I}. This variability is usually studied in the frequency domain with the Fourier transform~\citep[e.g.][]{1989ASIC..262...27V}. The power density spectrum (PDS) describes the strength of the variability as a function of Fourier frequency~\citep{1983ApJ...266..160L, 1989ASIC..262...27V}. 
The PDS is usually decomposed into several incoherent (additive) variability components~\citep[e.g.][]{2000MNRAS.318..361N, 2002ApJ...572..392B, 2004astro.ph.10551V}, which can be described by Lorentzian functions. 
In the frequency range of $\sim0.01-30$ Hz, the most common components in black-hole systems include the low-frequency quasi-periodic oscillations (QPOs) and their harmonics~\citep{2000MNRAS.318..361N, 2002ApJ...572..392B, 2004astro.ph.10551V, 2015MNRAS.448.1298P, 2020MNRAS.496.5262V, 2020MNRAS.494.1375Z, 2012MNRAS.423..694R, 2023AdSpR..71.3508D}, as well as the band-limited noise component (BLN) ~\citep{2002ApJ...572..392B, 2004astro.ph.10551V} with a zero centroid frequency. Sometimes, a high-frequency peak noise component (HF-PN), correlated with the low-frequency QPOs through the PBK correlation (Psaltis, Belloni and van der Klis~\citeyear[][]{1999ApJ...520..262P}), is also observed~\citep{2002ApJ...572..392B, 2024MNRAS.529.4624Y}. In recent years, some new components have been discovered in these systems, such as the shoulder of the QPO~\citep{2002ApJ...572..392B} and the so-called imaginary QPO~\citep{2024MNRAS.527.9405M, 2025A&A...696A.128B, 2025arXiv250303078F}, indicating the diversity of timing variability components.

Another powerful tool to study the timing properties of BHXBs is the cross spectrum (CS) of the correlated light curves of a source in two different energy bands~\citep{1996MNRAS.280..227N, Bendat-2010}. The CS is a complex function of Fourier frequency. The phase angle of the CS in the complex plane represents the phase lag at a given Fourier frequency.
For a sinusoidal wave with a given frequency~$\nu$, the time lag equals to the phase lag divided by $2\pi\nu$.
There are many models to explain the phase/time lags of QPO and BLN, including reverberation~\citep{2018MNRAS.475.4027M, 2019MNRAS.488..348M, 2021MNRAS.507...55M}, propagating mass accretion rate fluctuations~\citep{2006MNRAS.367..801A, 2023MNRAS.519.4434K}, Comptonization in a relativistic jet/outflow~\citep{2018A&A...614L...5K}, Lense-Thirring precession~\citep{2009MNRAS.397L.101I, 2016MNRAS.461.1967I}, time-dependent Comptonization~\citep{2020MNRAS.492.1399K, 2021MNRAS.501.3173G, 2022MNRAS.515.2099B} and jet precession~\citep{2021NatAs...5...94M}.

In the past decade, the study of the amplitudes and lags of the QPOs and BLN has improved our understanding of the physical and geometrical properties of the accretion flow around black-holes~\citep{2011MNRAS.415.2323I, 2016MNRAS.461.1967I, 2018MNRAS.475.4027M, 2020A&A...640L..16K, 2020MNRAS.492.1399K, 2022NatAs...6..577M}. 
Building on the success of this approach, it is further assumed that those Lorentzian components are not just a useful empirical description of the power density spectrum, but that they represent physical processes occurring within the system~\citep{2024MNRAS.527.9405M}.

While the Lorentzian decomposition has been widely used in the last twenty years, there was no reliable way to separate the lags of the individual components, until \citet{2024MNRAS.527.9405M} proposed a joint-fitting approach. This method simultaneously fits the PDS and the real and imaginary parts of the CS, under the assumptions that the individual Lorentzian functions are incoherent with one another but each of them is fully coherent with themselves measured in different energy bands.
In the framework introduced by \citet{2024MNRAS.527.9405M}, the lag of an individual variability component is a function of Fourier frequency.
Among all the possible functions, \citet{2024MNRAS.527.9405M} studied the two simplest cases, the constant phase-lag and the constant time-lag models.
For the constant phase-lag model, the real or imaginary part of the CS can be decomposed into the same combination of Lorentzian functions that fits the PDS.
For the constant time-lag model, the models of the real and imaginary parts of the CS are the same as the one of the PDS, but with each Lorentzian function multiplied by, respectively, $\cos(2\pi t_cv)$ and $\sin(2\pi t_cv)$, where $t_c$ is the constant time lag. Therefore, the real and imaginary parts of the CS show oscillations as a function of Fourier frequency with a period of $1/t_c$.

\citet{2024MNRAS.527.9405M} also proposed a understanding of the intrinsic coherence function~\citep{1996MNRAS.280..227N, 1997ApJ...474L..43V} of the correlated light curves of a source in two different energy bands.
If the coherence function at a given frequency range is high, it must be that a single Lorentzian dominates the variability.
If the coherence function at a given frequency range is low, it is the case of the superposition of two or more uncorrelated Lorentzians in that frequency range.

The new galactic X-ray transient, Swift~J1727.8–1613, was discovered on August 24, 2023 by Swift/BAT~\citep{2023GCN.34537....1P}, and was identified as a black-hole X-ray binary candidate by subsequent observations~\citep{2023ATel16208....1C, 2023ATel16207....1O, 2023ATel16211....1M}.
Low-frequency QPOs have been detected with Swift/XRT~\citep{2023ATel16215....1P}, NICER~\citep{2023ATel16219....1D}, AstroSat/LAXPC~\citep{2023ATel16235....1K}, INTEGRAL~\citep{2024MNRAS.531.4893M} and \textit{Insight}-HXMT~\citep{2024MNRAS.529.4624Y}.

Based on the Insight-HXMT light-curve and hardness-intensity diagram of the source, \citet{2024MNRAS.529.4624Y} divided the observations in which prominent type-C QPOs~\citep{2004A&A...426..587C} were detected into two sub-states: the Normal State and the Flare State.
In the Normal State, the QPO frequency first increases rapidly and then remains more or less constant at 1~Hz, whereas in the Flare State the QPO frequency varies with the flares~\citep[see Figure 4 in][]{2024MNRAS.529.4624Y}.

In this paper, we will use the same \textit{Insight}-HXMT observations in \cite{2024MNRAS.529.4624Y} and focus on the Normal State, in which we find an interesting dip-like feature in the real part of the cross spectrum.
The paper is organized as follows: In Section~\ref{sec:OBSERVATION AND DATA REDUCTION} we describe the data reduction and analysis with \textit{Insight}-HXMT. In Section~\ref{sec:results} and \ref{sec:dipdip} we show the results from the timing analysis, and in Section~\ref{sec:discussion} we discuss our results.

\section{Observations and data reduction}
\label{sec:OBSERVATION AND DATA REDUCTION}

Following the detection of Swift J1727.8–1613 by MAXI/GSC and Swift/BAT, \textit{Insight}-HXMT observations started on August 25, 2023 (MJD 60181) and ended on October 6, 2023 (MJD 60222).
Launched on 2017 June 15, \textit{Insight}-HXMT~\citep{2020SCPMA..6349502Z} is the first Chinese X-ray observatory. \textit{Insight}-HXMT carries three instruments: the Low Energy (LE; 1--15 keV), the Medium Energy (ME; 5--30 keV), and the High Energy (HE; 20--250 keV) telescopes.
In this paper we use the same observations of Swift J1727.8–1613 reported by~\cite{2024MNRAS.529.4624Y}.

We use the \textit{Insight}-HXMT data analysis software \textit{Insight}-HXMTDAS (version 2.06) and the latest CALDB files (version 2.07) to process the data, using only data from the small-FOV detectors. We create good time intervals (GTI) with the following selection criteria: pointing offset angle $< 0.04^{\circ}$, elevation angle $> 10^{\circ}$, geomagnetic cutoff rigidity $> 8$~GV, and the satellite being at least $300$~s away from the crossing of the South Atlantic Anomaly (SAA). 
We estimate the background with the {\sc lebkgmap, mebkgmap}, and {\sc hebkgmap} tasks in \textit{Insight}-HXMTDAS.

We use GHATS\footnote{http://astrosat-ssc.iucaa.in/uploads/ghats\_home.html} to compute the Fast Fourier Transform to produce PDS, the real and imaginary parts of the CS, the phase lag spectrum and the intrinsic coherence function for each individual observation. 
To compute the PDS and CS, we select time segments of $\sim$ 64~s and a time resolution of 1 ms (Nyquist frequency of 500~Hz), except for the analysis in Section~\ref{sec:Energy dependent of the dip} where we use a time resolution of 3~ms (Nyquist frequency of 166.7~Hz) and segments of 49.152~s.
We compute PDS, real and imaginary parts of the CS in each segment, and then average those to get a PDS, real and imaginary parts of the CS per observation. 
We rebin the averaged PDS and real and imaginary parts of the CS in frequency by a factor $\approx 1.047 = 10^{1/50}$ or $\approx 1.023 = 10^{1/100}$ (see details in each section) to increase the signal-to-noise ratio further.
Finally, we use the averaged and rebinned PDS and real and imaginary parts of the CS to compute the phase lag spectrum and the coherence function for each observation.
The PDS and the real and imaginary parts of the CS are normalized to fractional rms-squared units~\citep{2002ApJ...572..392B}. The Poisson noise is subtracted and the background is considered when we compute the rms normalization.
To increase the signal-to-noise ratio to satisfy our studies, we combine PDS and CS with similar properties (see Table~\ref{tab:group}).

We use Xspec~v.12.14.0~\citep{1996ASPC..101...17A} to fit the PDS and the real and imaginary parts of the CS.
To improve the stability of fitting procedures, we rotate the cross spectra counterclockwise by 45$^\circ$~\citep{2024MNRAS.527.9405M} such that cross spectra with a near-zero phase lag would end up having more or less equal real and imaginary parts.
We always fit the PDS and the real and imaginary parts of the rotated CS simultaneously, with the method proposed by~\cite{2024MNRAS.527.9405M}.
We consider that a Lorentzian is significantly needed if the normalization of that Lorentzian divided by its 1-$\sigma$ error is larger than 3 in either the PDS or the CS.
All errors represent the 68\% confidence range for a single parameter unless otherwise stated.

\section{General analysis and a dip-like feature in the real part of the cross spectrum}
\label{sec:results}

\subsection{HID and power spectrum}

\begin{figure}
	\includegraphics[width=\columnwidth]{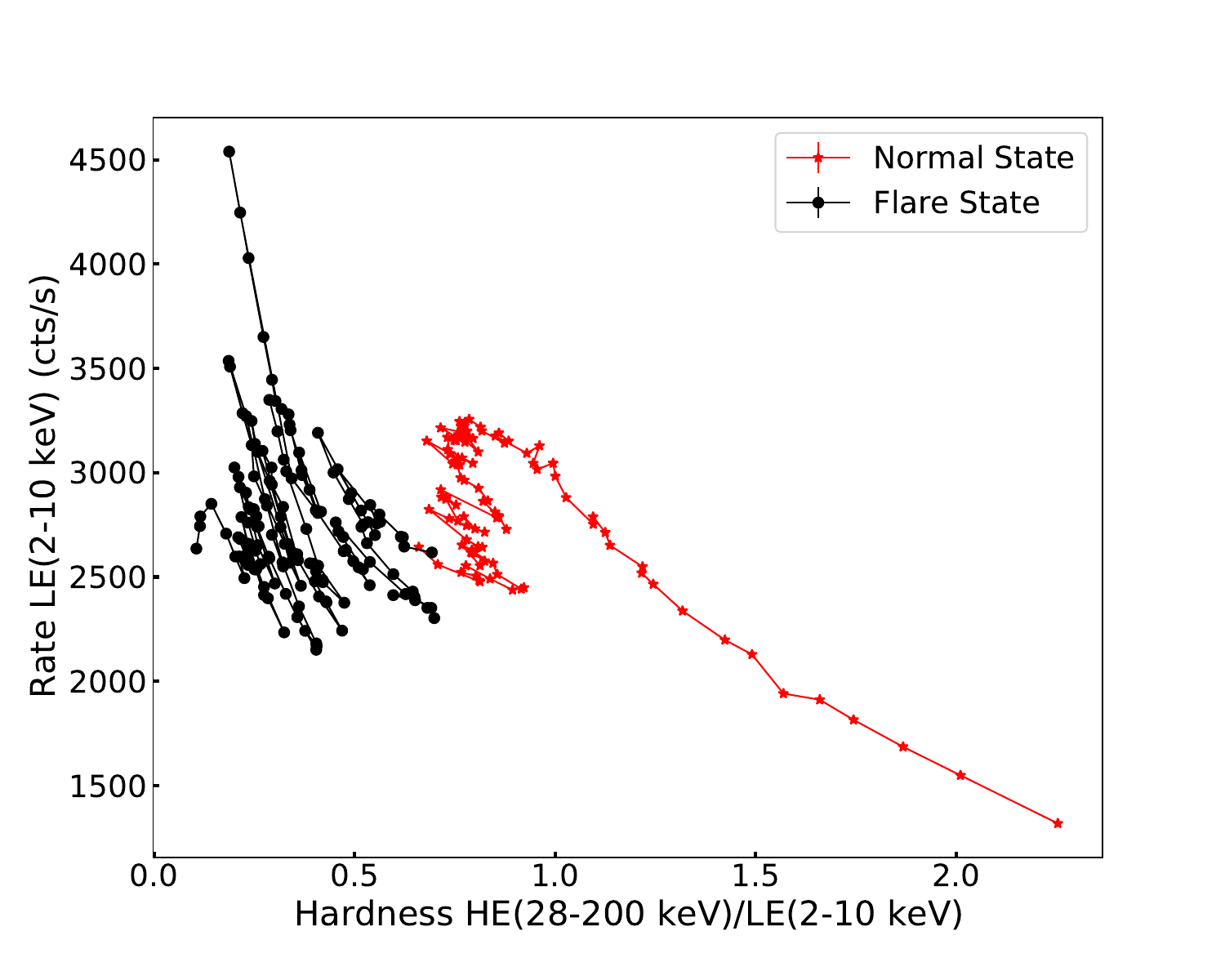}
    \caption{Hardness intensity diagram of the 2023 outburst of Swift~J1727.8$-$1613, observed with \textit{Insight}-HXMT. The red asterisks are observations in the Normal State defined by~\citet{2024MNRAS.529.4624Y}. The black circles are observations in the Flare State. This paper focuses on the observations in the Normal State.}    
    \label{fig:hid}
\end{figure}

\begin{figure*}
	\includegraphics[width=\textwidth]{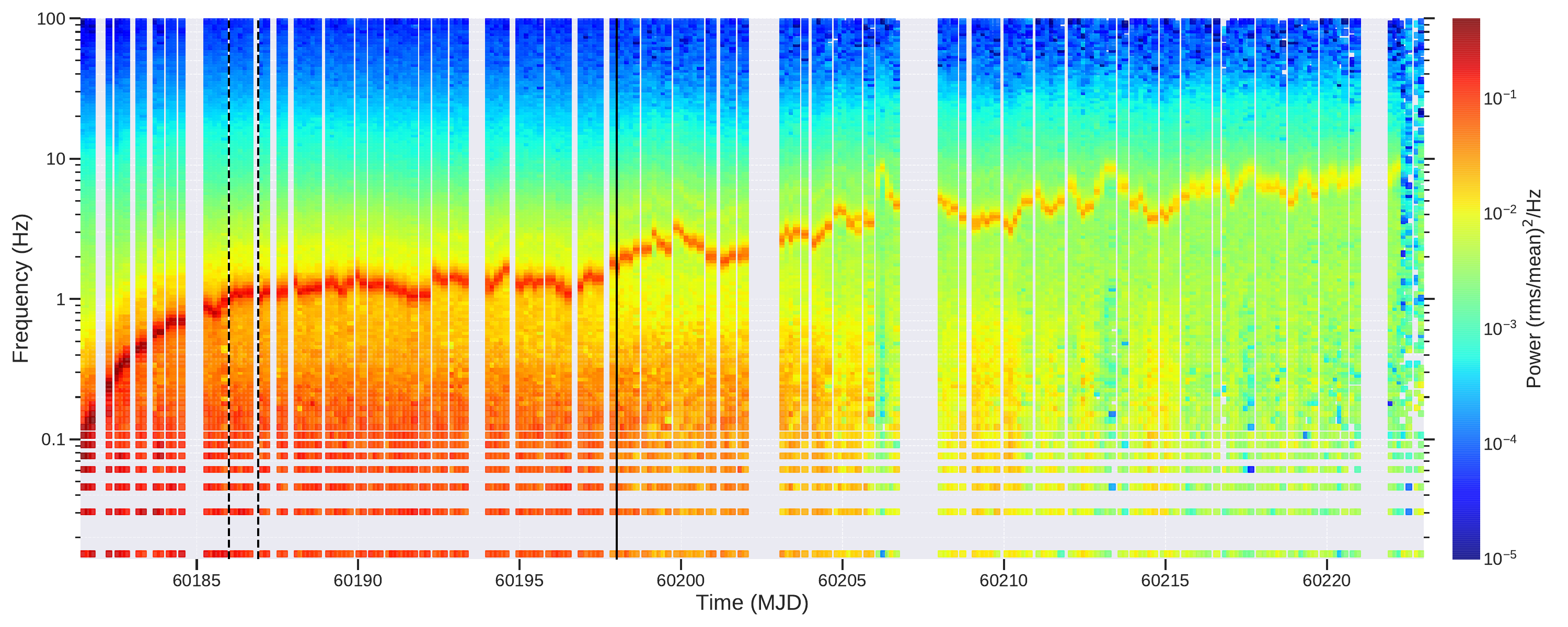}
    \caption{Dynamical power density spectrum in the HE 28$-$200~keV band of Swift~J1727.8$-$1613 throughout the \textit{Insight}-HXMT observation period. The color scale represents the Poisson-noise-subtracted power in fractional rms-squared units. The black solid vertical line marks the transition from the Normal State to the Flare State (MJD=60198). The dotted vertical lines indicate the time interval of the observation shown in Fig.~\ref{fig:pds_cs},~\ref{fig:plot_plcon_time}, \ref{fig:plot_plgau_time} and \ref{fig:plot_tlag_time}.}  
    \label{fig:pds_evolution}
\end{figure*}

\begin{figure}
	\includegraphics[width=\columnwidth]{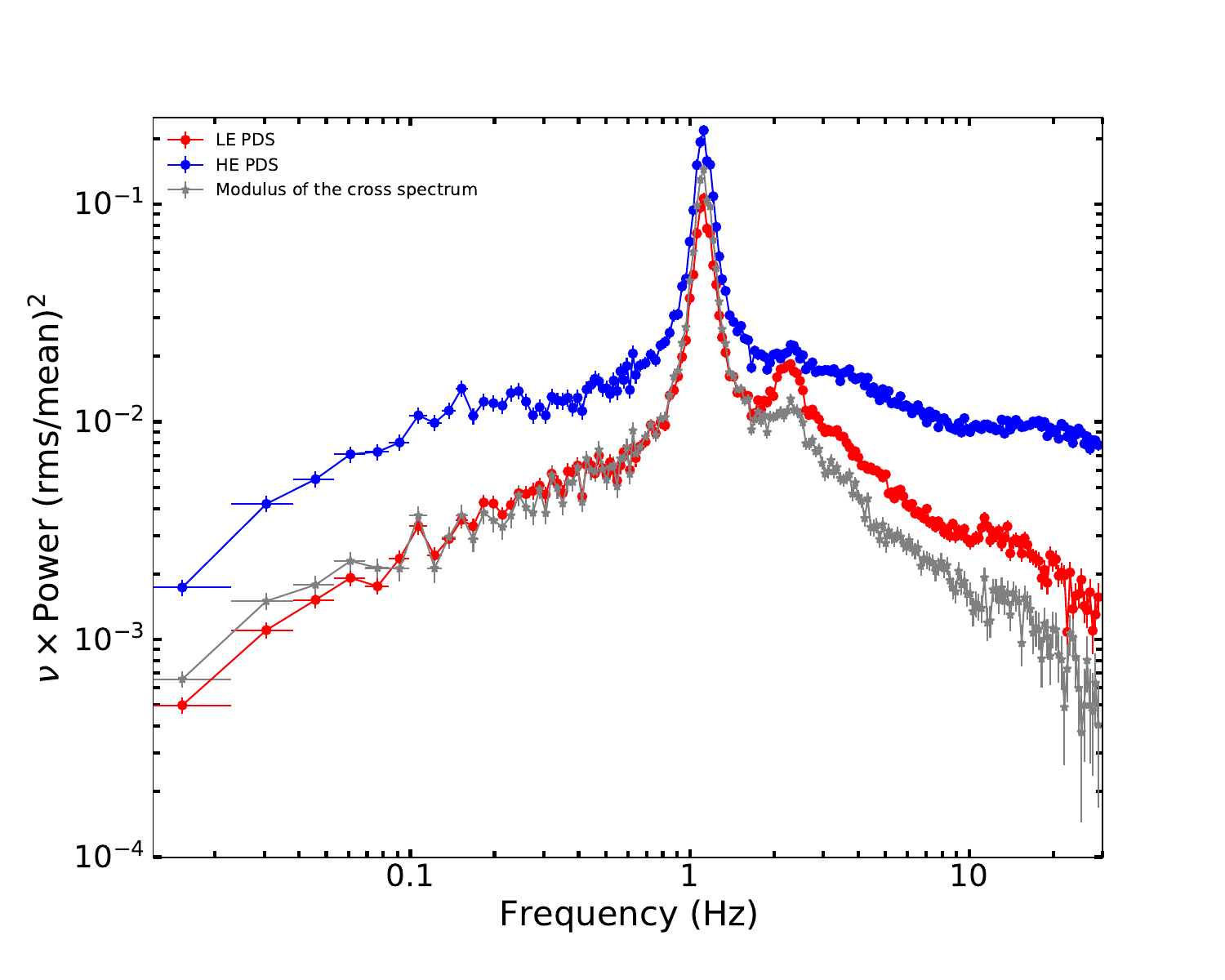}
	\vspace{-5mm}
    \caption{Representative power spectra and cross spectrum of Swift~J1727.8$-$1613 in the Normal State. The LE 2$-$10 keV PDS (red), the HE 28$-$200 keV PDS (blue), and the modulus of the cross spectrum of the HE data with respect to the LE data (gray), corresponding to the observations between the black dotted vertical lines in Fig.\ref{fig:pds_evolution}.} 
    \label{fig:pds_cs}
\end{figure}

\begin{figure}
	\includegraphics[width=\columnwidth]{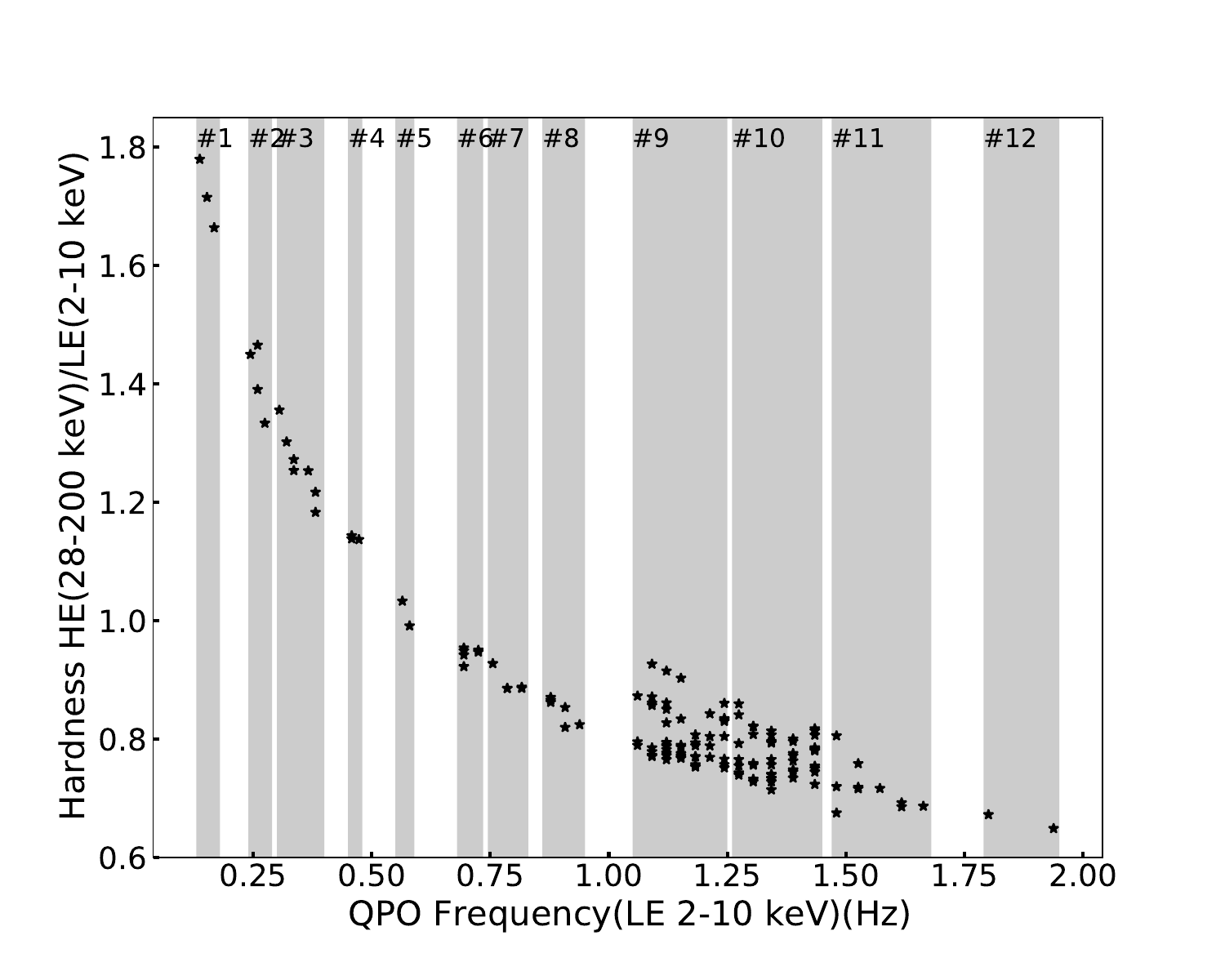}
	\vspace{-5mm}
    \caption{Hardness versus QPO frequency for the observations in the Normal State of Swift J1727.8$-$1613. The observations are combined into 12 groups, shown by the shadowed regions (see Table~\ref{tab:group} for the detailed information of the groups).}  
    \label{fig:fhd}
\end{figure}

In Fig.~\ref{fig:hid} we present the hardness intensity diagram of the 2023 outburst of Swift J1727.8$-$1613 observed with \textit{Insight}-HXMT.
The hardness ratio is defined as the ratio of the background-subtracted count rate in the HE 28$-$200~keV and the LE 2$-$10~keV bands, and the intensity is the background-subtracted count rate in the LE 2$-$10~keV band.

Following \citet[][]{2024MNRAS.529.4624Y}, we divided the \textit{Insight}-HXMT observations of Swift J1727.8$-$1613 into the Normal State (red asterisks) and the Flare State (black circles) in Fig.~\ref{fig:hid}.
From the broadband fractional-rms amplitude in Figure 2 of~\citet{2024MNRAS.529.4624Y}, we conclude that the so-called Normal State in Swift J1727.8$-$1613 corresponds to the LHS and the start of the HIMS~\citep{2005AIPC..797..197B, 2011MNRAS.410..679M}.
The trace of the source in the Normal State looks like the right upper branch of a "q" as seen in other sources~~\citep[e.g.][]{2004MNRAS.355.1105F,2005Ap&SS.300..107H,2011BASI...39..409B}, whereas in the Flare State the intensity varies as the hardness ratio decreases, so the trajectory of the source in the HID shows a sawtooth shape.

In Fig.~\ref{fig:pds_evolution} we illustrate the time evolution of the PDS in the HE 28$-$200~keV band throughout the \textit{Insight}-HXMT observation period. A low-frequency QPO is prominently visible across all observations. The black vertical solid line in the figure marks the transition from the Normal State to the Flare State. Initially, the frequency of the QPO rises rapidly, then levels off at around $1.0-1.5$~Hz during the Normal State. However, in the Flare State the QPO frequency varies, increasing to approximately 8~Hz by the end of the observations.

In Fig.~\ref{fig:pds_cs} we present the LE 2-10 keV PDS, the HE 28-200 keV PDS and the modulus of the cross spectrum of the HE data with respect to the LE data for a representative observation in the Normal State extending between MJD 60186 and 60187; the time interval of this observation is marked with black dotted vertical lines in Fig.~\ref{fig:pds_evolution}. 
The strongest components in the LE and HE PDS are a QPO at $\sim1$~Hz and its harmonic at $\sim2$ Hz.
The total variability of the HE 28$-$200 keV data is always higher than that of the LE 2$-$10 keV data, with the largest difference at low and high frequencies. The modulus of the cross spectrum presents a similar shape as the LE PDS, albeit with lower values at high frequencies.

We plot the hardness ratio versus the QPO frequency for the observations in the Normal State in Fig.~\ref{fig:fhd}.
Each point represents an observation from an individual good time interval, in which we discard the good time intervals that are shorter than 320 s (5$\times$64 s).
The QPO frequency is defined as the frequency at which the ``Power$\times$Frequency'' is maximum~\citep{2002ApJ...572..392B}, where ``Power'' is the power of the Poisson-noise-subtracted PDS in the LE 2$-$10 keV band (also in the HE 28$-$200 keV band) in fractional rms-squared units.
The hardness ratio and QPO frequency show an anti-correlation: the hardness ratio decreases from $\sim$~1.8 to $\sim$~0.6 as the QPO frequency increases from 0.13 Hz to 1.93 Hz. This is seen also in other sources~\citep[e.g.][]{2022NatAs...6..577M, 2022MNRAS.514.2891Z, 2022MNRAS.513.4196G, 2024MNRAS.527.7136B}. We observe that as the QPO frequency exceeds $\sim1.0$ Hz the relation between QPO frequency and hardness broadens, such that at a given QPO frequency the hardness can change by a factor of up to $\sim1.3$. A similar, but more extreme, behavior is seen in the case of GRS~1915$+$105~\citep{2022NatAs...6..577M}, and to a lesser extent in GX~339$-$4~\citep{2024MNRAS.527.5638Z}.

\begin{table}
	\centering
	\caption{The QPO frequency range and number of segments averaged for each group in Fig.~\ref{fig:fhd}.}
	\label{tab:group}
	\begin{tabular}{lll}
        \hline
        Group index & QPO frequency (Hz) & Number of segments \\
        \hline
        \#1 & 0.13$-$0.16 & 35 \\
        \#2 & 0.24$-$0.27 & 48 \\
        \#3 & 0.30$-$0.38 & 57 \\
        \#4 & 0.45$-$0.47 & 49 \\
        \#5 & 0.56$-$0.57 & 11 \\
        \#6 & 0.69$-$0.72 & 51 \\
        \#7 & 0.75$-$0.81 & 54 \\
        \#8 & 0.87$-$0.93 & 50 \\        
        \#9 & 1.06$-$1.24 & 547 \\
        \#10 & 1.27$-$1.43 & 490 \\
        \#11 & 1.48$-$1.66 & 109 \\
        \#12 & 1.80$-$1.93 & 13 \\
        \hline
	\end{tabular}
\end{table}

To obtain data with good-enough statistics, we combine the observations into 12 groups, marked with gray shaded area in Fig.~\ref{fig:fhd}.
We check that the PDS and the CS of the observations in the same group are consistent with each other.
The definition of the groups does not take into account the hardness ratio, because observations with close QPO frequencies but different hardness ratios have PDS and CS that are consistent with each other. 
For each group, we generate a separate LE 2$-$10 keV PDS, a HE 28$-$200 keV PDS and a CS of photons in the HE 28$-$200 keV band with respect to those in the LE 2$-$10 keV band.
We rebin the averaged PDS and CS of a group logarithmically in frequency by a factor $\approx 1.047 (= 10^{1/50})$ to increase the signal-to-noise ratio further, except for Group~\#9, \#10 and \#11 that have enough number of PDS and CS to reach a good signal-to-noise ratio with a rebin factor $\approx 1.023 (= 10^{1/100})$. In Table~\ref{tab:group} we record the frequency range of the QPO and the number of segments of each group.

\subsection{Dip-like feature in the real part of the cross spectrum}
\label{sec:dipdip}

\begin{figure*}
    \centering
	\vspace{-5mm}
	\includegraphics[width=\columnwidth]{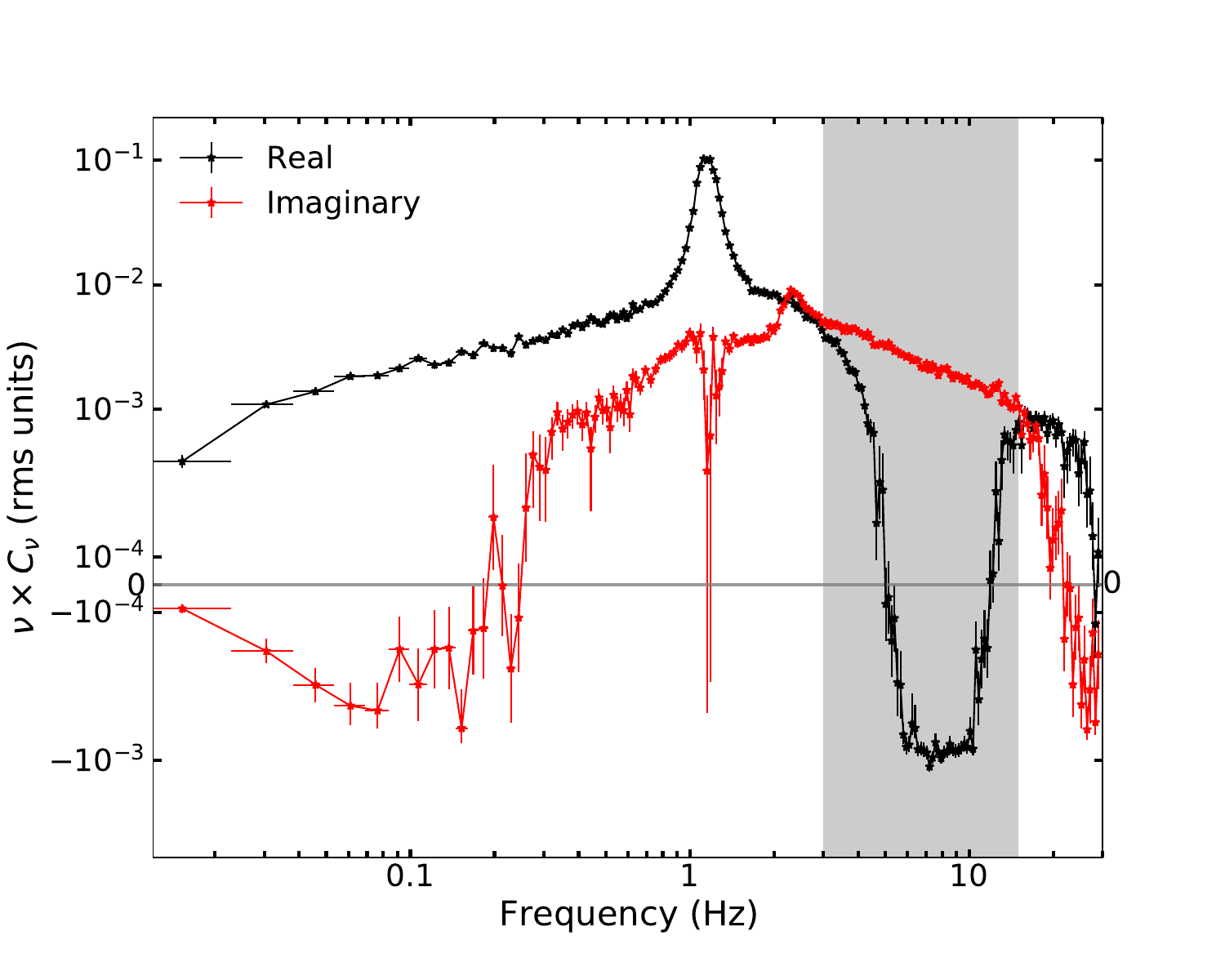}\vspace{-5mm}
    \includegraphics[width=\columnwidth]{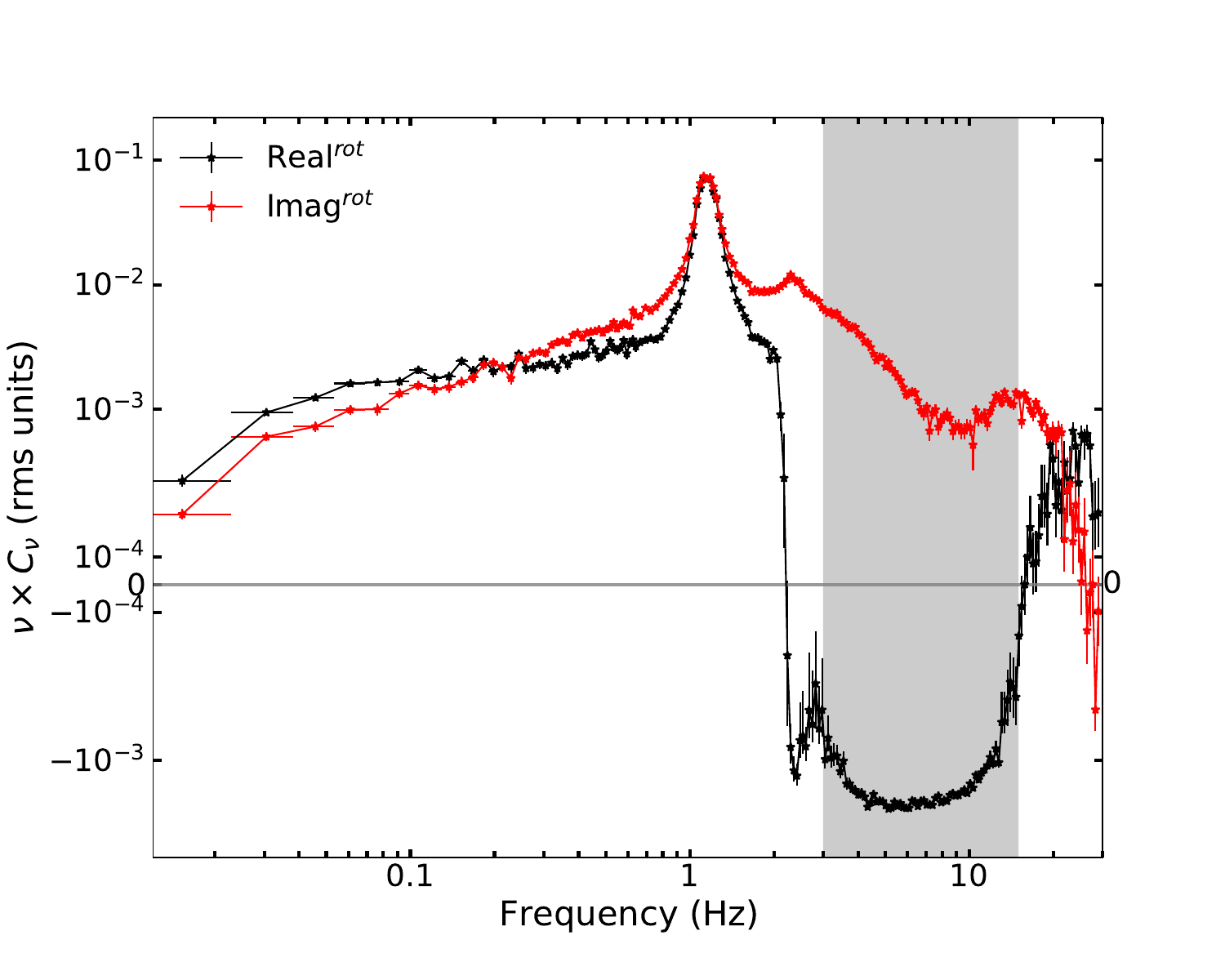}\vspace{-5mm}
	\includegraphics[width=\columnwidth]{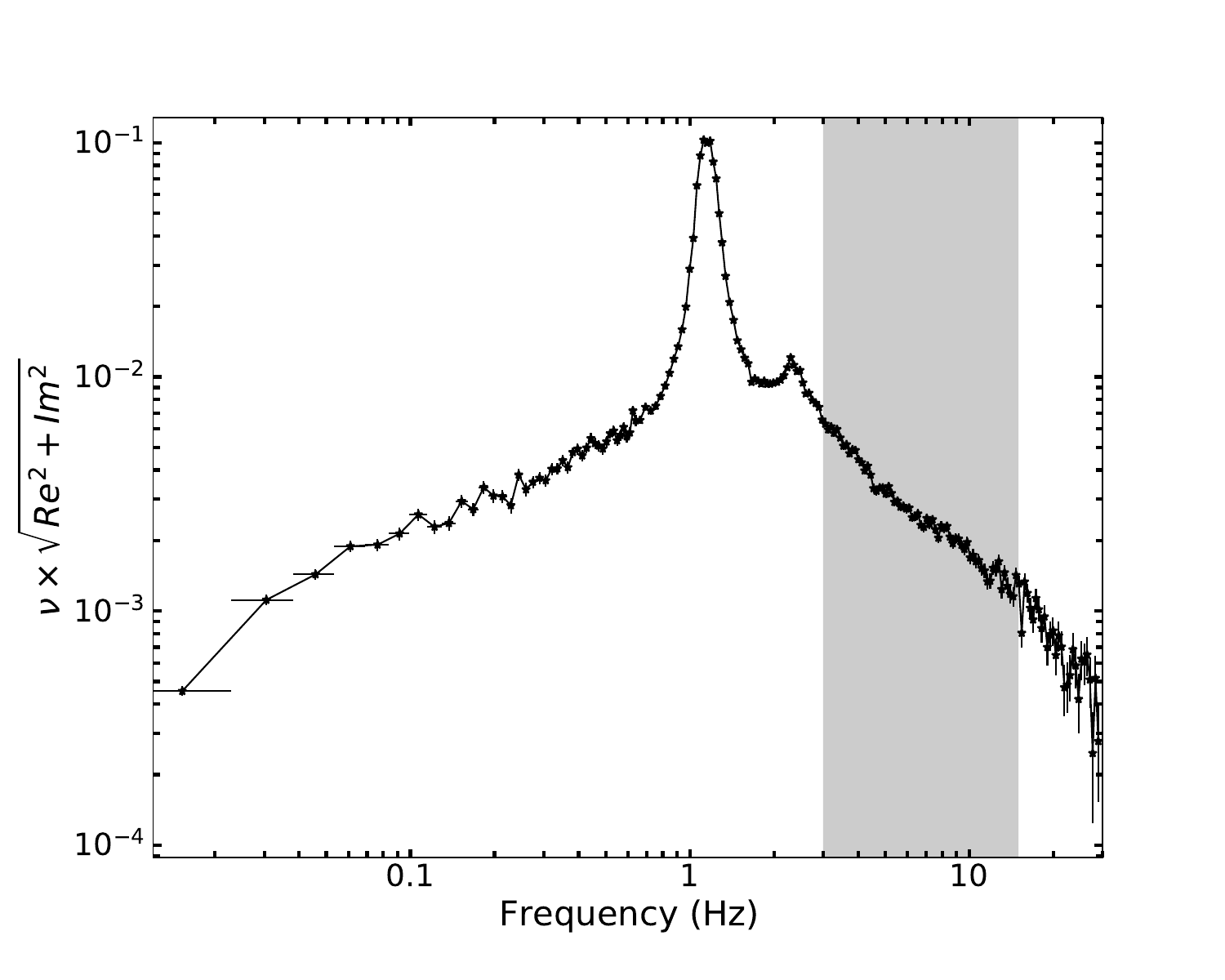}
    \includegraphics[width=\columnwidth]{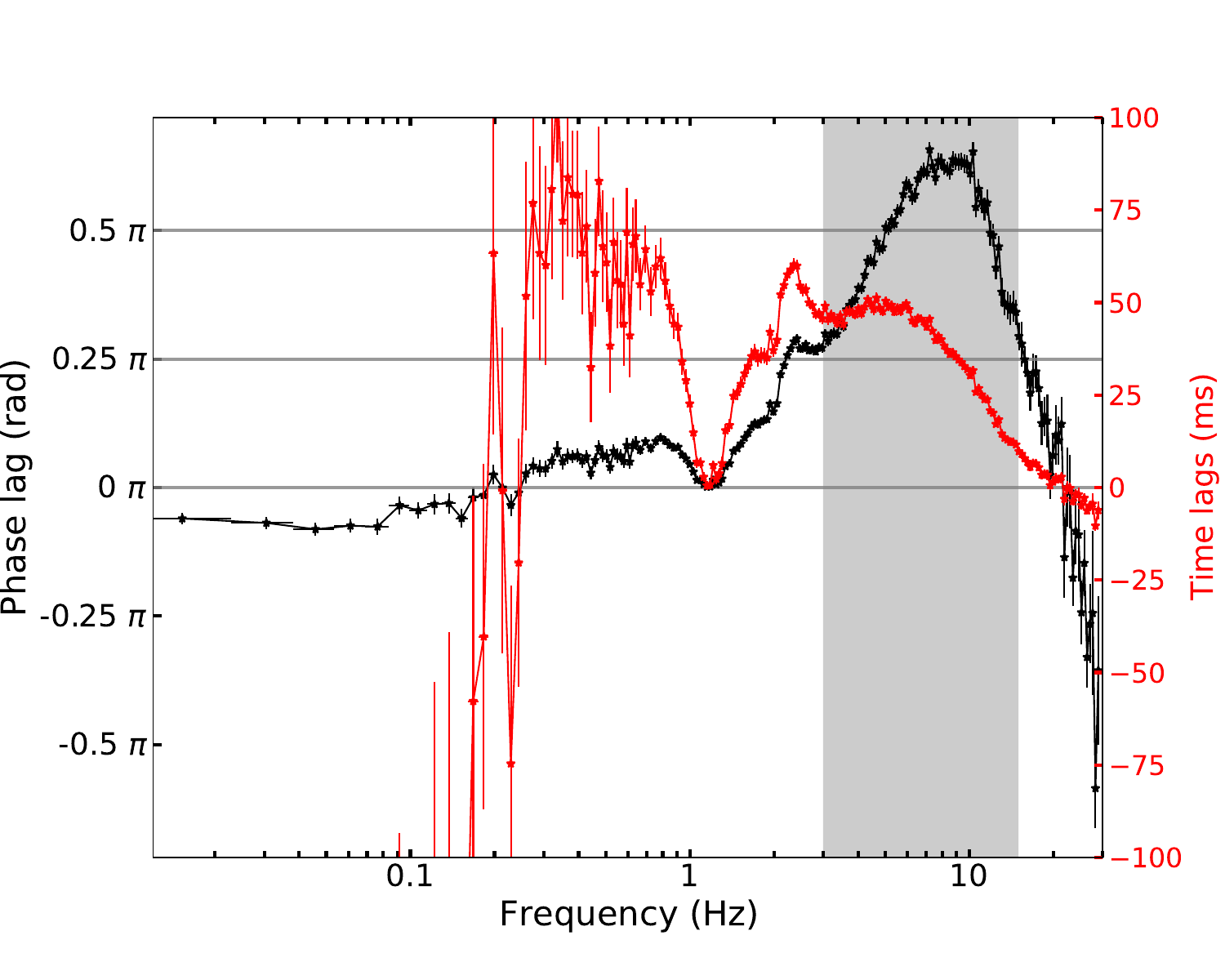}
 	\vspace{-5mm}
    \caption{Top Left: Real and imaginary parts of the cross spectrum of Swift~J1727.8$-$1613 for Group~\#9 in Fig.~\ref{fig:fhd}.
    Top Right: Real and imaginary parts of the cross spectrum rotated counterclockwise by 45$^\circ$.
    Bottom Left: Modulus of the cross spectrum.
    Bottom Right: Phase and time lags. For a clear presentation of the real and imaginary parts of the cross spectrum, data in the cross spectrum with absolute values greater than or equal to $5\times10^{-4}$ are scaled logarithmically, while values with absolute values smaller than  $5\times10^{-4}$ are scaled linearly.}
    \label{fig:fig_1.06_1.27_real_imag}
\end{figure*}

In the Normal State we detect a prominent dip-like feature at $\sim$3.0$-$15.0 Hz in the real part of the CS of photons in the HE 28$-$200 keV band with respect to those in the LE 2$-$10 keV band. 
A representative observation is shown in Fig.~\ref{fig:fig_1.06_1.27_real_imag} in which we plot the real and imaginary parts of the CS of Group~\#9 in the top left panel. 
In Group~\#9 we average 547 individual segments to compute the power and cross spectra.
To clearly display the high-frequency part of the data, we plot $\nu P(\nu)$ in the $y$ axis.
The main features in the real part are a significant peak at the QPO frequency and a dip at $\sim$3.0$-$15.0 Hz (grey shaded region in Fig.~\ref{fig:fig_1.06_1.27_real_imag}) that reaches negative values near the minimum. 
At the beginning and end  of the dip, the real and imaginary parts have similar values, indicating a phase lag of $\sim\pi/4$. At the dip the real part decreases rapidly and then increases, while the imaginary part decreases smoothly.
Since at the frequencies of the dip in the real part of the CS the imaginary part is always positive, negative values of the real part mean a phase lag between $\pi/2$ and $\pi$.

Similar to~\cite{2024MNRAS.527.9405M}, in the top right panel of Fig.~\ref{fig:fig_1.06_1.27_real_imag} we plot the real and imaginary parts of the cross spectrum rotated counterclockwise by 45$^\circ$, $2^{-1/2}$(Re $-$ Im) and $2^{-1/2}$(Re $+$ Im), respectively. The dip occurs when the real part begins to become lower than the imaginary part. Since for the rotated cross spectrum a real part equal to zero means a phase lag of $\pi/4$, the dip corresponds to the frequency range where the real part is negative for the rotated cross spectrum.
As shown in the bottom right panel of Fig.~\ref{fig:fig_1.06_1.27_real_imag}, the dip in the real part of  the CS leads to a hump in the phase-lag spectrum, with a phase lag larger than $\pi/2$.
A similar hump, together with a strong narrow peak at the frequency of the second harmonic of the QPO, dominates the time lags, which reach the maximum value at slightly lower frequencies than those in the phase-lag spectrum.

We also plot the modulus of the cross spectrum in the bottom left panel, which exhibits a smooth decline in the $\sim$3.0$-$15.0 Hz range. Therefore, the dip in the real part is dominated by changes of the phase lags instead of changes of the modulus of the cross spectrum.

The real part of the rotated CS also exhibits a sharp negative peak at the frequency of the second harmonic of the QPO, while the imaginary part of the rotated CS shows a positive peak. This indicates a phase lag greater than $\pi/4$ at the frequency of the second harmonic of the QPO, as shown in the bottom right panel of Fig.~\ref{fig:fig_1.06_1.27_real_imag}.

\section{Joint-fit of the power and cross spectra}

\subsection{Joint-fit for data collected during a continuous monitoring interval}
\label{sec:Joint-fit of power and cross spectra}

In this section we analyze the representative observation marked with black dotted vertical lines in Fig.~\ref{fig:pds_evolution}. 
This observation includes 138 segments of 64 s.
We rebin the averaged PDS and real and imaginary parts of the CS in frequency by a factor $\approx 1.023 (= 10^{1/100})$.
We then fit jointly the PDS and the CS for this observation with the method proposed by~\cite{2024MNRAS.527.9405M}.
We fit simultaneously the LE 2$-$10 keV PDS and the HE 28$-$200 keV PDS, as well as the real and imaginary parts of the CS (HE 28$-$200 keV with respect to LE 2$-$10 keV).
We use the same number of Lorentzians for the LE 2$-$10 keV PDS and the HE 28$-$200 keV PDS and link the frequency and full width at half maximum (FWHM) of each Lorentzian across the different spectra. For each Lorentzian in the PDS, we include the corresponding Lorentzian to the real and imaginary parts of the CS:
\begin{equation}
\begin{aligned}
& \mathrm{Re} = \sum_i C_i\;L(\nu;\nu_{0,i},\Delta_i) \cos{[g_i(\nu;p_j)]}\\
& \mathrm{Im} = \sum_i C_i\;L(\nu;\nu_{0,i},\Delta_i) \sin{[g_i(\nu;p_j)]},\\
\end{aligned}
\label{eq1}
\end{equation}
where $L(\nu;\nu_{0,i},\Delta_i)$ are the same Lorentzians we fit to the PDS, with the same frequency and FWHM, and the $g_i(\nu;p_j)$ are functions of frequency with $j$ parameters $p_j$ that represent the lag spectrum of the each Lorentzian.
In this paper, we explore three assumptions for these functions $g_i(\nu;p_j)$:
(1) The phase lags are constant, 
$g_i(\nu;p_j) = 2\pi\;k_i$, where the quantity $2\pi k_i$ describes the frequency-independent phase-lags of the $i^{th}$ Lorentzian.
(2) The time lags are constant,
$g_i(\nu;p_j) = 2\pi\;k_i\nu$, where the quantity $k_i$ are the frequency-independent time-lags of the $i^{th}$ Lorentzian.
(3) The phase lags reach a maximum or minimum value at the Lorentzian centroid frequency, so appearing as a peak. Following \cite{2022MNRAS.515.1914Z},  we use a Gaussian function to describe the frequency dependence of the phase lags (so the name Gaussian phase-lag model),
\begin{equation}
\begin{aligned}
& g_i(\nu;p_j) = 2\pi\;k_i\;e^{-\frac{1}{2} ( \frac{\nu-{\nu_{0,i}}}{\sigma_i} )^2},\\
\end{aligned}
\label{eq4}
\end{equation}
where $v_{0,i}$ is the same centroid frequency of each Lorentzian, and we take $\sigma_i = \Delta_i$. In all three cases, for each Lorentzian, the $k_i$ are free parameters of the model that need to be fitted. We do not fit the phase-lag frequency spectrum or the intrinsic coherence function, as they are not independent.
If we get a good fit of the PDS and the real and imaginary parts of the CS, instead we verify that the predicted model of the phase lags and the intrinsic coherence ~\citep[Equation 9 and 10 in][]{2024MNRAS.527.9405M}, are consistent with the data.

\subsubsection{Constant phase-lag model}
\label{sec:Constant phase-lag model}

\begin{figure*}
    \centering
    \vspace{-5mm}
  	\includegraphics[width=\columnwidth]{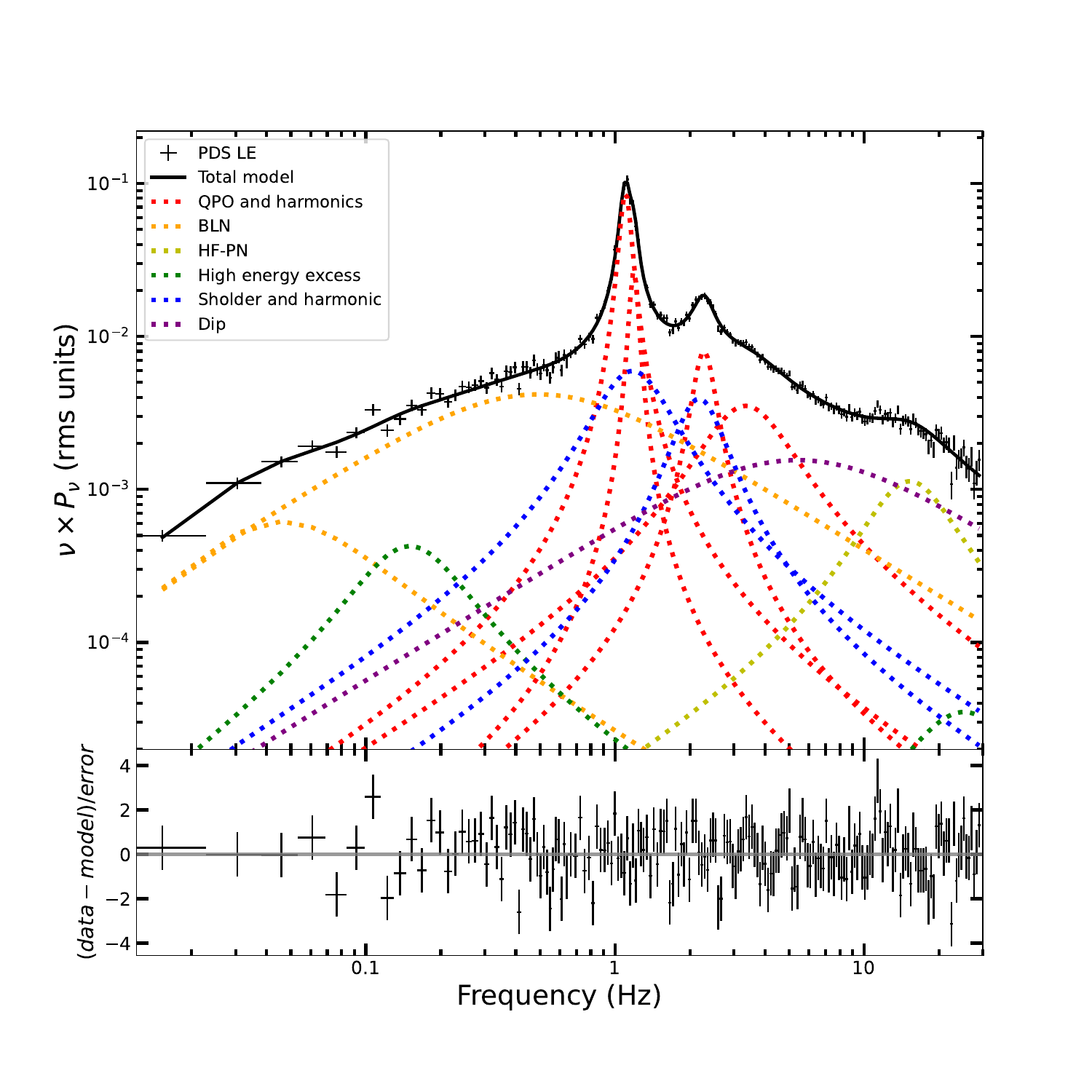}\vspace{-5mm}
    \includegraphics[width=\columnwidth]{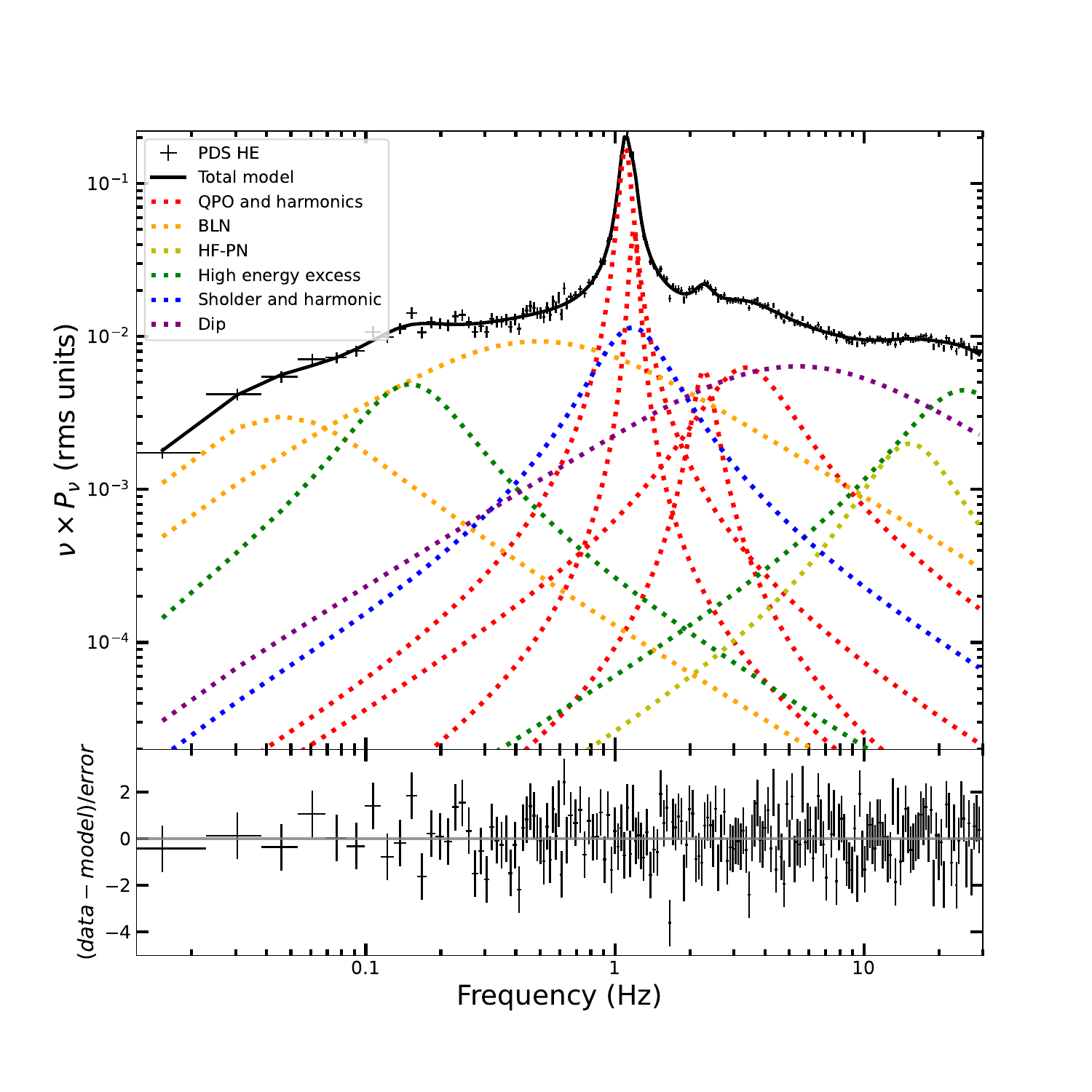}\vspace{-5mm}
	\includegraphics[width=\columnwidth]{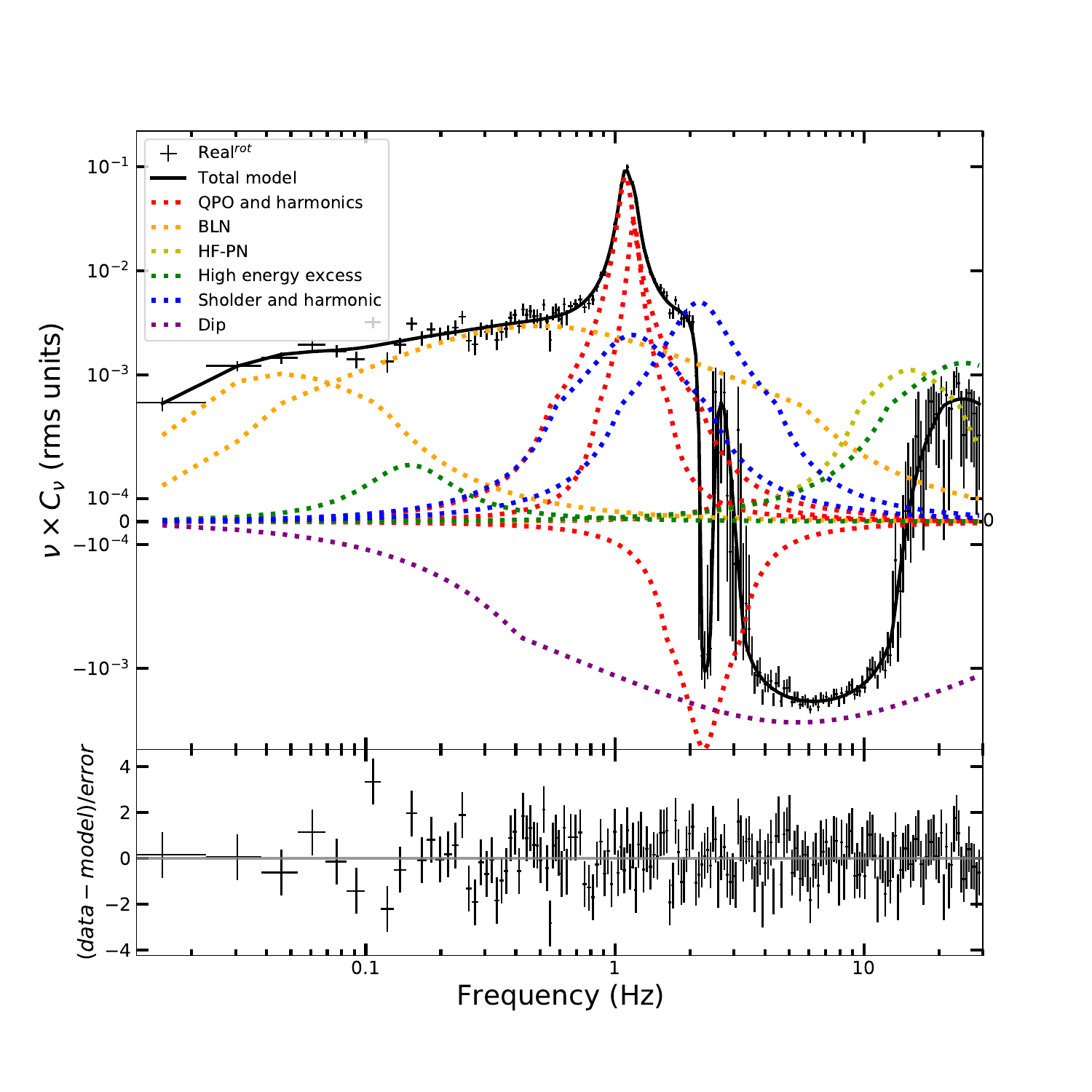}
 	\includegraphics[width=\columnwidth]{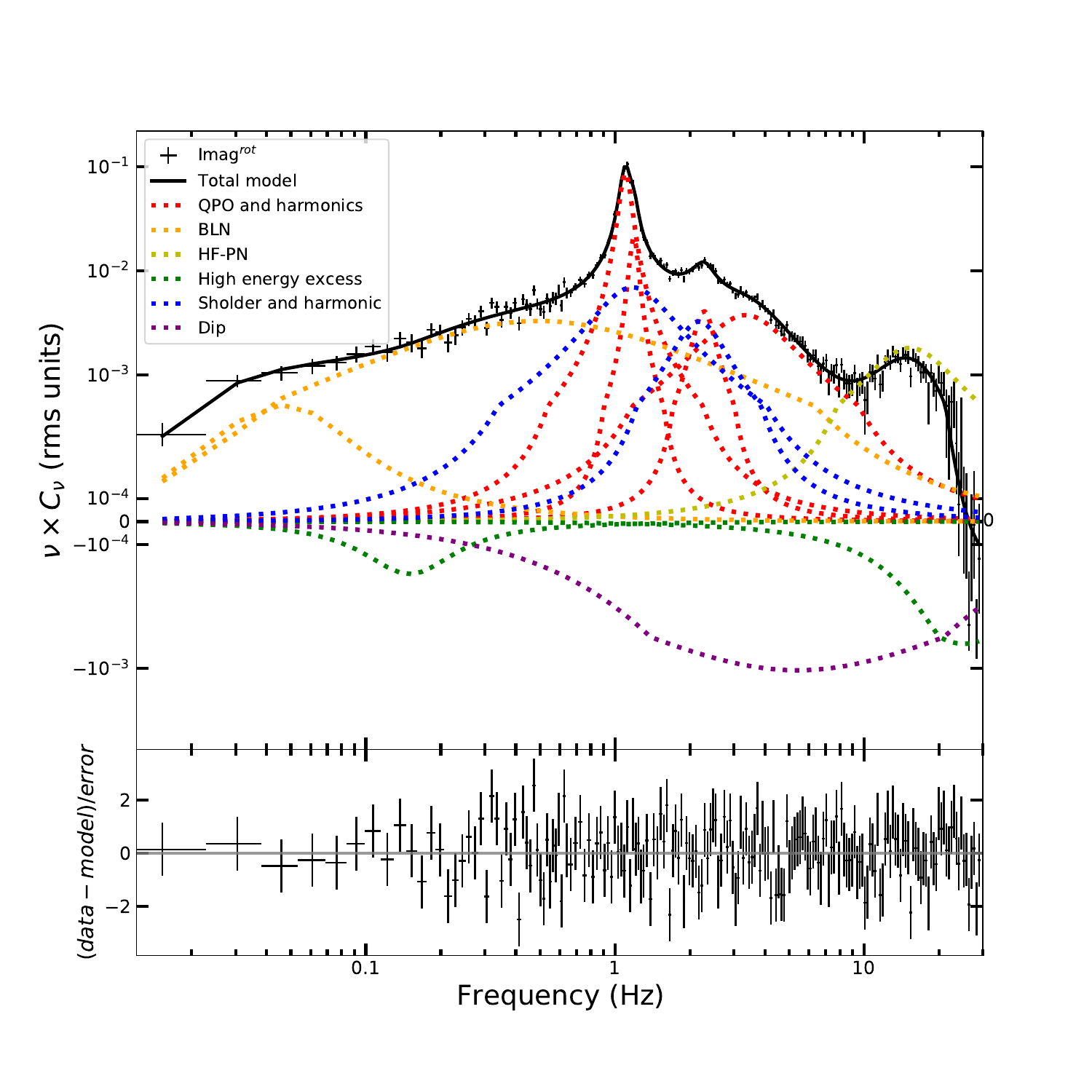}
    \vspace{-5mm}
    \caption{LE 2$-$10 keV (upper left panel), HE 28$-$200 keV PDS (upper right panel), the real (lower left panel) and imaginary (lower right panel) parts of the rotated CS of Swift~J1727.8$-$1613 for the data shown in Fig.~\ref{fig:pds_cs}, fitted with a model (solid lines) consisting of 12 Lorentzians. Individual components are plotted with colored dotted lines.
    The fits were done assuming the constant phase-lag model. For a clear presentation of the real and imaginary parts of the cross spectrum, data with absolute values greater than or equal to $5\times10^{-4}$ are scaled logarithmically, while values with absolute values smaller than  $5\times10^{-4}$ are scaled linearly.}
    \label{fig:plot_plcon_time}
\end{figure*}

\begin{table*}
	\centering
	\caption{The parameters of the Lorentzians of the data shown in Fig.~\ref{fig:pds_cs} obtained with three different lag models.}
    \label{tab:parameters_time}
    \scalebox{0.82}{
    \setlength{\tabcolsep}{0.7mm}{ 
    \renewcommand\arraystretch{1.5}
	\begin{tabular}{l@{\hspace{0mm}}lllllllll}
        \hline
        & \multicolumn{3}{c}{Constant phase lag} & \multicolumn{3}{c}{Gaussian phase lag} & \multicolumn{3}{c}{Constant time lag} \\
        &  Frequency (Hz) & FWHM (Hz ) & Phase lag (rad)& Frequency (Hz) & FWHM (Hz ) & Phase lag (rad)$^{\dagger}$ & Frequency (Hz) & FWHM (Hz ) & Time lag (ms)\\
        \hline
         QPO and harmonics  & & & & & & & & &\\
         QPO  & $1.101^{+0.002}_{-0.003}$& $0.12 \pm 0.01$& $0.03 \pm 0.03$& $1.101 \pm 0.003$& $0.12^{+0.006}_{-0.003}$& $0.02 \pm 0.03$& $1.101^{+0.003}_{-0.002}$& $0.123^{+0.005}_{-0.004}$& $5 \pm 4$\\
         & $1.189^{+0.006}_{-0.005}$& $0.1 \pm 0.01$& $-0.2 \pm 0.1$& $1.188 \pm 0.006$& $0.11 \pm 0.01$& $-0.15 \pm 0.07$& $1.189^{+0.006}_{-0.003}$& $0.11 \pm 0.01$&$-19 \pm 10$\\
          Second harmonic  & $2.26 \pm 0.01$& $0.48 \pm 0.05$& $1.8^{+0.1}_{-0.2}$& $2.27 \pm 0.01$& $0.48^{+0.06}_{-0.04}$& $1.5 \pm 0.1$& $2.26 \pm 0.01$& $0.52^{+0.04}_{-0.03}$& $96 \pm 4$\\
          High-order harmonics  & $3.0 \pm 0.1$& $2.8^{+0.3}_{-0.1}$& $0.76^{+0.08}_{-0.05}$& $2.8^{+0.2}_{-0.1}$& $3.4^{+0.4}_{-0.3}$& $1.0 \pm 0.1$& $3.0 \pm 0.1$& $3.0 \pm 0.2$& $58^{+1}_{-3}$\\
        \hline
        BLN   & & & & & & & & & \\ 
        BLN1  & $0.024^{+0.004}_{-0.005}$& $0.07^{+0.02}_{-0.01}$& $-0.3 \pm 0.1$& $0.024^{+0.003}_{-0.004}$& $0.07^{+0.02}_{-0.01}$& $-0.25^{+0.08}_{-0.07}$& $0.022^{+0.004}_{-0.006}$& $0.09 \pm 0.01$&$-690^{+30}_{-60} $\\
        BLN2  & $<0.2$& $1.0 \pm 0.2$& $0.06^{+0.05}_{-0.04}$& $<0.2$& $0.8^{+0.15}_{-0.04}$& $0.03^{+0.04}_{-0.05}$& $0.12^{+0.03}_{-0.06}$& $0.7^{+0.09}_{-0.08}$& $96^{+4}_{-6}$\\
        \hline
        HF-PN   & $13.9 \pm 0.4$& $12.6^{+0.9}_{-1.6}$& $0.2 \pm 0.1$& $12.8^{+0.4}_{-0.9}$& $14^{+1}_{-2}$& $-0.6^{+0.2}_{-0.3}$& $12.4 \pm 0.3$& $9.5^{+0.7}_{-0.9}$& $9 \pm 2$\\
        \hline
        High-energy excess  & & & & & & & & & \\
        Low-frequency  & $0.13^{+0.01}_{-0.02}$& $0.15^{+0.04}_{-0.03}$& $-1.5^{+1.2}_{-0.9}$$^{\ast}$& $0.12 \pm 0.02$ & $0.15 \pm 0.02$& $5.5^{+0.5}_{-0.6}$$^{\ast}$& $0.13^{+0.01}_{-0.02}$& $0.14^{+0.04}_{-0.03}$& $-6100^{+500}_{-400}$$^{\ast}$\\
        High-frequency  & $21^{+1}_{-3}$& $27^{+6}_{-5}$& $-1.2 \pm 0.1$& $19^{+1}_{-4}$& $35^{+8}_{-6}$& $-2.8^{+0.1}_{-0.2}$& $15 \pm 2$& $33^{+3}_{-2}$& $-10^{+2}_{-1}$$^{\ast}$\\
        \hline
        Shoulder and harmonic  & & & & & & & & & \\
        Shoulder  & $1.1^{+0.03}_{-0.02}$& $0.9^{+0.2}_{-0.1}$& $0.5 \pm 0.1$& $1.05^{+0.04}_{-0.05}$& $1.0^{+0.2}_{-0.1}$& $0.48^{+0.06}_{-0.08}$
& $0.98^{+0.05}_{-0.03}$& $1.2^{+0.2}_{-0.1}$& $121^{+4}_{-5}$\\
        Harmonic  & $2.11^{+0.06}_{-0.05}$& $1.1^{+0.3}_{-0.2}$& $-0.2 \pm 0.1$& $2.1 \pm 0.1$& $0.9^{+0.3}_{-0.1}$& $-0.3^{+0.3}_{-0.2}$& $1.4 \pm 0.1$& $2.6^{+0.2}_{-0.1}$& $-44 \pm 3$\\
        \hline
        Dip & $<2.3$& $11^{+2}_{-1}$& $2.66^{+0.05}_{-0.08}$& $3.7^{+0.7}_{-0.2}$& $10.5^{+0.8}_{-0.7}$& $2.7 \pm 0.1$& $5.3 \pm 0.3$& $7.0^{+0.7}_{-0.6}$& $36 \pm 3$\\  
        \hline
        $\chi^2$/dof. & \multicolumn{3}{c}{793/684} & \multicolumn{3}{c}{791/684} & \multicolumn{3}{c}{784/684} \\
        \hline
        \multicolumn{10}{l}{\small $^{\dagger}$ Phase lag at the centroid frequency of the Lorentzian.}\\
        \multicolumn{10}{l}{\small $^{\ast}$ Lorentzian is less than 3$\sigma$ significant in the CS.}\\
	\end{tabular}}}
\end{table*}

In this section we assume the constant phase-lag model and we rotate the cross spectrum counterclockwise by 45$^\circ$ before the fitting.
Therefore, the models we use are $\sum_{i=1}^n A_iL(\nu;\nu_{0,i},\Delta_i)$ to fit the LE 2$-$10 keV PDS, $\sum_{i=1}^n B_iL(\nu;\nu_{0,i},\Delta_i)$ to fit the HE 28$-$200 keV PDS, $\sum_{i=1}^n C_iL(\nu;\nu_{0,i},\Delta_i)\cos{[2\pi k_i + \pi/4]}$ to fit the real part and $\sum_{i=1}^n C_iL(\nu;\nu_{0,i},\Delta_i)\sin{[2\pi k_i + \pi/4]}$ to fit the imaginary part of the rotated CS. The frequency and FWHM of each Lorentzian for the PDS and CS are linked. 
A fit with 12 Lorentzian ($n = 12$) yields $\chi^2 = $ 793 for 684 degrees of freedom (dof) and all Lorentzians are at least 3 $\sigma$ significant in either the PDS or the CS.
In Fig.~\ref{fig:plot_plcon_time} we plot the fit results: in the upper left panel we plot the LE PDS,  in the upper right panel the HE PDS, in the bottom left panel the real part and in the bottom right panel the imaginary part of the rotated CS.

In Table~\ref{tab:parameters_time} we list the centroid frequencies, FWHM, and phase lags of the Lorentzians, distinguishing them with different colored dashed lines in Fig.~\ref{fig:plot_plcon_time}. Based on their parameters and correlations with the QPO, we divide the Lorentzians into six groups. The first group includes the QPO, the second and the high-order harmonics, which are all plotted with red dotted lines. We use two Lorentzians to model the QPO because the QPO frequency moves slightly in the period of the observation.
The second group is the BLN, including two Lorentzians plotted with orange dotted lines that dominate at frequencies lower than $\sim$1~Hz. 
The third group is the high-frequency peak noise (HF-PN) plotted with yellow line, correlated with the low-frequency QPOs through the PBK correlation, and also reported by \cite{2024MNRAS.529.4624Y}. 
The fourth group includes two Lorentzians that are only significant in the HE band, so we name them high energy excess, plotted with green dotted lines. 
The fifth group contains the shoulder of the QPO and its second harmonic plotted with blue dotted lines. These components exhibit significantly different phase lags compared to those of the fundamental and second harmonics of the QPO. 
The sixth group is the Lorentzian that fits the dip feature at $\sim 3.0-15.0$ Hz, plotted with the purple dotted line.

In the case of the constant phase-lag model, both the real and imaginary parts of the CS have the same Lorentzian shape as the PDS, multiplied by, respectively, the cosine or sine of a parameter. The Lorentzian that fits the dip feature in the real part is broad with a centroid frequency of $<2.3$ Hz, a FWHM of $\sim 11$ Hz and a phase lag of $\sim 2.7$ rad.  As can be seen in Fig.~\ref{fig:plot_plcon_time}, the Lorentzian that fits the dip feature (negative for the real part, and positive for the imaginary part due to the phase lag being greater than $\pi/2$) is very broad, and therefore tends to overestimate the contribution of the dip-like feature to both the real and imaginary parts of the CS at low and high frequencies.  To compensate for this, the model requires other Lorentzians to offset the differences. This situation raises concerns about possible degeneracy of the model. To address this issue, we will introduce the Gaussian phase-lag model in the following sub-subsection.

\subsubsection{Gaussian phase-lag model}

\begin{figure*}
    \centering
	\vspace{-5mm}
  	\includegraphics[width=\columnwidth]{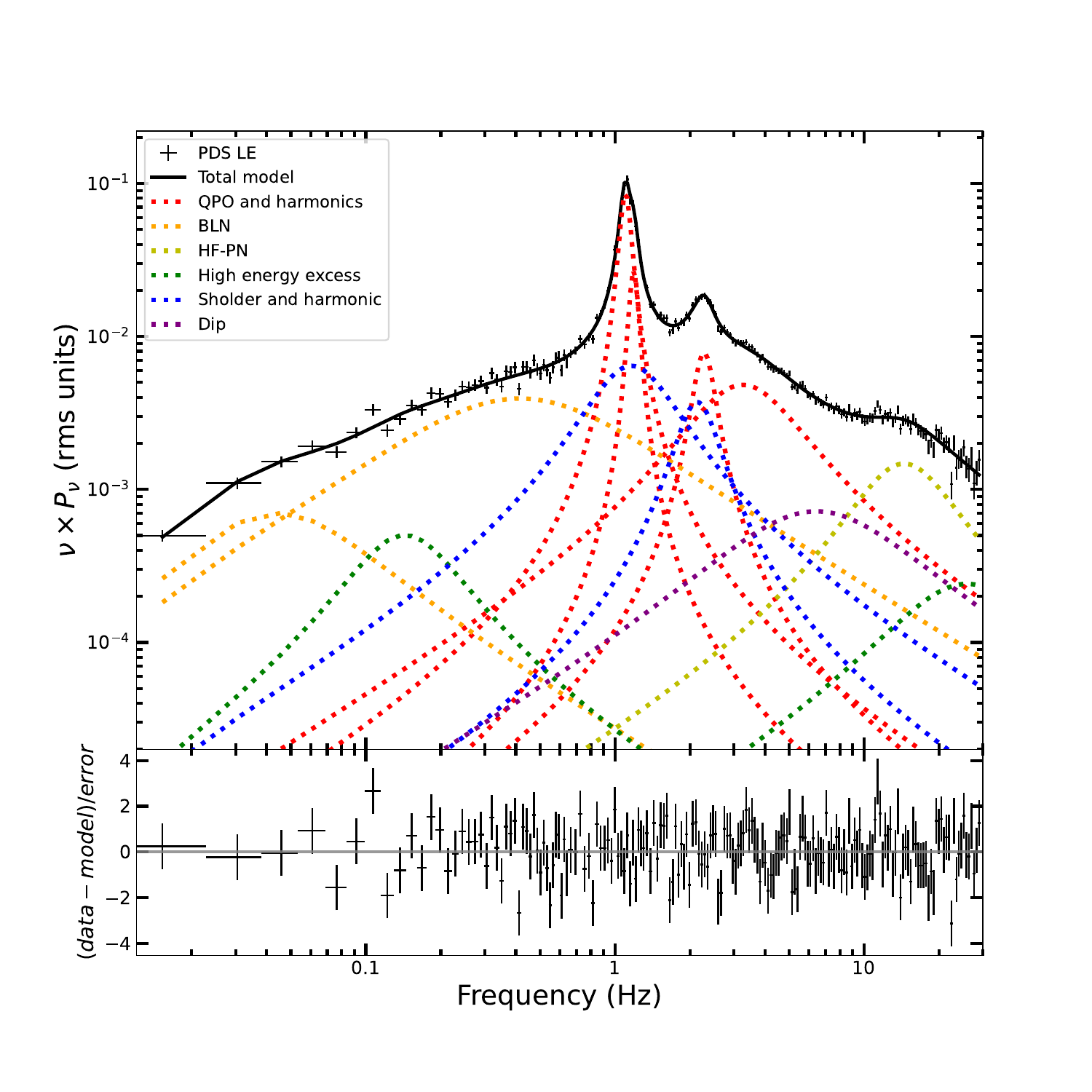}\vspace{-5mm}
    \includegraphics[width=\columnwidth]{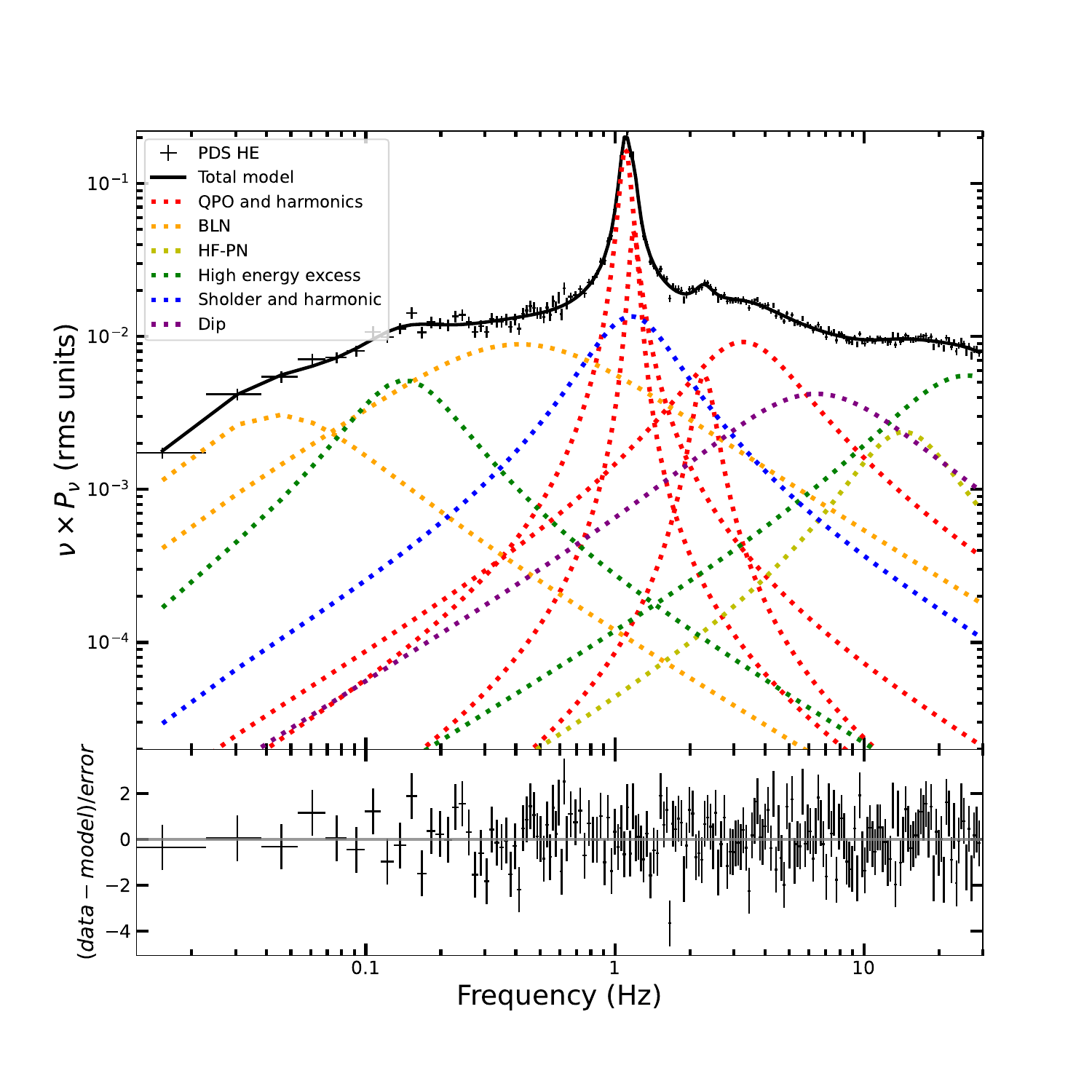}\vspace{-5mm}
	\includegraphics[width=\columnwidth]{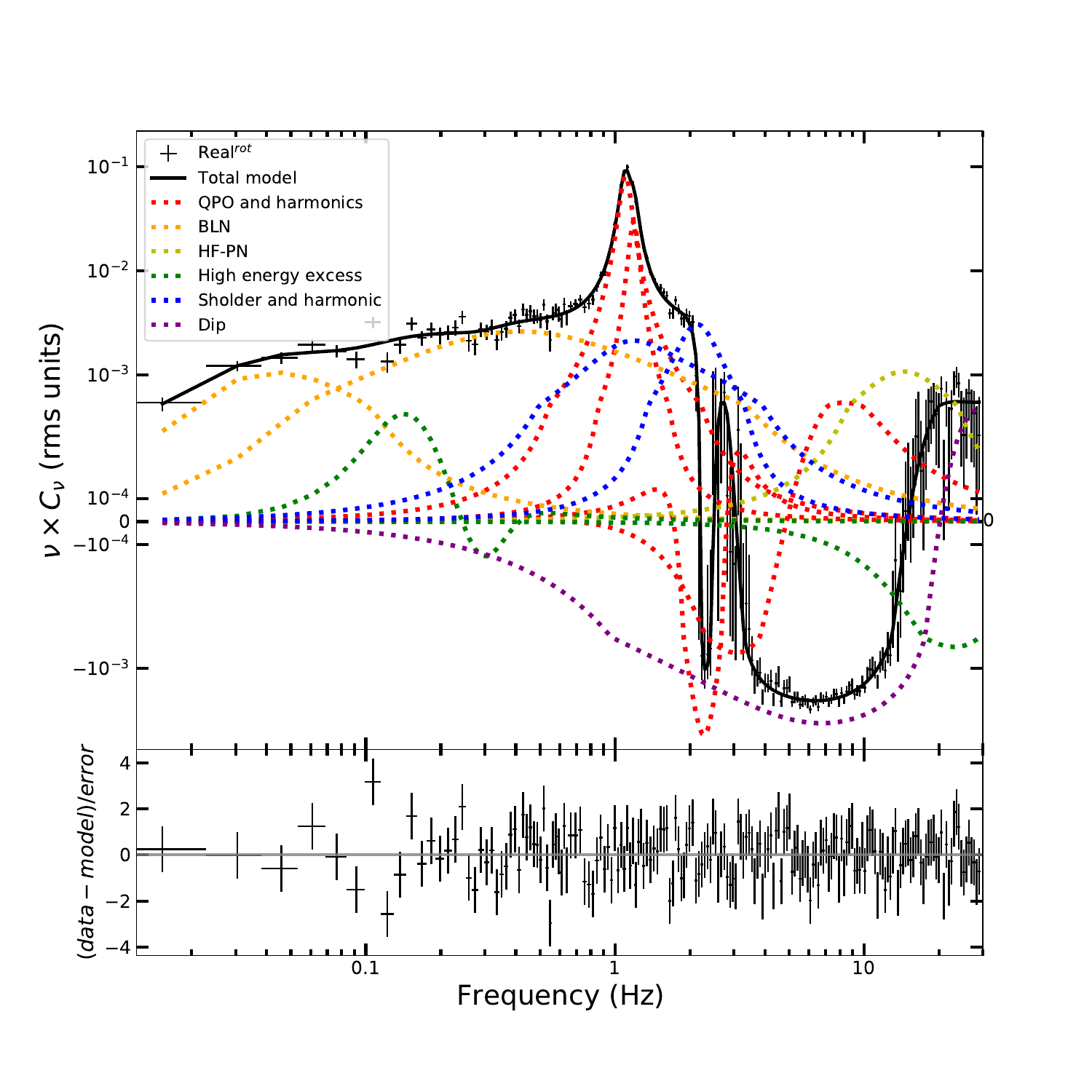}
 	\includegraphics[width=\columnwidth]{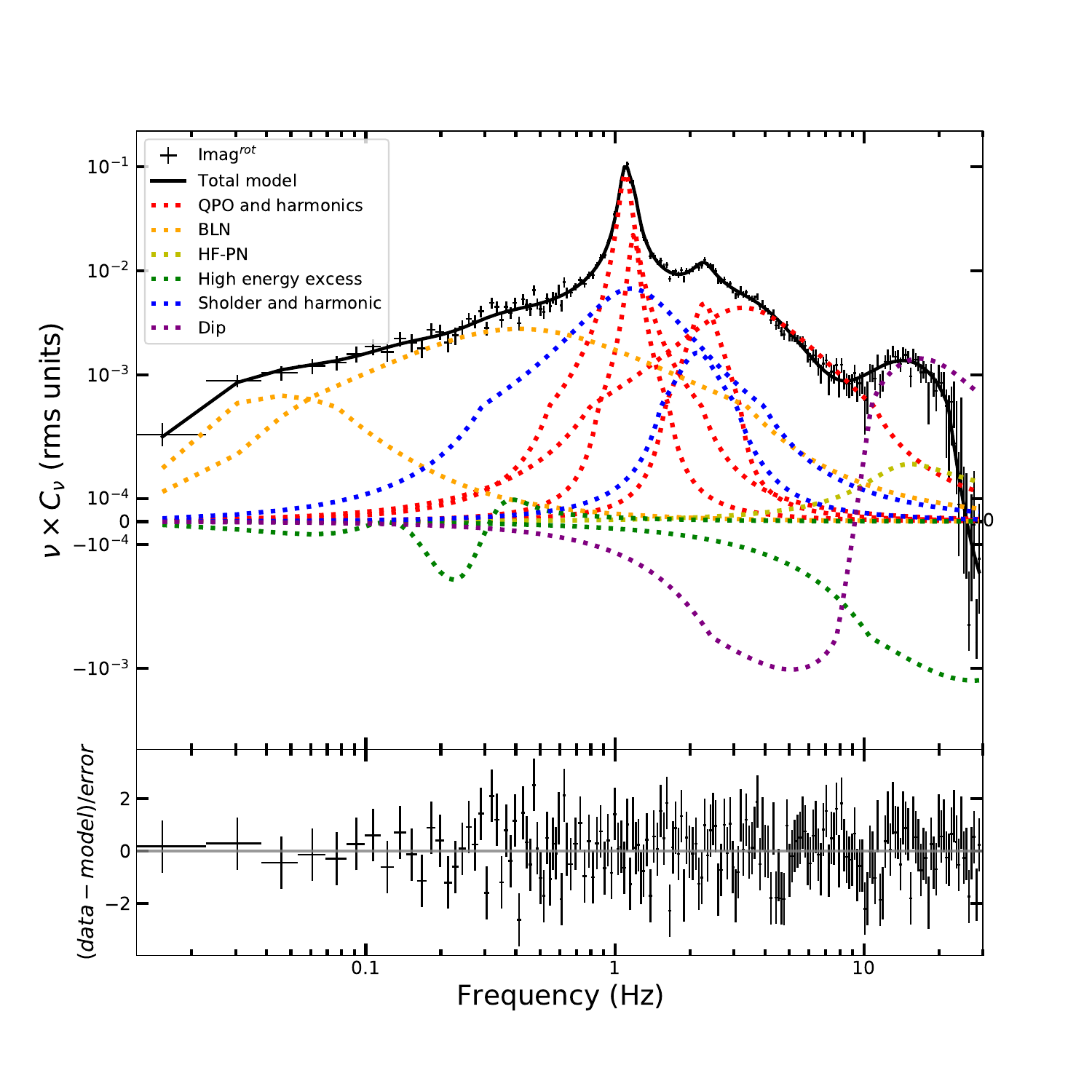}
 	\vspace{-5mm}
    \caption{Same as Fig.~\ref{fig:plot_plcon_time}, but assuming the Gaussian phase-lag model.}
    \label{fig:plot_plgau_time}
\end{figure*}

Similar to~\cite{2022MNRAS.515.1914Z}, in this section we assume that the phase lag frequency spectrum of each Lorentzian is a Gaussian function of Fourier frequency that reaches the maximum or minimum at the centroid frequency of the Lorentzian. 
Therefore, the models of the real and imaginary parts of the rotated CS are, respectively, $\sum_{i=1}^n C_iL(\nu;\nu_{0,i},\Delta_i)\cos{(2\pi\;k_i\;e^{-\frac{1}{2}(\frac{\nu-{\nu_0}_i}{\sigma_i})^2}  + \pi/4)}$ and $\sum_{i=1}^n C_iL(\nu;\nu_{0,i},\Delta_i)\sin{(2\pi\;k_i\;e^{-\frac{1}{2}(\frac{\nu-{\nu_0}_i}{\sigma_i})^2}  + \pi/4)}$.
We link the $\sigma$ of the Gaussian function to the FWHM of the Lorentzian in the fitting and limit $2\pi k_i$ between $-2\pi$ and $2\pi$.
The fit results with 12 Lorentzians are shown in Fig.~\ref{fig:plot_plgau_time}. All Lorentzians are at least 3~$\sigma$ significant in either the PDS or the rotated CS, and the parameters of the Lorentzians are reported in Table~\ref{tab:parameters_time}. The fit yields $\chi^2 = $ 791 for 684 dof.

As we explained, the Gaussian phase-lag model peaks at the centroid frequency of the Lorentzian.
Away from that frequency, the absolute value of the phase lag gradually decreases and finally approaches zero. 
In this case, the Lorentzian that fits the dip has a centroid frequency of $3.7^{+0.7}_{-0.2}$~Hz and a FWHM of $10.5^{+0.8}_{-0.7}$~Hz, as well as a phase lag of $2.7 \pm 0.1$~rad at the centroid frequency~\footnote{Since the phase lag are defined except an additive factor $2n\pi$, for the Gaussian phase-lag model we can write that $g_i(\nu;p_j) = 2\pi\;k_i\;e^{-\frac{1}{2} ( \frac{\nu-{\nu_{0,i}}}{\sigma_i} )^2} + 2n\pi$. At the centroid frequency of the Lorentzian, $g_i(\nu_0;p_j) = 2\pi\;k_i\; + 2n\pi$, which we call $\Delta\Phi_D$, the phase lag of the Gaussian model at the centroid frequency of the dip. Notice that, if we add or subtract $2\pi$ to $2\pi\;k_i$ when we plot $\Delta\Phi_D$, this only applies at the centroid frequency of the dip.}.
Away from the centroid frequency of the dip, the phase lags decrease to less than $\pi/2$ at some frequencies, where the real part of the Lorentzian that fits the dip will change from negative to positive, rather than being consistently negative as with the constant phase-lag model.
This characteristic of the Gaussian phase-lag model means that the Lorentzian accounting for the dip effectively captures the main feature of the real and imaginary parts of the CS at $\sim$3.0$-$15.0 Hz, without the need of other Lorentzians in the model in that frequency range. In particular, the phase lags of the HF-PN and the high-frequency high-energy excess change significantly for the Gaussian phase-lag model, exhibiting lower phase lags than the ones obtained from the constant phase-lag model.

In Appendix~\ref{sec:Fit result with the constant time-lag model} we describe the fit using the constant time-lag model.

\subsection{Joint-fit for data combined in Table~\ref{tab:group}}
\label{sec:Joint-fits}

\begin{figure*}
    \centering
	\vspace{-5mm}
  	\includegraphics[width=\columnwidth]{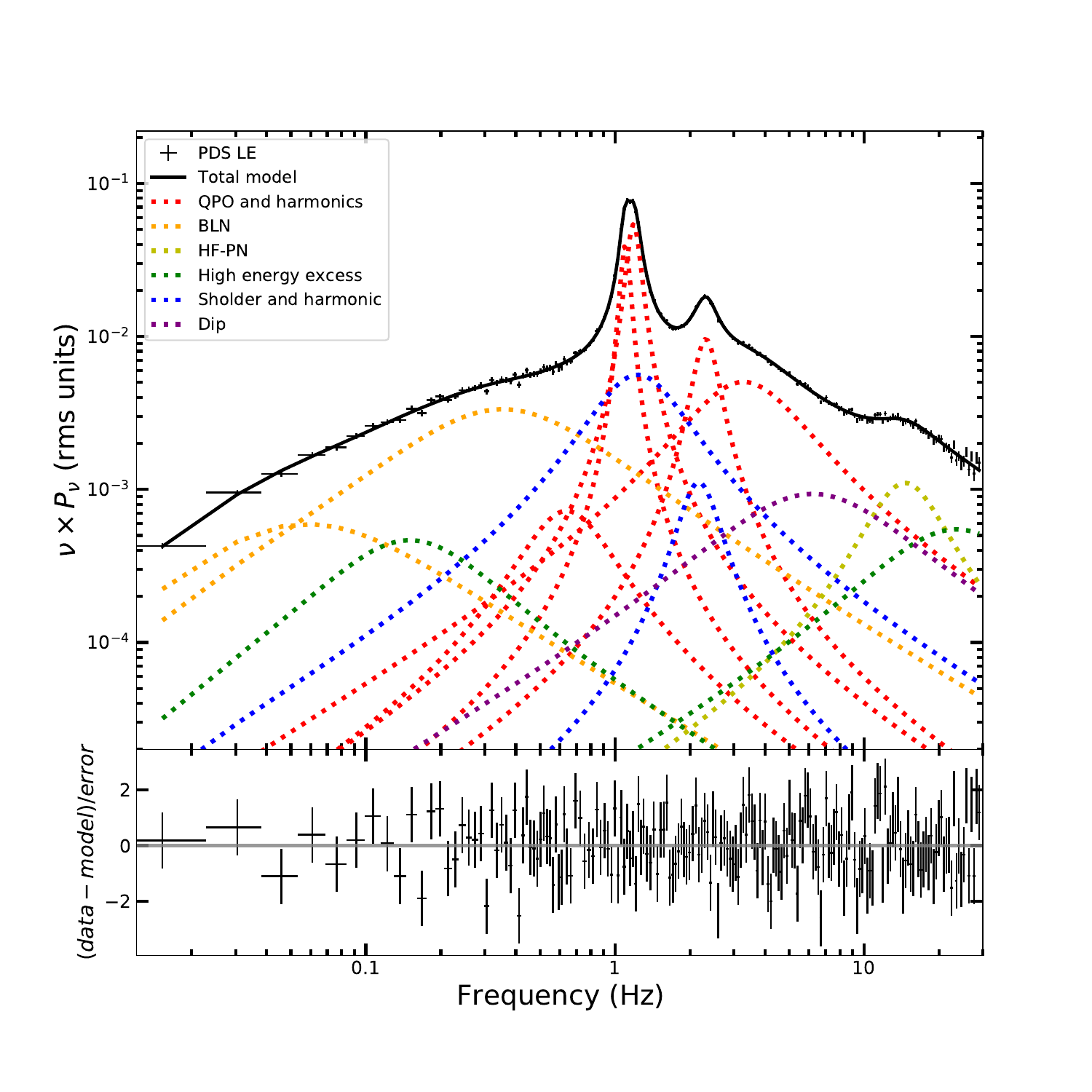}\vspace{-5mm}
    \includegraphics[width=\columnwidth]{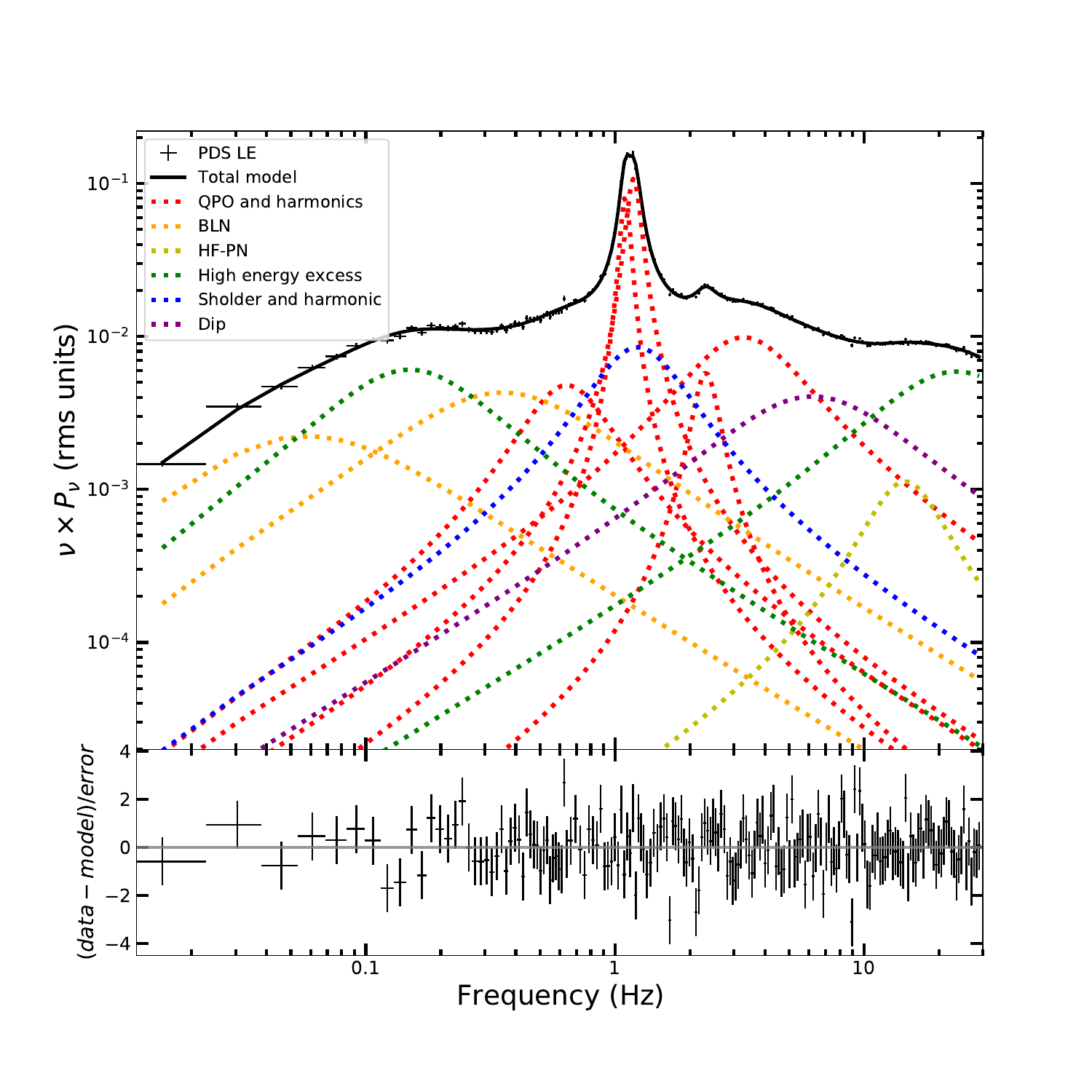}\vspace{-5mm}
	\includegraphics[width=\columnwidth]{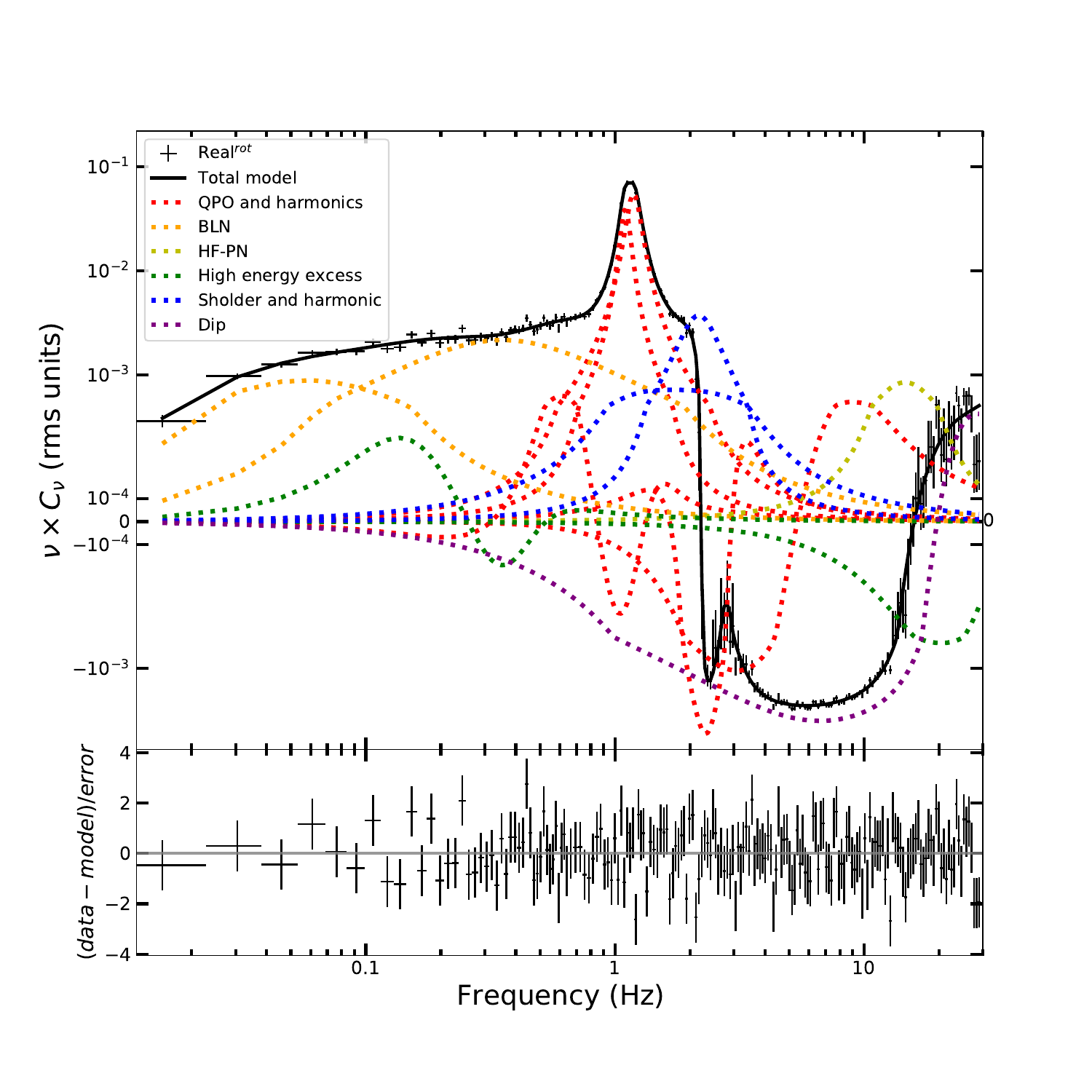}
 	\includegraphics[width=\columnwidth]{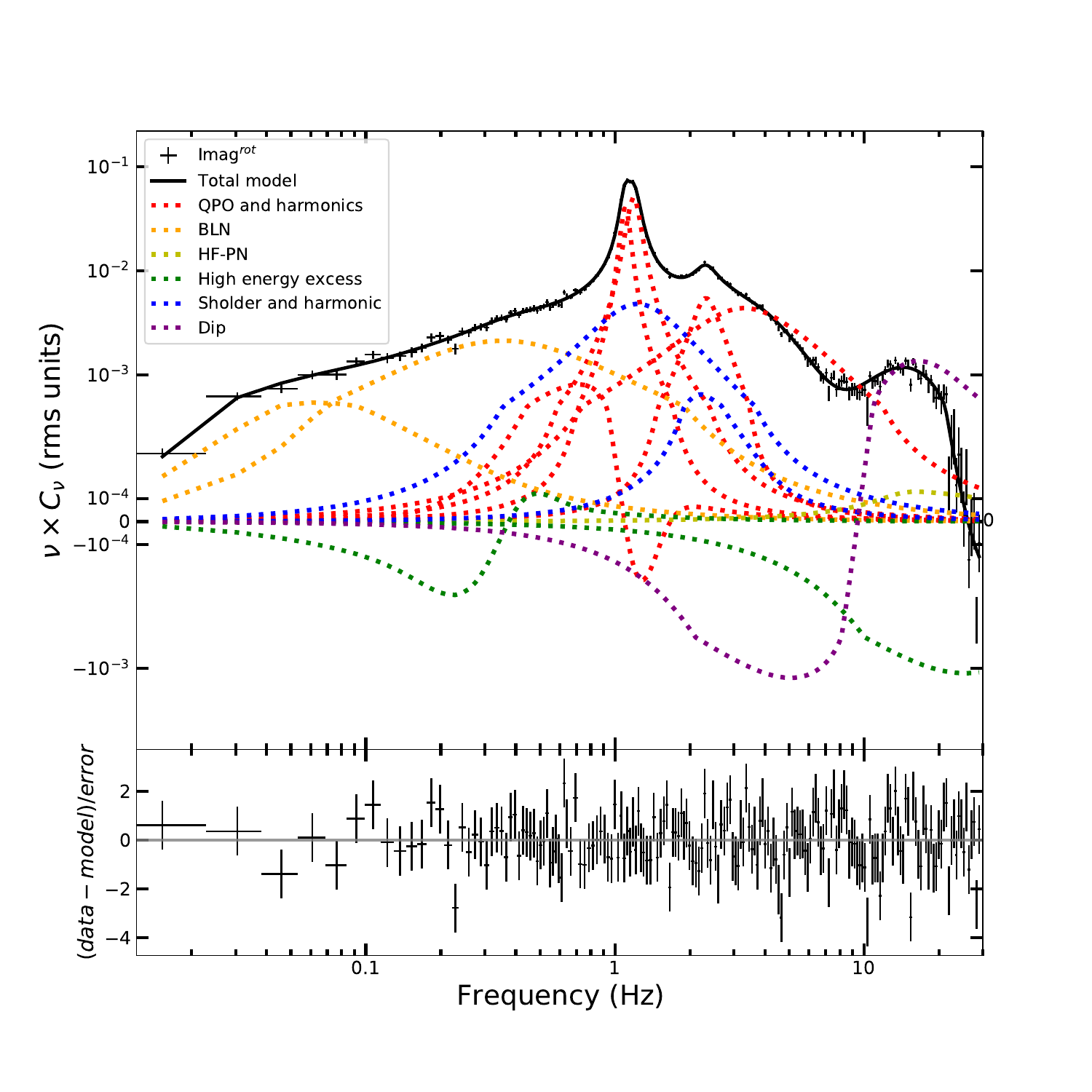}
 	\vspace{-5mm}
    \caption{LE 2$-$10 keV (upper left panel) and HE 28$-$200 keV PDS (upper right panel) and the real (lower left panel) and imaginary (lower right panel) parts of the rotated CS of Swift~J1727.8$-$1613 for Group~$\#$9 in Fig.~\ref{fig:fhd}, fitted with a model (solid lines) consisting of 13 Lorentzians. Individual components are plotted with colored dotted lines.
    The fits were done assuming the Gaussian phase-lag model.}
    \label{fig:plot_13lor}
\end{figure*}

\begin{table*}
	\centering
	\caption{Parameters of the Lorentzian accounting for the dip in the real part of the CS of Group~\#9 for the HE $28-200$ keV band with respect to the LE $2-10$ keV band light curves, obtained with different lag models.}
	\label{tab:comparison_lag_models}
	\begin{tabular}{l|l|l|l}
        \hline
         & Constant phase lag & Gaussian phase lag & Constant time lag  \\
        \hline
        Frequency (Hz)  & $<$ 0.81 & $3.6 \pm 0.5$ & $4.7 \pm 0.6$ \\
        FWHM (Hz)  & $9.3_{-0.6}^{+1.1}$ & $10.1 \pm 0.3$ &  $9.7_{-0.9}^{+0.8}$ \\
        Phase lag (rad)  & $2.74_{-0.05}^{+0.03}$ & ${2.78_{-0.07}^{+0.08}}^{\dagger}$ &  \\
        Time lag (ms)  & & & $33_{-1}^{+2}$ \\
        $\chi^2$/dof. & 810/678 & 781/678 & 841/678 \\
        \hline
        \multicolumn{4}{l}{\small $^{\dagger}$Phase lag at the centroid frequency of the dip.}\\
	\end{tabular}
\end{table*}

\begin{figure*}
    \centering
	\vspace{-5mm}
	\includegraphics[width=\columnwidth]{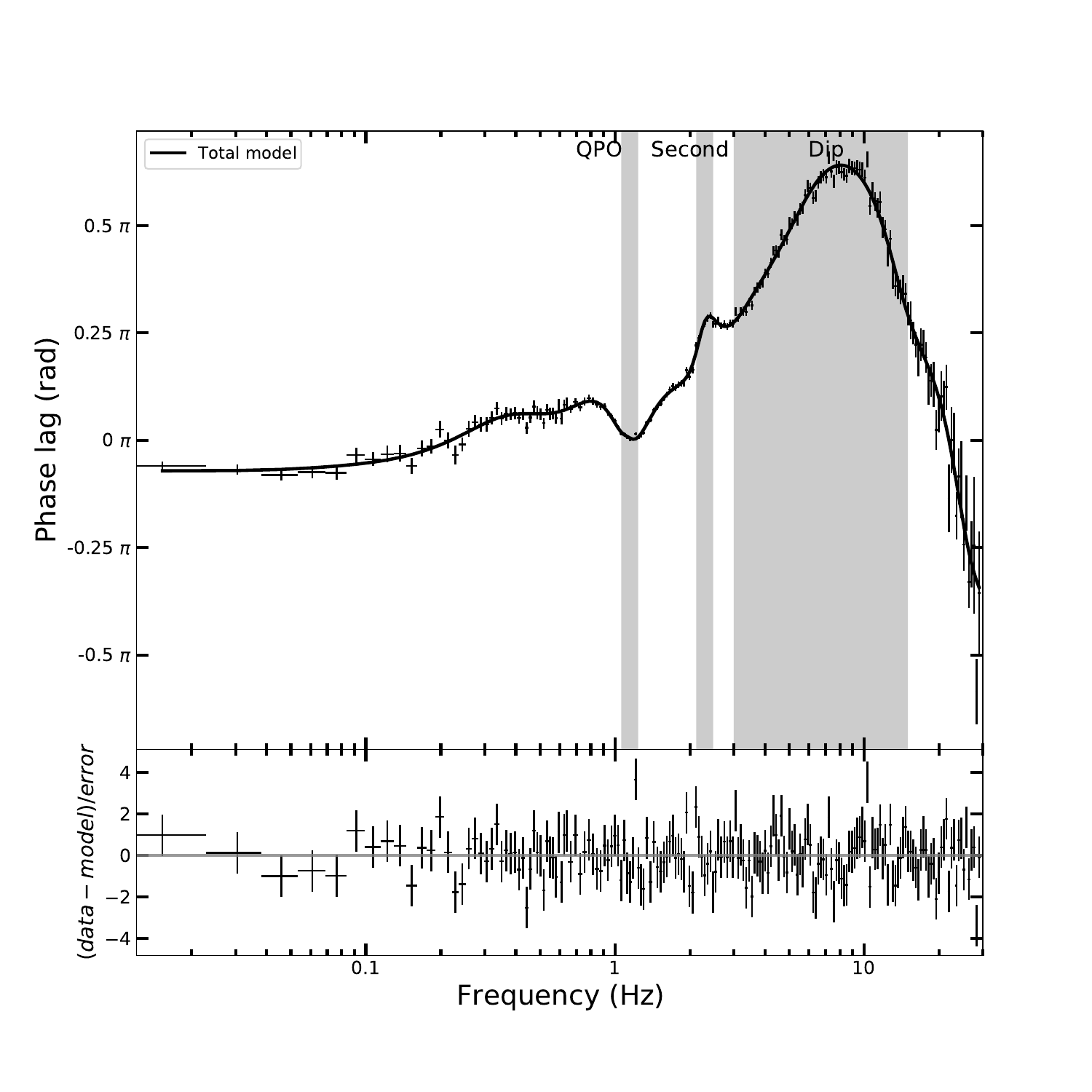}
 	\includegraphics[width=\columnwidth]{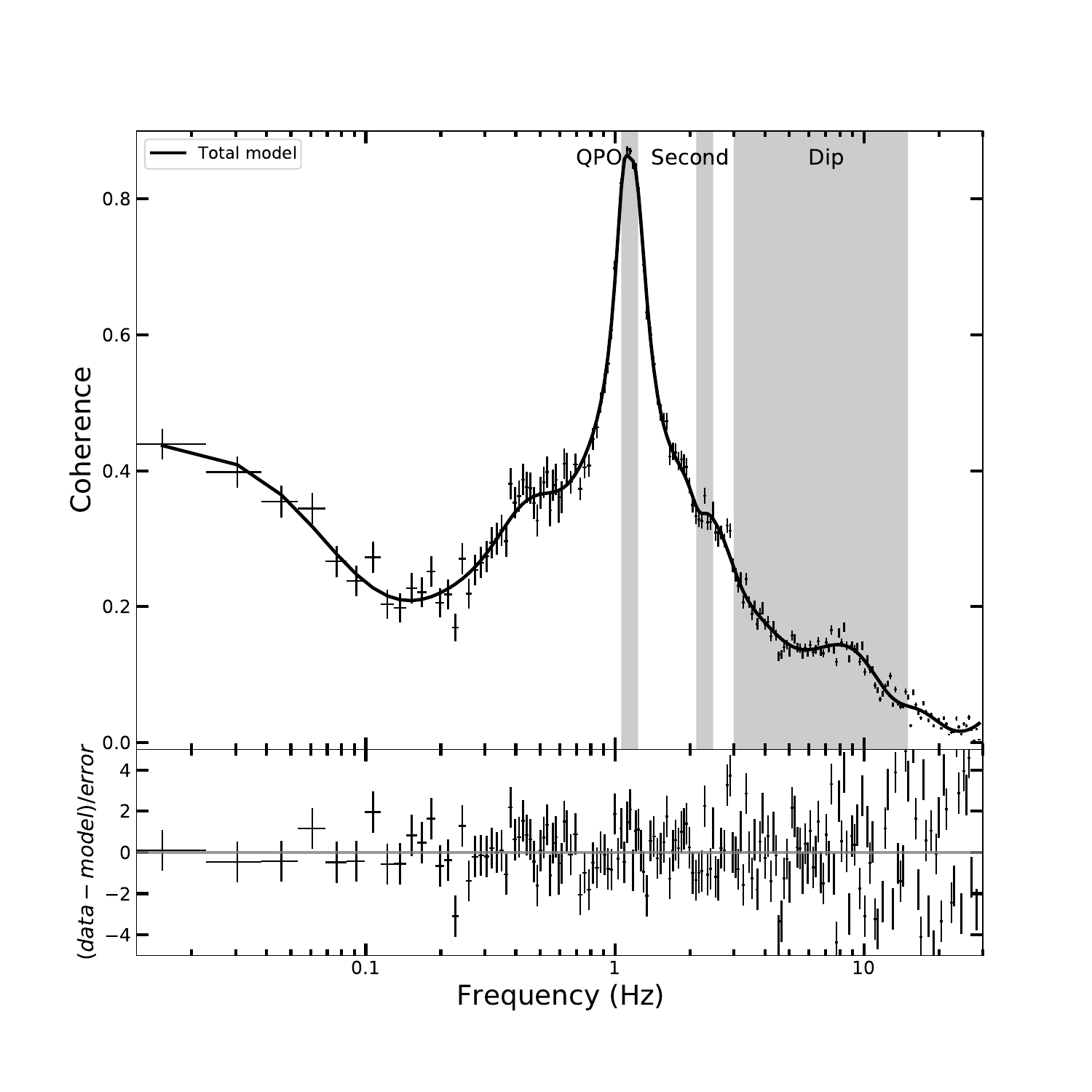}
 	\vspace{-5mm}
    \caption{The phase-lag spectrum (left panel) and intrinsic coherence function (right panel) of Swift~J1727.8$-$1613 for Group~\#9. The models are not fitted to the data, but predicted on the basis of the parameters of the Lorentzians fitted to the LE 2-10 keV and HE 28-200 keV PDS and the real and imaginary parts of the CS of those same two energy bands assuming the Gaussian phase-lag model (Fig.~\ref{fig:plot_13lor}). The vertical gray bands indicate the QPO, the QPO second harmonic and the dip feature, respectively.}
    \label{fig:fig_1.06_1.24_coh}
\end{figure*}

In this section we fit jointly the PDS and the CS, using the same methods in Section~\ref{sec:Joint-fit of power and cross spectra} but for different observations that are recorded in Table~\ref{tab:group}. 
We first fit the data of Group~$\#$9 in Fig.~\ref{fig:fhd} with the three different lag models. In Fig.~\ref{fig:plot_13lor} we show a fit with 13 Lorentzian ($n = 13$), using the Gaussian phase-lag model.
Compared to the data shown in Fig.~\ref{fig:pds_cs}, ~\ref{fig:plot_plcon_time} and \ref{fig:plot_plgau_time}, Group~$\#$9 has more time segments to be averaged. Because of this we are more sensitive to weaker features, and an additional Lorentzian with a centroid frequency that is at around half of that of the QPO significantly reduces the $\chi^2$. We identify this new Lorentzian as the sub-harmonic of the QPO and plot it with a red dotted line in Fig.~\ref{fig:plot_13lor}. 
The sub- and high-order harmonics of the QPO show a broad shape in Fig.~\ref{fig:plot_13lor}, which could partly result from the (sub)harmonics shifting during the observation. However, the definition of a QPO having Q$>$2 (Q=F/FWHM, where F is the frequency of QPO.) is somewhat arbitrary. As showed by~\citet{2002ApJ...572..392B}, the broad features follow the same correlation as the narrow QPOs, and indeed become narrow QPOs as their characteristic frequencies increase.
The other components have parameters that, within one sigma error, are consistent to those obtained from the fit in Fig.~\ref{fig:plot_plgau_time}. We therefore conclude that the combination of the data as shown in Fig.~\ref{fig:fhd} preserves the information in the individual observations, but gives a better signal-to-noise ratio.

We initially performed a joint fit of the two PDS without including the CS, achieving a good fit with a combination of 10 Lorentzians. When we then attempted a simultaneous fit of the PDS and the real and imaginary parts of the CS, keeping the centroid frequencies and FWHM of the Lorentzians fixed to the values obtained from the PDS-only fit, the fits to the CS were poor. We next allowed the parameters of the Lorentzians to vary, with the frequencies and FWHM tied across all the spectra, while keeping their number fixed to 10, which improved the fit but still left significant residuals in the real and imaginary parts of the CS. Finally, a successful fit of all features was achieved after adding three additional Lorentzians, resulting in a total of 13.

The fit with the Gaussian phase-lag model yields $\chi^2 = $ 781 for 678 degrees of freedom (dof), slightly better than the constant phase-lag model, which yields a $\chi^2$ of 810 for 678 degrees of freedom, and the constant time-lag model, which yields a $\chi^2$ value of 841 for 678 degrees of freedom.
With the constant phase-lag model, the dip appearing in the real part of the CS is modeled by a broad Lorentzian with a centroid frequency $<$ 0.81~Hz and a FWHM of $9.3_{-0.6}^{+1.1}$~Hz. The phase lag of this Lorentzian is $2.74_{-0.05}^{+0.03}$ rad.  With the Gaussian phase-lag model, the Lorentzian that fits the dip has a frequency of $3.6 \pm 0.5$~Hz and a FWHM of $10.1 \pm 0.3$~Hz, as well as a phase lag $\Delta\Phi_D =
 2.78_{-0.07}^{+0.08}$~rad. We give the parameters of the Lorentzian obtained with the three different lag models in Table~\ref{tab:comparison_lag_models}.

In Fig.~\ref{fig:fig_1.06_1.24_coh} we plot the predicted models of the phase-lag spectrum (left panel) and the intrinsic coherence function (right panel). The predicted models are consistent with the data.

\subsection{Evolution of the dip}

\begin{figure}
    \centering
	\vspace{-5mm}
	\includegraphics[width=\columnwidth]{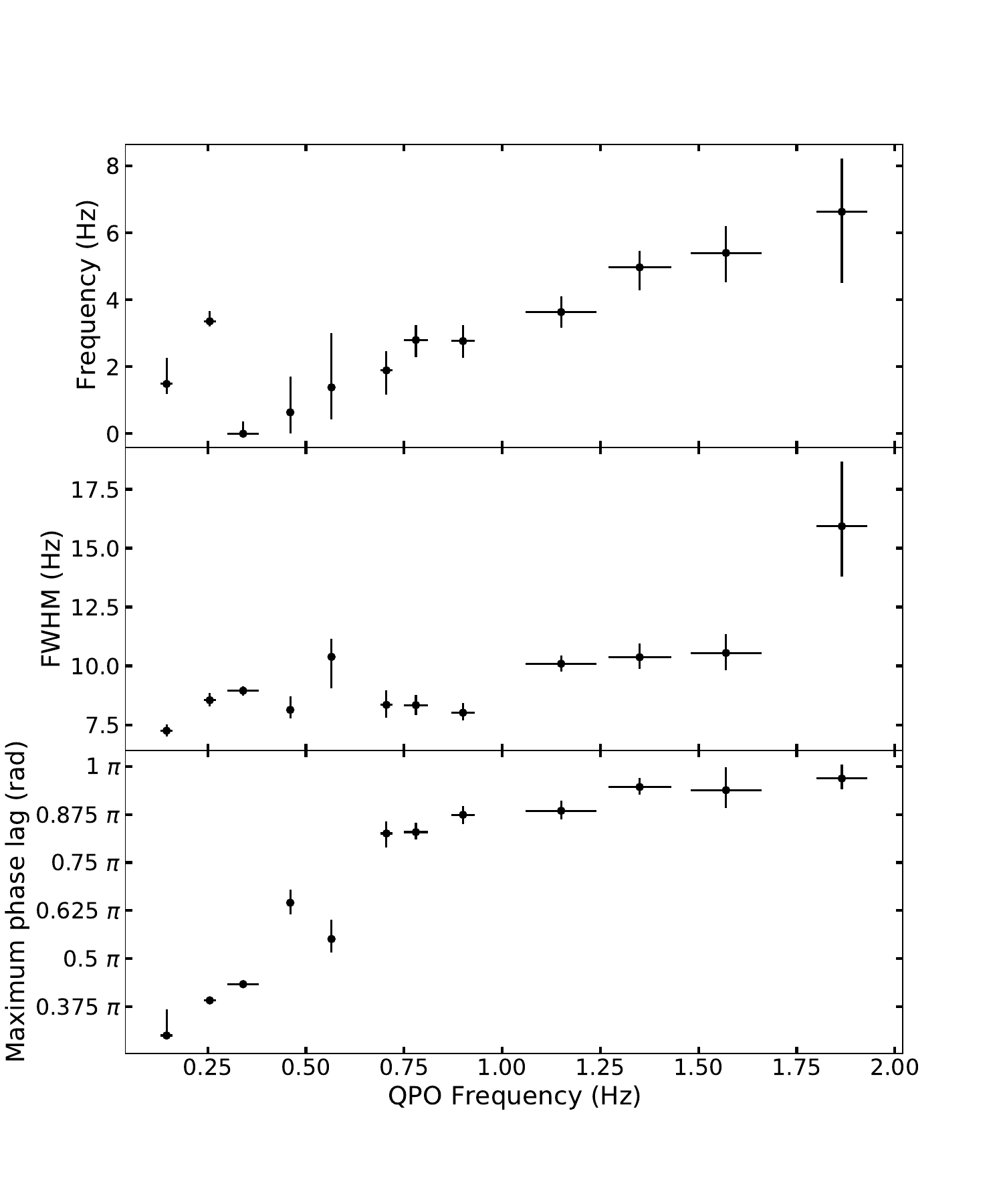}
 	\vspace{-10mm}
    \caption{The centroid frequency, FWHM and phase lags at the centroid frequency of the Lorentzian accounting for the dip in the real part of the cross spectrum of Swift~J1727.8$-$1613  as a function of the QPO frequency. The lags are for the HE 28$-$200 keV with respect to the LE 2$-$10 keV data, and were obtained using the Gaussian phase-lag model.}
    \label{fig:evolution}
\end{figure}

\begin{figure}
    \centering
	\vspace{-5mm}
	\includegraphics[width=\columnwidth]{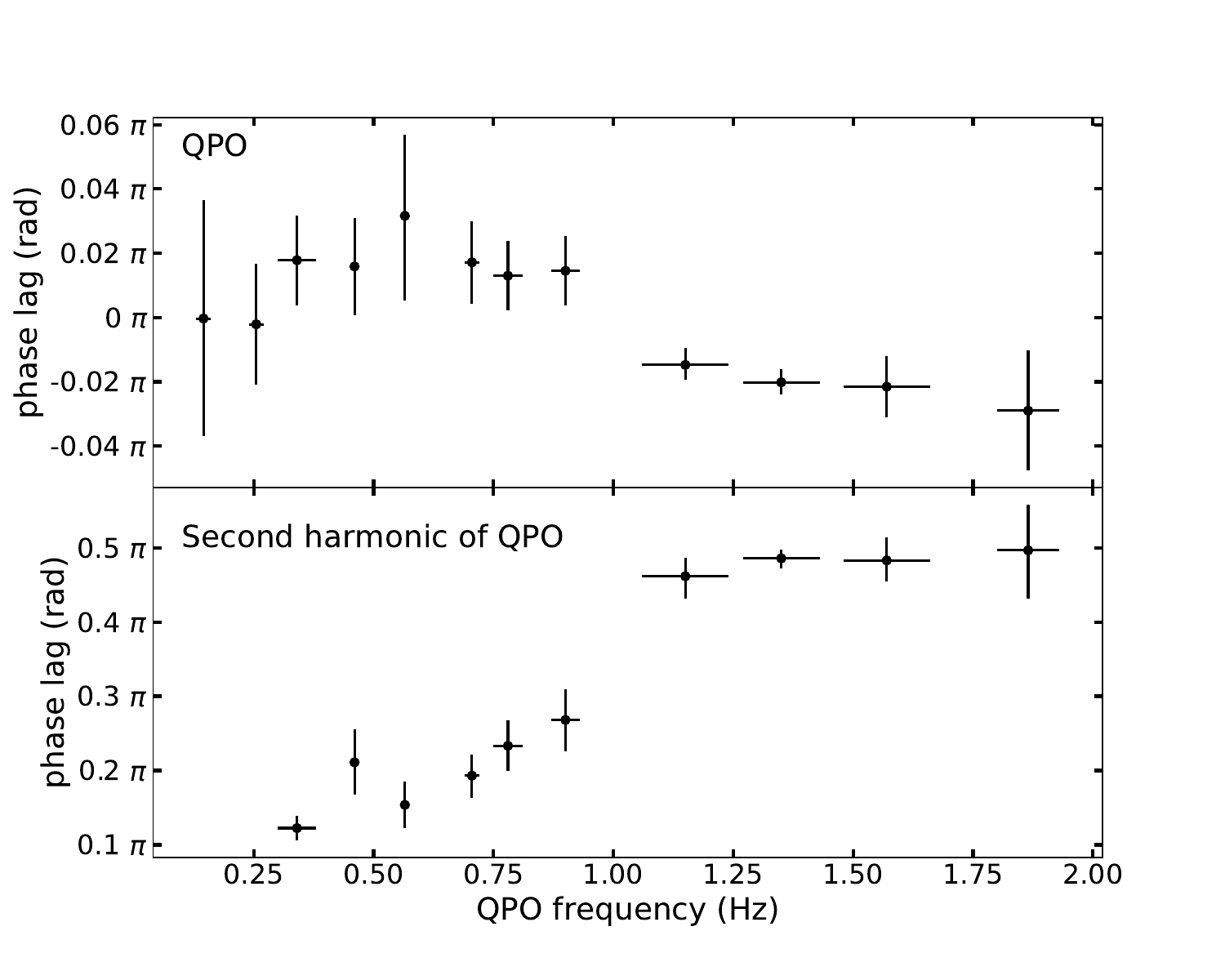}
 	\vspace{-5mm}
    \caption{The phase lags at the centroid frequency of the QPO (upper panel) and second harmonic (bottom panel) of Swift~J1727.8$-$1613 as a function of the QPO frequency. We used the Gaussian phase-lag model for the fits. The second harmonic is not detected in the first two groups.}
    \label{fig:evolution_qpo_harmonic}
\end{figure}

In this section we fit the PDS and the CS of each group shown in Table~\ref{tab:group} and study the evolution of the parameters of the Lorentzian accounting for the dip in the real part of the CS.
We assume that the phase lags are a Gaussian function of Fourier frequency (eq.~\ref{eq4}).
We plot the evolution of the centroid frequency, FWHM and the phase lags at the centroid frequency of the Lorentzian in Fig.~\ref{fig:evolution}.
As the QPO frequency increases, the frequency of the Lorentzian that fits the dip increases from $<1$~Hz to $\sim6$~Hz,
while the FWHM of the dip increases from $\sim7.5$ Hz to $\sim15$~Hz.
As we see in the bottom panel of this figure, the phase lags at the centroid frequency of the Lorentzian first increase rapidly from $<0.4\pi$ to $\sim0.9\pi$ when the QPO frequency increases from $\sim0.13$ Hz to $\sim1$ Hz, and then remain more or less constant when the QPO frequency increases further up to $\sim2$~Hz.

We also plot the phase lags of the QPO and its second harmonic in Fig.~\ref{fig:evolution_qpo_harmonic}. Notably, both phase lags show significant changes when the QPO frequency is around $1.0$~Hz. The phase lags of the QPO transition from positive to negative, while the phase lags of the second harmonic increase rapidly from $\sim 0.2\pi$ to $\sim0.5\pi$.

\subsection{Energy dependence of the dip}
\label{sec:Energy dependent of the dip}

In this section we focus on the data of Group~\#9 and divide the full energy range into 12 bands to study the energy dependence of the dip.
The separate energy bands are LE 1.0$-$2.6 keV, LE 2.6$-$4.8 keV, LE 4.8$-$7 keV, LE 7$-$11 keV, ME 7$-$11 keV, ME 11$-$15 keV, ME 15$-$23 keV, ME 23$-$35 keV, HE 25$-$35 keV, HE 35$-$48 keV, HE 48$-$67 keV and HE 67$-$100 keV.
We select a time resolution of 3~ms (therefore, the Nyquist frequency is $\sim$ 167 Hz) and a segment length of 49.125~s to produce the PDS and the CS.
We use the LE 1.0$-$2.6 keV band as reference to produce the CS.
Finally, we obtain 12 power density spectra and 11 cross spectra.
The bands LE 7$-$11 keV and ME 7$-$11 keV, and ME 23$-$35 keV and HE 25$-$35 keV cover, respectively, the same energy ranges but, since the measurements are from different instruments, we can compare the results. 
We rebin the averaged PDS and CS logarithmically in frequency by a factor $\approx 1.047 (= 10^{1/50})$ to increase the signal-to-noise ratio further.

\begin{figure*}
    \centering
	\vspace{-5mm}
	\includegraphics[width=\columnwidth]{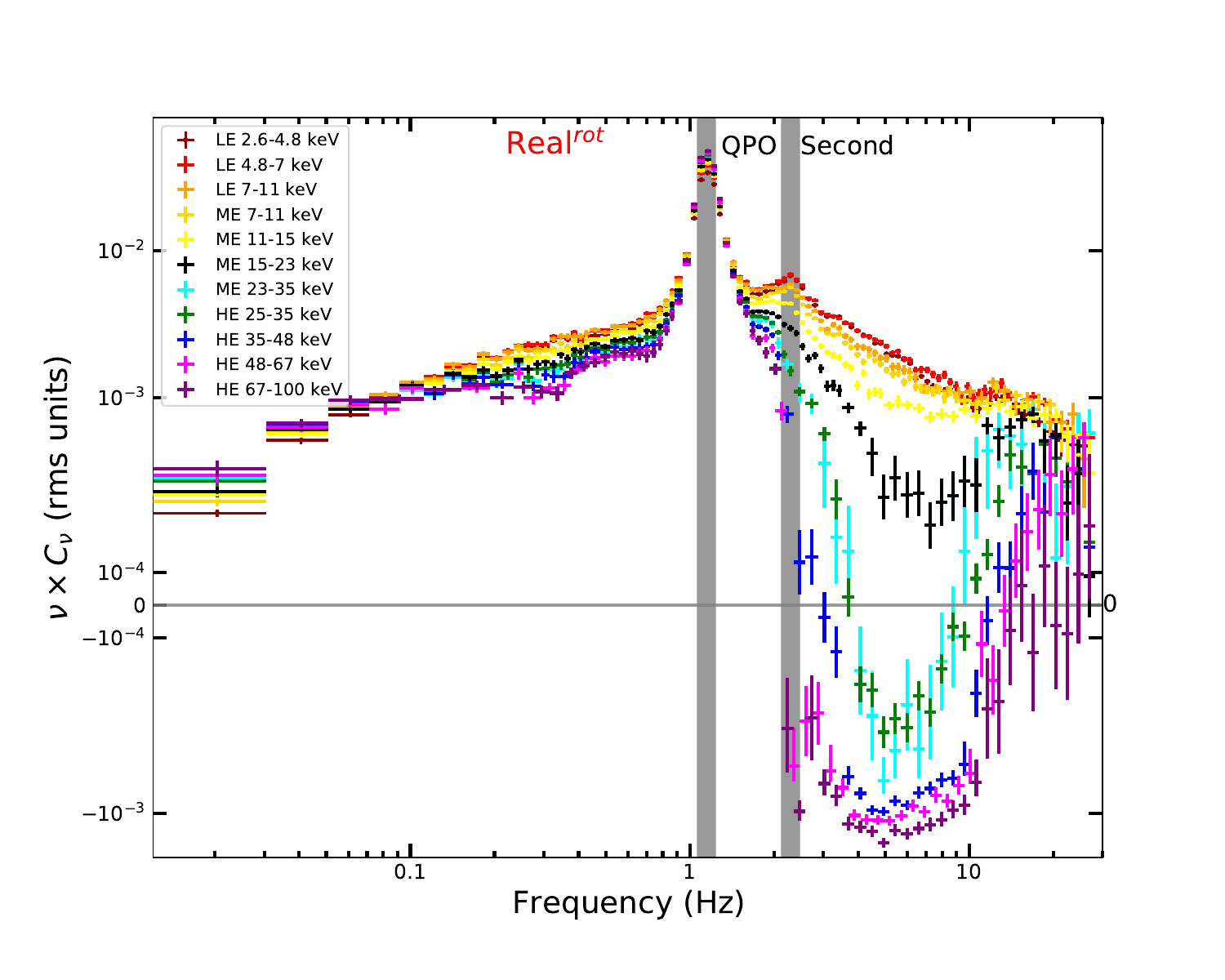}
    \includegraphics[width=\columnwidth]{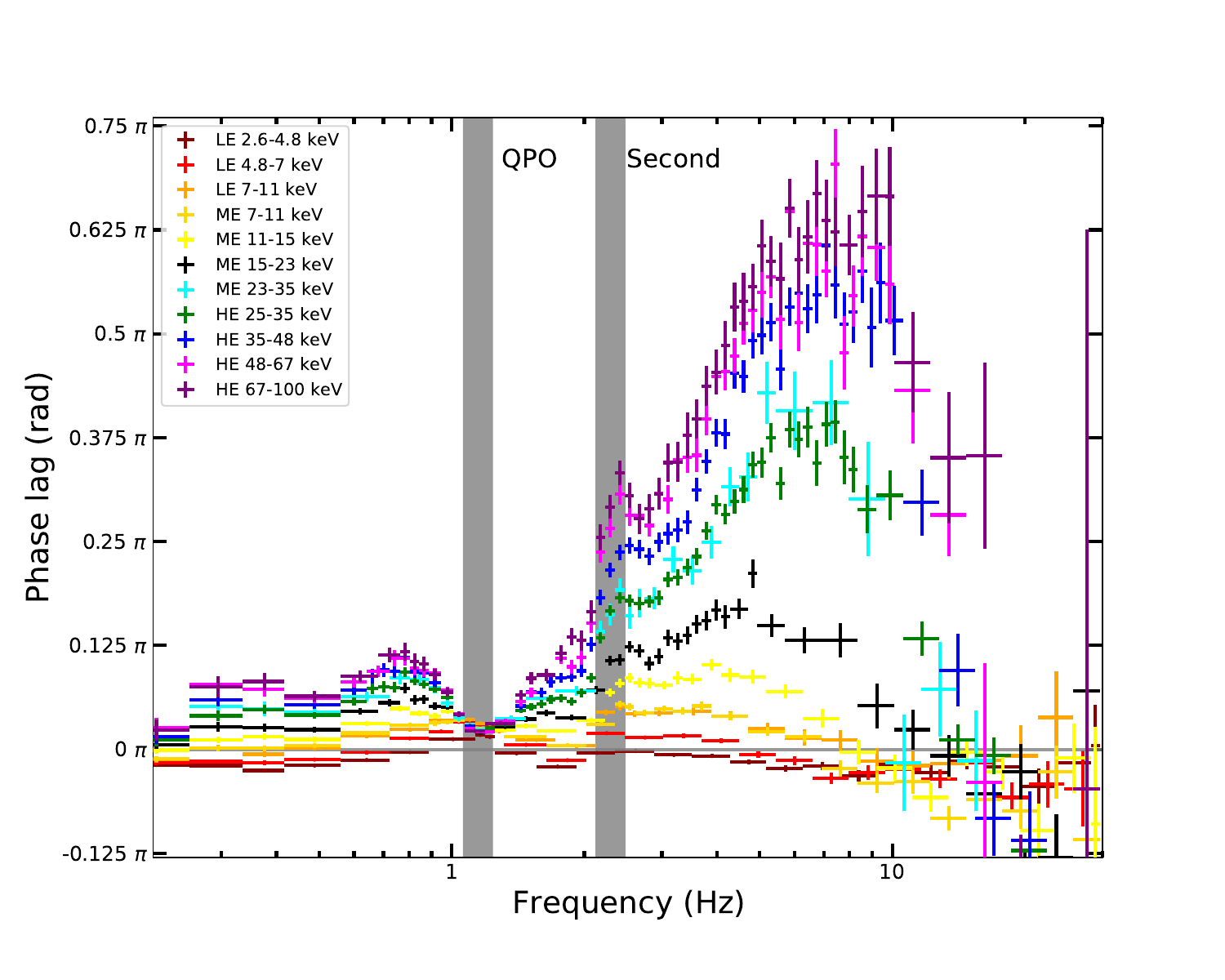}
 	\vspace{-5mm}
    \caption{Energy dependence of the rotated real part and the phase lags of the cross spectrum of Swift~J1727.8$-$1613 for Group~\#9 in Table~\ref{tab:group}. The reference energy band is LE 1.0$-$2.6 keV. The vertical gray bands indicate the QPO and the QPO second harmonic.}
    \label{fig:ene_dependent}
\end{figure*}

\begin{figure}
    \centering
	\vspace{-5mm}
	\includegraphics[width=\columnwidth]{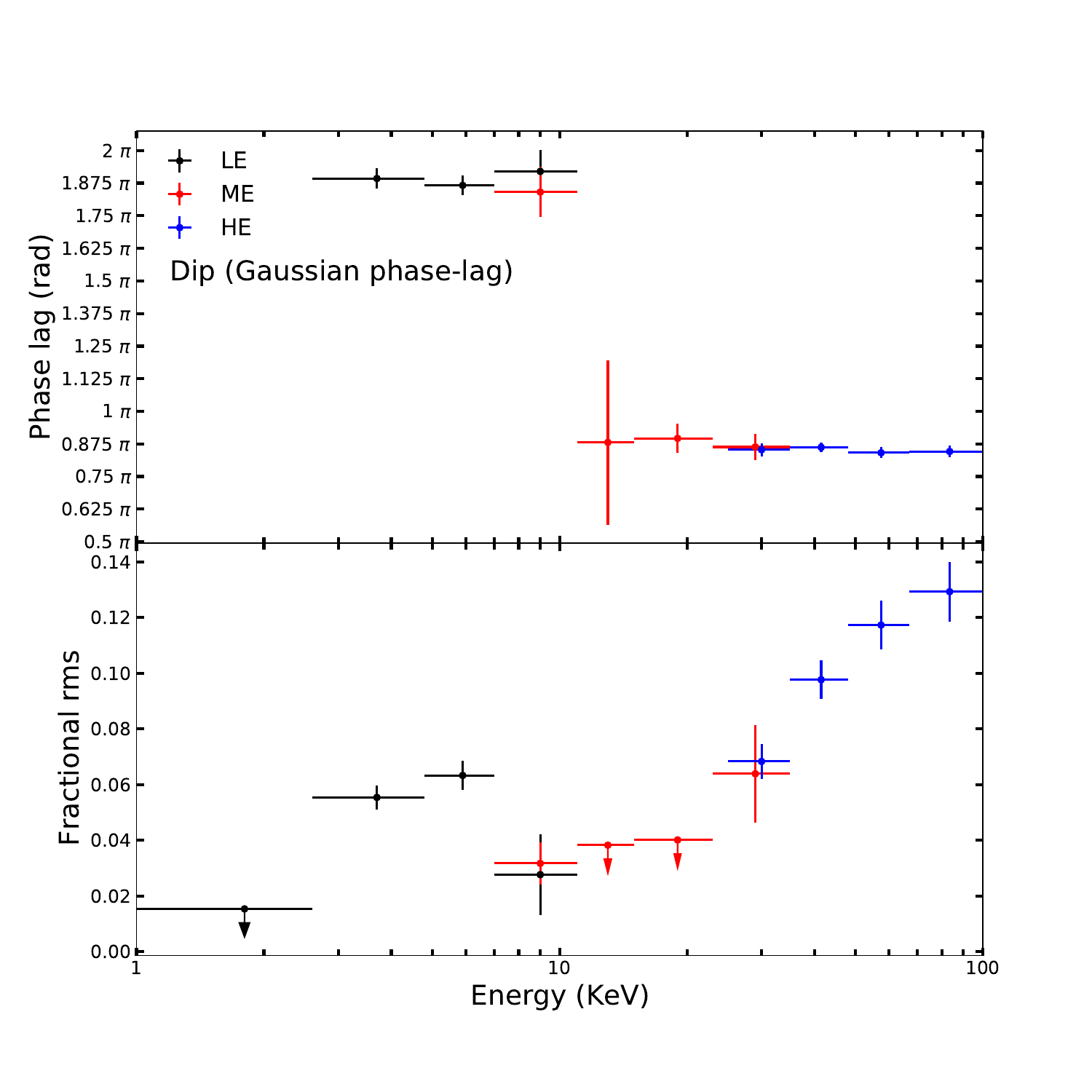}
 	\vspace{-5mm}
    \caption{Energy dependence of the phase lags at the centroid frequency of the dip, $\Delta\Phi_D$, and the fractional rms of the dip of Swift~J1727.8$-$1613 for Group~\#9 in Table~\ref{tab:group} derived from the Gaussian phase-lag model. The errors correspond to the 3-$\sigma$ confidence ranges of the parameters calculated with MCMC. The x-error bars indicate the width of the corresponding energy band.}
    \label{fig:ene_dependent_par}
\end{figure}

In Fig.~\ref{fig:ene_dependent}, we plot the rotated real parts (left panel) and the phase lags (right panel) of the cross spectra generated from the different energy bands. The dip, appearing in the $\sim 3-15$~Hz in the rotated real part, starts to become apparent above 15 keV, and the minimum starts to be negative above 23 keV; this is the case for both the ME 23$-$35 keV and HE 25$-$35 keV bands. 
The right panel of Fig.~\ref{fig:ene_dependent} shows that the phase lags in the $\sim 3-15$~Hz frequency range increase significantly above 23 keV.
To obtain the parameters of the dip as a function of energy, we fit jointly the 12 power density spectra and the real and imaginary parts of the 11 rotated cross spectra assuming either the Gaussian phase-lag or the constant phase-lag model.
We subsequently calculate the 3-$\sigma$ confidence range of the parameters using the Markov chain Monte-Carlo algorithm (MCMC). 
The Goodman-Weare algorithm is applied for a total of 200000 samples and a burn-in phase of 200000 to let the chain reach a steady state.

With the Gaussian phase-lag model, the 3-$\sigma$ confidence ranges of the centroid frequency and the FWHM of the Lorentzian that fits the dip are $3.66-4.58$ Hz and $7.70-8.50$ Hz, respectively. The centroid frequency is consistent with that obtained from the joint-fit of the LE 2$-$10 and HE 28$-$200 keV PDS and the corresponding CS in Table~\ref{tab:comparison_lag_models} with the Gaussian phase-lag model, whereas the FWHM is slightly lower.
We plot the phase lag at the centroid frequency of the dip, $\Delta\Phi_D$, and the fractional rms amplitude of the dip in Fig.~\ref{fig:ene_dependent_par}.
The phase lag at the centroid frequency of the dip is $\Delta\Phi_D \sim 1.9 \pi$ when the energy is below 11 keV, but drops rapidly to 0.9$\pi$ when the energy is above 11 keV.
The fractional rms of the dip initially increases, peaking in the $4.8-7.0$ keV range, before decreasing to a local minimum around $10-20$ keV, where the drop of $\Delta\Phi_D$ happens, and it then increases again at higher energies.

\begin{figure}
    \centering
	\vspace{-5mm}
	\includegraphics[width=\columnwidth]{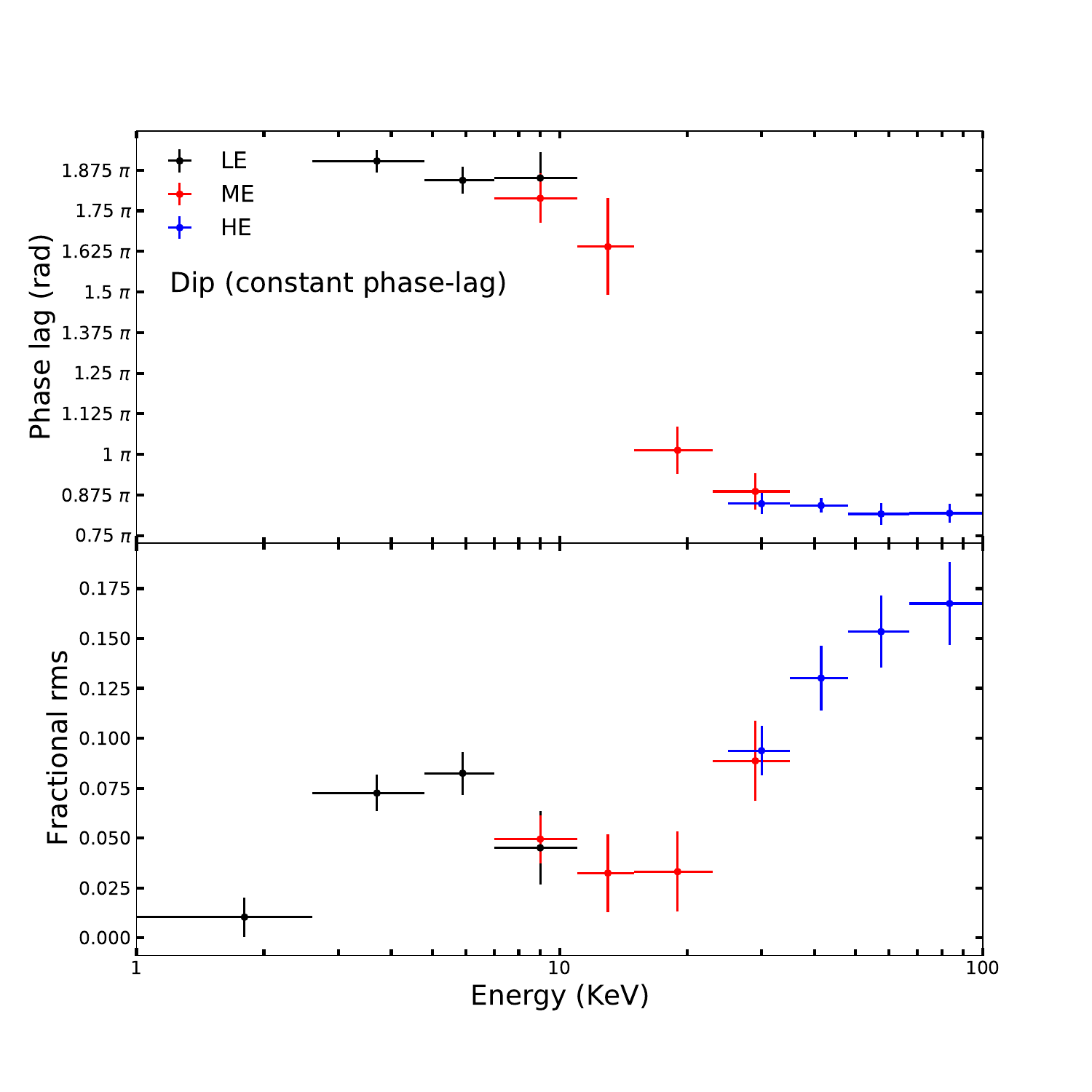}
 	\vspace{-5mm}
    \caption{Same as Fig.~\ref{fig:ene_dependent_par}, but assuming the constant phase-lag model.}
    \label{fig:ene_dependent_par_plc}
\end{figure}

With the constant phase-lag model, the 3-$\sigma$ confidence ranges of the centroid frequency and the FWHM of the Lorentzian that fits the dip are $2.00-3.03$ Hz and $6.70-8.20$ Hz, respectively. The centroid frequency is higher, and the FWHM is lower, than those in Table~\ref{tab:comparison_lag_models}.
We also plot the phase lags and the fractional rms amplitude of the dip in Fig.~\ref{fig:ene_dependent_par_plc}.
The phase lags decrease from $\sim1.9\pi$ to $\sim0.9\pi$ with energy, experiencing a rapid drop near $\sim10-20$~keV.
Similar to the results with the Gaussian phase-lag model, the fractional rms of the corresponding Lorentzian reaches a local minimum near $\sim10-20$~keV, where the drop of the phase lag happens, and it then increases with energy.
In summary, the Gaussian and constant phase-lag models give similar energy dependence of the parameters of the Lorentzian that fits the dip.

\section{Discussion}
\label{sec:discussion}

Taking advantage of the broad energy coverage of \textit{Insight}-HXMT, we find a prominent previously unnoticed feature in the cross spectrum (CS) of Swift~J1727.8$-$1613 during its Normal State as defined by~\citet{2024MNRAS.529.4624Y}, while the QPO frequency increases from 0.13 to 1.93 Hz.
The so-called dip-like feature appears in the 3-15 Hz frequency range of the real part of the cross spectrum obtained between soft data from the LE detector, and the hard photons from the HE instrument.
We find that the real part of the CS reaches negative values near the minimum of the dip, and the phase lags are dominated by a hump with a phase lag between $\pi/2$ and $\pi$ at the corresponding frequency range.
Our analysis suggests that the dip in the real part of the CS is due to the phase lag instead of the modulus of the cross spectrum.

We perform joint fits of the power density spectra (PDS) and CS, with the method proposed by~\citet{2024MNRAS.527.9405M}. In addition to the constant phase-lag and constant time-lag models, we also explore the Gaussian phase-lag model that gives a reduced $\chi^2$ slightly better than those obtained with the other two lag models (Section~\ref{sec:Joint-fits}).
Finally, we study the evolution and energy dependence of the parameters of the Lorentzian that fits the dip.
The phase lags of the dip first increase rapidly and then stay more or less constant as the QPO frequency increases (Fig.~\ref{fig:evolution}, bottom panel).
Surprisingly, the phase lags and the fractional rms amplitude of the Lorentzian that fits the dip show significant changes simultaneously at around 15~keV (Fig.~\ref{fig:ene_dependent_par} and~\ref{fig:ene_dependent_par_plc}).

\subsection{Lag models}

We assume three lag models to fit jointly the LE 2$-$10~keV and HE 28$-$200~keV PDS and the real and imaginary parts of the CS of the photons in the HE 28$-$200~keV with respect to those in the LE 2$-$10~keV: the constant phase-lag, the constant time-lag and the Gaussian phase-lag models.
When we fit the data of Group~$\#$9 with good signal-to-noise ratio in Section~\ref{sec:Joint-fits}, among these models, the Gaussian phase-lag model gives a slightly better fit, with a $\chi^2$ of 781 for 678 degrees of freedom, compared to the constant phase-lag model, which yields a $\chi^2$ of 810 for 678 degrees of freedom, and the constant time-lag model, which yields a $\chi^2$ value of 841 for 678 degrees of freedom.
Regardless of the lag model used, the fitting process involves an additive combination of 13 Lorentzians. Each Lorentzian is always significant at a level of at least 3~$\sigma$ in either the PDS or the CS.

For an individual Lorentzian, $L(\nu;\nu_{0,i},\Delta_i)$, the model to fit the LE 2$-$10~keV PDS is $A_i L(\nu;\nu_{0,i},\Delta_i)$, the model to fit the HE 28$-$200~keV PDS is $B_i L(\nu;\nu_{0,i},\Delta_i)$, the model to fit the real part of the CS is $C_i L(\nu;\nu_{0,i},\Delta_i) \cos{[g_i(\nu;p_j)]}$ and the model to fit the imaginary part of the CS is $C_i L(\nu;\nu_{0,i},\Delta_i) \sin{[g_i(\nu;p_j)]}$.
In the constant phase-lag model the $g_i(\nu;p_j)$ is constant.
This implies that the PDS and both the real and imaginary parts of the CS have the same Lorentzian shape, each of them multiplied by a different constant factor.
With the constant phase-lag model, the Lorentzian that fits the dip appears to overestimate the contribution of the dip-like feature both in the real and imaginary parts of the CS. Especially at low and high frequencies, there are significant differences between this Lorentzian and the data. The model therefore requires other Lorentzians to offset those differences. This raises concerns about possible degeneracy between the Lorentzian components.

On the contrary, for the Gaussian phase-lag model, $g_i(\nu;p_j) = 2\pi\;k_i\;e^{-\frac{1}{2}(\frac{\nu-{\nu_0}_i}{\sigma_i})^2}$, the absolute values of the phase lags decrease and approach gradually to zero as the frequency moves away from the centroid frequency of the Lorentzian. This model implies that the real and imaginary parts of the CS have different shapes. The phase-lag spectra in the black-hole X-ray binaries usually exhibit small phase lags, typically between $-0.3\pi$ and $0.3\pi$~\citep[e.g.][]{2020MNRAS.494.1375Z, 2023MNRAS.519.4434K}. If the phase lag of the Gaussian phase-lag model is between $-\pi/2$ and $\pi/2$, compared to those of the constant phase-lag model, the real part of the CS is broader due to the broadening effect of the cosine function that is multiplied to the Lorentzian function, while the imaginary part becomes narrower.
In our fits the phase lag at the centroid frequency of the Lorentzian that fits the dip falls between $\pi/2$ and $\pi$. This indicates that the real part of the CS is negative only over a narrow frequency range centered around the dip's centroid frequency, rather than consistently negative as it would be for the constant phase-lag model.
As we can see in the right panel of Fig.~\ref{fig:plot_plgau_time} and~\ref{fig:plot_13lor}, the close alignment of the highlighted Lorentzian that fits the dip with the data reduces concerns regarding a potential degeneracy.

Several physical models have been proposed to explain the lags of the short time-scale X-ray variability observed in black hole binaries. In the model of propagating mass accretion rate fluctuations~\citep[e.g.][]{2006MNRAS.367..801A, 2023MNRAS.519.4434K}, the hard lags of the broad-band variability arise from the propagation time of these fluctuations. In the model of Comptonization in a relativistic jet/outflow~\citep{2018A&A...614L...5K}, the hard lags of the broad-band variability are due to the difference in travel time between photons emitted from the disc that reach the observer directly and photons that are inverse-Compton scattered in the jet/outflow before reaching the observer. Both models provide a physical explanation for the constant time-lag model.

Most models of the QPO, such as Lense-Thirring precession~\citep{2009MNRAS.397L.101I, 2016MNRAS.461.1967I}, time-dependent Comptonization~\citep{2020MNRAS.492.1399K, 2021MNRAS.501.3173G, 2022MNRAS.515.2099B} and jet precession~\citep{2021NatAs...5...94M}, consider that the physical or geometrical parameters of the system are modulated by a sinusoidal signal at the QPO frequency. As a result, the observed flux is modulated by a sinusoidal signal at that frequency. A sinusoidal signal is represented as a single point in the Fourier frequency domain, such that its lag is also a single value, not a function of frequency.
However, in real observations, the variability component appears over a range of frequencies rather than at a single frequency. To model its shape, we typically use a Lorentzian function characterized by two parameters: the centroid frequency and the FWHM, the latter of which describes the width of the component. For the constant or Gaussian phase-lag models, there is less justification for why the phase lag at other frequencies should either remain the same as at the centroid frequency or vary according to a specific function of frequency. It is indeed difficult to envisage a mechanism to explain the lags in terms of simple travel times, either of photons or perturbations in the flow, that would lead to a constant or Gaussian phase-lag profile. 

\subsection{Number of Lorentzians}

The PDS of low-mass X-ray binaries are typically described as a superposition of multiple Lorentzians~\citep[e.g.][]{2000MNRAS.318..361N, 2002ApJ...572..392B, 2004astro.ph.10551V}. These Lorentzians are classified into different groups based on their centroid frequencies, FWHM, and correlations with each other. In the frequency range of $\sim$0.01-30 Hz, several key components have been identified in black-hole systems. One of the primary features in this range is the low-frequency QPO~\citep{2000MNRAS.318..361N, 2002ApJ...572..392B, 2004astro.ph.10551V}. Sometimes, harmonics of the low-frequency QPO are also observed, including the sub and second harmonic~\citep{2015MNRAS.448.1298P, 2020MNRAS.496.5262V, 2020MNRAS.494.1375Z}, as well as higher-order harmonics, such as third and fourth harmonics~\citep{2012MNRAS.423..694R, 2023AdSpR..71.3508D}. Another significant component within this frequency range is the band-limited noise component~\citep[BLN;][]{2002ApJ...572..392B, 2004astro.ph.10551V} with a zero centroid frequency. Furthermore, a high-frequency peaked noise component (therefore, HF-PN), which is correlated with the low-frequency QPOs through the PBK correlation, is also observed~\citep{1999ApJ...520..262P, 2000MNRAS.318..361N, 2024MNRAS.529.4624Y}.
In addition, \citet{2024MNRAS.527.9405M} have identified new components, the shoulder of the low-frequency QPO and the second harmonic of the shoulder~\citep[see also][]{1997A&A...322..857B, 2002ApJ...572..392B}.

In Section~\ref{sec:Joint-fits} we successfully fit all features of the LE 2$-$10 keV and HE 28$-$200 keV PDS of Group~\#9, along with the real and imaginary parts of the corresponding CS, using 13 Lorentzian functions.
We classify these 13 Lorentzians into six groups: the QPO and its harmonics, the BLN, the HF-PN, the high-energy excess, the shoulder of the QPO and harmonic, and the Lorentzian that fits the dip. The ``high-energy excess'' group includes two components: a low-frequency excess with a centroid frequency of $\sim$0.1 Hz and a FWHM of $\sim$0.2 Hz, and a high-frequency excess with a centroid frequency of $\sim$16 Hz and a FWHM of $\sim$35 Hz. These two components are only significant in the high-energy PDS. Additionally, the coherence function shows minima at the corresponding frequency ranges, indicating that the variability components dominating the low- and high-energy PDS likely originate from different parts of the system. The Lorentzian that fits the dip is a newly discovered component, further highlighting the diversity of timing phenomena involved. 

The coherence function in Fig.~\ref{fig:fig_1.06_1.24_coh} shows a striking shape, peaking at the QPO frequency. A similar shape has been observed in the source XTE~J1550$-$564~\citep{2000ApJ...531L..45C}, but with a different phase-lag spectrum (see Fig. 1, right panel in their paper). We present the coherence function from different energy bands with respect to the LE 1$-$2.6~keV energy band. As we show in Fig.~\ref{fig:coherence}, the coherence function declines rapidly at $\sim$10$-$25~keV both at low and high frequencies, suggesting that at those frequencies different variability components, likely originating from distinct radiative regions, dominate the low and high energy variability properties of the source~\citep{1996MNRAS.280..227N, 2000MeScT..11.1825B}. The spectral analysis from the broad X-ray energy band suggests the presence of two Comptonization regions~\citep{2024ApJ...970L..33Y}, with cut-off energies of $\sim$20 keV and >50~keV, respectively. These two distinct Comptonization regions may explain the energy dependence of the coherence function. 

\subsection{The dip-like feature in the real part of the CS}

A hump dominating the phase-lag spectrum at frequencies slightly above the QPO frequency, was previously observed in the sources XTE~J1550$-$564~\citep{2000ApJ...531L..45C} and MAXI~J1820$+$070~\citep{2023MNRAS.519.4434K}.
In these sources the phase lag of the hump increases with energy but is always below $\pi/2$, therefore, the real part of the CS does not reach negative values at the corresponding frequency range.
In comparison, the dip we find in the real part of the CS in Swift~J1727.8$-$1613 is much stronger than in other sources and reaches negative values near the minimum. At the same time, the phase lag of the hump is between $\pi/2$ and $\pi$.

The fractional rms amplitude of the Lorentzian that fits the dip increases with energy above 15~keV (Fig.~\ref{fig:ene_dependent_par} and \ref{fig:ene_dependent_par_plc}, bottom panels), indicating that this component is closely associated with the Comptonization region, either an X-ray corona or a jet. 
In Fig.~\ref{fig:evolution} we plot the evolution of the parameters of the Lorentzian that fits the dip. If the time lags reflect travel time in the corona, we roughly estimate the size of the X-ray corona using the parameters of the Lorentzian that fits the dip, applying the formula $L \sim \tau c$, where $\tau=\Delta\phi_{D}/2\pi\nu_{0}$, $\Delta\phi_D$ is the phase lag at the centroid frequency $\nu_0$, and $c$ is the speed of light; this relation assumes that the time lag corresponds to the light travel time across the size of the corona~\citep{2003PhR...377..389R}. We calculate the size in kilometers and in units of R$_g$ ($GM/c^2$) assuming an 8 M$_{\odot}$ black hole.  The size of the Comptonization region first increases to above 10$^5$~km or 10$^4$ R$_g$ as the frequency of the QPO increases from 0.13~Hz to 0.30~Hz, and then decreases to 10$^4$~km or 10$^3$ R$_g$ as the QPO frequency increases further. A bright and very extended compact jet, observed around MJD 60185-60195, was discovered through radio observations by \citet{2024ApJ...971L...9W}. The physical extent of the approaching resolved jet is around 10$^9$~R$_g$ assuming a black hole mass of 8 M$_{\odot}$ and a distance of 2.7~kpc.
Distance measurements reported by~\citet{2025A&A...693A.129M} and~\citet{2025arXiv250206448B} indicated a larger distance, which in turn resulted in a more extended jet.
Remarkably, during that period, the extended continuous jet gradually became fainter and less extended.

Interestingly, we notice that the energy-dependent rms-amplitude and phase-lag spectra in Fig.~\ref{fig:ene_dependent_par} and \ref{fig:ene_dependent_par_plc} closely resemble the shapes predicted by the time-dependent Comptonization model, vKompth~\citep{2020MNRAS.492.1399K, 2022MNRAS.515.2099B}, under the assumption of a low feedback factor \citep[see Fig.~1 and~2 in][]{2022MNRAS.515.2099B}. In their model, a time perturbation with a sinusoidal shape around the time-averaged spectrum is introduced, and the numerical solution of the Kompaneets equation~\citep{1957JETP....4..730K} is computed to obtain the expected energy-dependent rms-amplitude and phase-lag spectra of this sinusoidal variable. This model offers a potential explanation for the rms-amplitude and phase-lag spectra of the Lorentzian that fits the dip. In their scenario, the seed photons are either black-body or disk black-body, generally with a low temperature below 1 keV. However, in our spectra, the pivot point occurs around 15 keV, suggesting that the seed photons may originate from a different corona with a lower temperature, as indicated by the energy spectral analysis  \citep{2024ApJ...970L..33Y}. In future work, we plan to use the vKompth model to further investigate the properties of the corona responsible for the Lorentzian that fits the dip.

\begin{acknowledgements}
We thank the referee for the insightful suggestions that have improved the clarity of our work. MM acknowledges the research programme Athena with project number 184.034.002, which is (partly) financed by the Dutch Research Council (NWO). PJ acknowledges support from the China Scholarship Council (CSC 202304910058). FG acknowledges support by PIBAA1275(CONICET). FG was also supported by grant PID2022-136828NB-C42 funded by the Spanish
 MCIN/AEI/ 10.13039/501100011033 and “ERDF A way of making Europe”.  SKR acknowledges the support of the COSPAR fellowship program for partially funding a visit to the University of Groningen.
\end{acknowledgements}

\bibliographystyle{aa}
\bibliography{aa54353-25}

\begin{thebibliography}{}

\bibitem[\protect\astroncite{{Ar{\'e}valo} \&
  {Uttley}}{2006}]{2006MNRAS.367..801A}
{Ar{\'e}valo} P., {Uttley} P.,  2006, \mnras
  \href{http://dx.doi.org/10.1111/j.1365-2966.2006.09989.x}{367, 801}

\bibitem[\protect\astroncite{{Arnaud}}{1996}]{1996ASPC..101...17A}
{Arnaud} K.A.,  1996,
\newblock In: {Jacoby} G.H., {Barnes} J. (eds.) Astronomical Data Analysis
  Software and Systems V, Vol. 101. Astronomical Society of the Pacific
  Conference Series, p.~17

\bibitem[\protect\astroncite{{Bellavita} et~al.}{2022}]{2022MNRAS.515.2099B}
{Bellavita} C., {Garc{\'\i}a} F., {M{\'e}ndez} M., {Karpouzas} K.,  2022,
  \mnras \href{http://dx.doi.org/10.1093/mnras/stac1922}{515, 2099}

\bibitem[\protect\astroncite{{Bellavita} et~al.}{2025}]{2025A&A...696A.128B}
{Bellavita} C., {M{\'e}ndez} M., {Garc{\'\i}a} F., et~al., 2025, \aap
  \href{http://dx.doi.org/10.1051/0004-6361/202453092}{696, A128}

\bibitem[\protect\astroncite{{Belloni}}{2005}]{2005AIPC..797..197B}
{Belloni} T.,  2005,
\newblock In: {Burderi} L., {Antonelli} L.A., {D'Antona} F., {di Salvo} T.,
  {Israel} G.L., {Piersanti} L., {Tornamb{\`e}} A., {Straniero} O. (eds.)
  Interacting Binaries: Accretion, Evolution, and Outcomes, Vol. 797. American
  Institute of Physics Conference Series, AIP,
  \href{http://dx.doi.org/10.1063/1.2130233}{p.197}

\bibitem[\protect\astroncite{{Belloni} et~al.}{2002}]{2002ApJ...572..392B}
{Belloni} T., {Psaltis} D., {van der Klis} M.,  2002, \apj
  \href{http://dx.doi.org/10.1086/340290}{572, 392}

\bibitem[\protect\astroncite{{Belloni} et~al.}{1997}]{1997A&A...322..857B}
{Belloni} T., {van der Klis} M., {Lewin} W.H.G., et~al., 1997, \aap 322, 857

\bibitem[\protect\astroncite{{Belloni}}{2010}]{2010LNP...794...53B}
{Belloni} T.M.,  2010,
\newblock {States and Transitions in Black Hole Binaries}. In: {Belloni} T.
  (ed.) Lecture Notes in Physics, Berlin Springer Verlag, Vol. 794., p.~53

\bibitem[\protect\astroncite{{Belloni} et~al.}{2024}]{2024MNRAS.527.7136B}
{Belloni} T.M., {M{\'e}ndez} M., {Garc{\'\i}a} F., {Bhattacharya} D.,  2024,
  \mnras \href{http://dx.doi.org/10.1093/mnras/stad3639}{527, 7136}

\bibitem[\protect\astroncite{{Belloni} et~al.}{2011}]{2011BASI...39..409B}
{Belloni} T.M., {Motta} S.E., {Mu{\~n}oz-Darias} T.,  2011, Bulletin of the
  Astronomical Society of India
  \href{http://dx.doi.org/10.48550/arXiv.1109.3388}{39, 409}

\bibitem[\protect\astroncite{{Bendat} \& {Piersol}}{2000}]{2000MeScT..11.1825B}
{Bendat} J.S., {Piersol} A.G.,  2000, Measurement Science and Technology
  \href{http://dx.doi.org/10.1088/0957-0233/11/12/702}{11, 1825}

\bibitem[\protect\astroncite{Bendat \& Piersol}{2010}]{Bendat-2010}
Bendat J.S., Piersol A.G.,  2010,
\newblock Random data :,
\newblock Wiley, New Delhi, 4th ed. edition
  https://www.wiley.com/en-us/Random+Data

\bibitem[\protect\astroncite{{Burridge} et~al.}{2025}]{2025arXiv250206448B}
{Burridge} B.J., {Miller-Jones} J.C.A., {Bahramian} A., et~al., 2025, arXiv
  e-prints \href{http://dx.doi.org/10.48550/arXiv.2502.06448}{
  arXiv:2502.06448}

\bibitem[\protect\astroncite{{Casella} et~al.}{2004}]{2004A&A...426..587C}
{Casella} P., {Belloni} T., {Homan} J., {Stella} L.,  2004, \aap
  \href{http://dx.doi.org/10.1051/0004-6361:20041231}{426, 587}

\bibitem[\protect\astroncite{{Castro-Tirado}
  et~al.}{2023}]{2023ATel16208....1C}
{Castro-Tirado} A.J., {Sanchez-Ramirez} R., {Caballero-Garcia} M.D., et~al.,
  2023, The Astronomer's Telegram 16208, 1

\bibitem[\protect\astroncite{{Cui} et~al.}{2000}]{2000ApJ...531L..45C}
{Cui} W., {Zhang} S.N., {Chen} W.,  2000, \apjl
  \href{http://dx.doi.org/10.1086/312520}{531, L45}

\bibitem[\protect\astroncite{{Debnath} et~al.}{2023}]{2023AdSpR..71.3508D}
{Debnath} D., {Chatterjee} K., {Nath} S.K., et~al., 2023, Advances in Space
  Research \href{http://dx.doi.org/10.1016/j.asr.2022.12.011}{71, 3508}

\bibitem[\protect\astroncite{{Draghis} et~al.}{2023}]{2023ATel16219....1D}
{Draghis} P.A., {Miller} J.M., {Homan} J., et~al., 2023, The Astronomer's
  Telegram 16219, 1

\bibitem[\protect\astroncite{{Fender} et~al.}{2004}]{2004MNRAS.355.1105F}
{Fender} R.P., {Belloni} T.M., {Gallo} E.,  2004, \mnras
  \href{http://dx.doi.org/10.1111/j.1365-2966.2004.08384.x}{355, 1105}

\bibitem[\protect\astroncite{{Fogantini} et~al.}{2025}]{2025arXiv250303078F}
{Fogantini} F.A., {Garc{\'\i}a} F., {M{\'e}ndez} M., et~al., 2025, arXiv
  e-prints \href{http://dx.doi.org/10.48550/arXiv.2503.03078}{
  arXiv:2503.03078}

\bibitem[\protect\astroncite{{Garc{\'\i}a} et~al.}{2022}]{2022MNRAS.513.4196G}
{Garc{\'\i}a} F., {Karpouzas} K., {M{\'e}ndez} M., et~al., 2022, \mnras
  \href{http://dx.doi.org/10.1093/mnras/stac1202}{513, 4196}

\bibitem[\protect\astroncite{{Garc{\'\i}a} et~al.}{2021}]{2021MNRAS.501.3173G}
{Garc{\'\i}a} F., {M{\'e}ndez} M., {Karpouzas} K., et~al., 2021, \mnras
  \href{http://dx.doi.org/10.1093/mnras/staa3944}{501, 3173}

\bibitem[\protect\astroncite{{Homan} \& {Belloni}}{2005}]{2005Ap&SS.300..107H}
{Homan} J., {Belloni} T.,  2005, \apss
  \href{http://dx.doi.org/10.1007/s10509-005-1197-4}{300, 107}

\bibitem[\protect\astroncite{{Ingram} \& {Done}}{2011}]{2011MNRAS.415.2323I}
{Ingram} A., {Done} C.,  2011, \mnras
  \href{http://dx.doi.org/10.1111/j.1365-2966.2011.18860.x}{415, 2323}

\bibitem[\protect\astroncite{{Ingram} et~al.}{2009}]{2009MNRAS.397L.101I}
{Ingram} A., {Done} C., {Fragile} P.C.,  2009, \mnras
  \href{http://dx.doi.org/10.1111/j.1745-3933.2009.00693.x}{397, L101}

\bibitem[\protect\astroncite{{Ingram} et~al.}{2016}]{2016MNRAS.461.1967I}
{Ingram} A., {van der Klis} M., {Middleton} M., et~al., 2016, \mnras
  \href{http://dx.doi.org/10.1093/mnras/stw1245}{461, 1967}

\bibitem[\protect\astroncite{{Ingram} \& {Motta}}{2019}]{2019NewAR..8501524I}
{Ingram} A.R., {Motta} S.E.,  2019, \nar
  \href{http://dx.doi.org/10.1016/j.newar.2020.101524}{85, 101524}

\bibitem[\protect\astroncite{{Karpouzas} et~al.}{2020}]{2020MNRAS.492.1399K}
{Karpouzas} K., {M{\'e}ndez} M., {Ribeiro} E.M., et~al., 2020, \mnras
  \href{http://dx.doi.org/10.1093/mnras/stz3502}{492, 1399}

\bibitem[\protect\astroncite{{Katoch} et~al.}{2023}]{2023ATel16235....1K}
{Katoch} T., {Antia} H.M., {Nandi} A., {Shah} P.,  2023, The Astronomer's
  Telegram 16235, 1

\bibitem[\protect\astroncite{{Kawamura} et~al.}{2023}]{2023MNRAS.519.4434K}
{Kawamura} T., {Done} C., {Axelsson} M., {Takahashi} T.,  2023, \mnras
  \href{http://dx.doi.org/10.1093/mnras/stad014}{519, 4434}

\bibitem[\protect\astroncite{{Kompaneets}}{1957}]{1957JETP....4..730K}
{Kompaneets} A.S.,  1957, Soviet Journal of Experimental and Theoretical
  Physics 4, 730

\bibitem[\protect\astroncite{{Kylafis} \& {Reig}}{2018}]{2018A&A...614L...5K}
{Kylafis} N.D., {Reig} P.,  2018, \aap
  \href{http://dx.doi.org/10.1051/0004-6361/201833339}{614, L5}

\bibitem[\protect\astroncite{{Kylafis} et~al.}{2020}]{2020A&A...640L..16K}
{Kylafis} N.D., {Reig} P., {Papadakis} I.,  2020, \aap
  \href{http://dx.doi.org/10.1051/0004-6361/202038468}{640, L16}

\bibitem[\protect\astroncite{{Leahy} et~al.}{1983}]{1983ApJ...266..160L}
{Leahy} D.A., {Darbro} W., {Elsner} R.F., et~al., 1983, \apj
  \href{http://dx.doi.org/10.1086/160766}{266, 160}

\bibitem[\protect\astroncite{{Ma} et~al.}{2021}]{2021NatAs...5...94M}
{Ma} X., {Tao} L., {Zhang} S.N., et~al., 2021, Nature Astronomy
  \href{http://dx.doi.org/10.1038/s41550-020-1192-2}{5, 94}

\bibitem[\protect\astroncite{{Mastroserio} et~al.}{2018}]{2018MNRAS.475.4027M}
{Mastroserio} G., {Ingram} A., {van der Klis} M.,  2018, \mnras
  \href{http://dx.doi.org/10.1093/mnras/sty075}{475, 4027}

\bibitem[\protect\astroncite{{Mastroserio} et~al.}{2019}]{2019MNRAS.488..348M}
{Mastroserio} G., {Ingram} A., {van der Klis} M.,  2019, \mnras
  \href{http://dx.doi.org/10.1093/mnras/stz1727}{488, 348}

\bibitem[\protect\astroncite{{Mastroserio} et~al.}{2021}]{2021MNRAS.507...55M}
{Mastroserio} G., {Ingram} A., {Wang} J., et~al., 2021, \mnras
  \href{http://dx.doi.org/10.1093/mnras/stab2056}{507, 55}

\bibitem[\protect\astroncite{{Mata S{\'a}nchez}
  et~al.}{2025}]{2025A&A...693A.129M}
{Mata S{\'a}nchez} D., {Torres} M.A.P., {Casares} J., et~al., 2025, \aap
  \href{http://dx.doi.org/10.1051/0004-6361/202451960}{693, A129}

\bibitem[\protect\astroncite{{McClintock} \&
  {Remillard}}{2006}]{2006csxs.book..157M}
{McClintock} J.E., {Remillard} R.A.,  2006,
\newblock {Black hole binaries}. In: {Lewin} W.H.G., {van der Klis} M. (eds.)
  Compact stellar X-ray sources, Vol. 39., p.157

\bibitem[\protect\astroncite{{M{\'e}ndez} et~al.}{2022}]{2022NatAs...6..577M}
{M{\'e}ndez} M., {Karpouzas} K., {Garc{\'\i}a} F., et~al., 2022, Nature
  Astronomy \href{http://dx.doi.org/10.1038/s41550-022-01617-y}{6, 577}

\bibitem[\protect\astroncite{{M{\'e}ndez} et~al.}{2024}]{2024MNRAS.527.9405M}
{M{\'e}ndez} M., {Peirano} V., {Garc{\'\i}a} F., et~al., 2024, \mnras
  \href{http://dx.doi.org/10.1093/mnras/stad3786}{527, 9405}

\bibitem[\protect\astroncite{{Mereminskiy} et~al.}{2024}]{2024MNRAS.531.4893M}
{Mereminskiy} I., {Lutovinov} A., {Molkov} S., et~al., 2024, \mnras
  \href{http://dx.doi.org/10.1093/mnras/stae1393}{531, 4893}

\bibitem[\protect\astroncite{{Miller-Jones} et~al.}{2023}]{2023ATel16211....1M}
{Miller-Jones} J.C.A., {Sivakoff} G.R., {Bahramian} A., {Russell} T.D.,  2023,
  The Astronomer's Telegram 16211, 1

\bibitem[\protect\astroncite{{Mu{\~n}oz-Darias}
  et~al.}{2011}]{2011MNRAS.410..679M}
{Mu{\~n}oz-Darias} T., {Motta} S., {Belloni} T.M.,  2011, \mnras
  \href{http://dx.doi.org/10.1111/j.1365-2966.2010.17476.x}{410, 679}

\bibitem[\protect\astroncite{{Nowak}}{2000}]{2000MNRAS.318..361N}
{Nowak} M.A.,  2000, \mnras
  \href{http://dx.doi.org/10.1046/j.1365-8711.2000.03668.x}{318, 361}

\bibitem[\protect\astroncite{{Nowak} \& {Vaughan}}{1996}]{1996MNRAS.280..227N}
{Nowak} M.A., {Vaughan} B.A.,  1996, \mnras
  \href{http://dx.doi.org/10.1093/mnras/280.1.227}{280, 227}

\bibitem[\protect\astroncite{{O'Connor} et~al.}{2023}]{2023ATel16207....1O}
{O'Connor} B., {Hare} J., {Younes} G., et~al., 2023, The Astronomer's Telegram
  16207, 1

\bibitem[\protect\astroncite{{Page} et~al.}{2023}]{2023GCN.34537....1P}
{Page} K.L., {Dichiara} S., {Gropp} J.D., et~al., 2023, GRB Coordinates Network
  34537, 1

\bibitem[\protect\astroncite{{Palmer} \&
  {Parsotan}}{2023}]{2023ATel16215....1P}
{Palmer} D.M., {Parsotan} T.M.,  2023, The Astronomer's Telegram 16215, 1

\bibitem[\protect\astroncite{{Pawar} et~al.}{2015}]{2015MNRAS.448.1298P}
{Pawar} D.D., {Motta} S., {Shanthi} K., et~al., 2015, \mnras
  \href{http://dx.doi.org/10.1093/mnras/stv024}{448, 1298}

\bibitem[\protect\astroncite{{Psaltis} et~al.}{1999}]{1999ApJ...520..262P}
{Psaltis} D., {Belloni} T., {van der Klis} M.,  1999, \apj
  \href{http://dx.doi.org/10.1086/307436}{520, 262}

\bibitem[\protect\astroncite{{Ratti} et~al.}{2012}]{2012MNRAS.423..694R}
{Ratti} E.M., {Belloni} T.M., {Motta} S.E.,  2012, \mnras
  \href{http://dx.doi.org/10.1111/j.1365-2966.2012.20906.x}{423, 694}

\bibitem[\protect\astroncite{{Reynolds} \& {Nowak}}{2003}]{2003PhR...377..389R}
{Reynolds} C.S., {Nowak} M.A.,  2003, \physrep
  \href{http://dx.doi.org/10.1016/S0370-1573(02)00584-7}{377, 389}

\bibitem[\protect\astroncite{{Tanaka} \&
  {Shibazaki}}{1996}]{1996ARA&A..34..607T}
{Tanaka} Y., {Shibazaki} N.,  1996, \araa
  \href{http://dx.doi.org/10.1146/annurev.astro.34.1.607}{34, 607}

\bibitem[\protect\astroncite{{van der Klis}}{1989}]{1989ASIC..262...27V}
{van der Klis} M.,  1989,
\newblock In: {{\"O}gelman} H., {van den Heuvel} E.P.J. (eds.) Timing Neutron
  Stars, Vol. 262. NATO Advanced Study Institute (ASI) Series C,
  \href{http://dx.doi.org/10.1007/978-94-009-2273-0_3}{p.~27}

\bibitem[\protect\astroncite{{van der Klis}}{2004}]{2004astro.ph.10551V}
{van der Klis} M.,  2004, arXiv e-prints
  \href{http://dx.doi.org/10.48550/arXiv.astro-ph/0410551}{ astro--ph/0410551}

\bibitem[\protect\astroncite{{van Doesburgh} \& {van der
  Klis}}{2020}]{2020MNRAS.496.5262V}
{van Doesburgh} M., {van der Klis} M.,  2020, \mnras
  \href{http://dx.doi.org/10.1093/mnras/staa1867}{496, 5262}

\bibitem[\protect\astroncite{{Vaughan} \& {Nowak}}{1997}]{1997ApJ...474L..43V}
{Vaughan} B.A., {Nowak} M.A.,  1997, \apjl
  \href{http://dx.doi.org/10.1086/310430}{474, L43}

\bibitem[\protect\astroncite{{Wood} et~al.}{2024}]{2024ApJ...971L...9W}
{Wood} C.M., {Miller-Jones} J.C.A., {Bahramian} A., et~al., 2024, \apjl
  \href{http://dx.doi.org/10.3847/2041-8213/ad6572}{971, L9}

\bibitem[\protect\astroncite{{Yang} et~al.}{2024}]{2024ApJ...970L..33Y}
{Yang} Z.X., {Zhang} L., {Zhang} S.N., et~al., 2024, \apjl
  \href{http://dx.doi.org/10.3847/2041-8213/ad60bd}{970, L33}

\bibitem[\protect\astroncite{{Yu} et~al.}{2024}]{2024MNRAS.529.4624Y}
{Yu} W., {Bu} Q.C., {Zhang} S.N., et~al., 2024, \mnras
  \href{http://dx.doi.org/10.1093/mnras/stae835}{529, 4624}

\bibitem[\protect\astroncite{{Zhang} et~al.}{2020a}]{2020MNRAS.494.1375Z}
{Zhang} L., {M{\'e}ndez} M., {Altamirano} D., et~al., 2020a, \mnras
  \href{http://dx.doi.org/10.1093/mnras/staa797}{494, 1375}

\bibitem[\protect\astroncite{{Zhang} et~al.}{2020b}]{2020SCPMA..6349502Z}
{Zhang} S.N., {Li} T., {Lu} F., et~al., 2020b, Science China Physics,
  Mechanics, and Astronomy 63, 249502

\bibitem[\protect\astroncite{{Zhang} et~al.}{2022}]{2022MNRAS.514.2891Z}
{Zhang} Y., {M{\'e}ndez} M., {Garc{\'\i}a} F., et~al., 2022, \mnras
  \href{http://dx.doi.org/10.1093/mnras/stac1050}{514, 2891}

\bibitem[\protect\astroncite{{Zhang} et~al.}{2024}]{2024MNRAS.527.5638Z}
{Zhang} Y., {M{\'e}ndez} M., {Motta} S.E., et~al., 2024, \mnras
  \href{http://dx.doi.org/10.1093/mnras/stad3623}{527, 5638}

\bibitem[\protect\astroncite{{Zhou} et~al.}{2022}]{2022MNRAS.515.1914Z}
{Zhou} D.K., {Zhang} S.N., {Song} L.M., et~al., 2022, \mnras
  \href{http://dx.doi.org/10.1093/mnras/stac1789}{515, 1914}

\end{thebibliography}

\newpage
\appendix

\section{Fit result with the constant time-lag model}
\label{sec:Fit result with the constant time-lag model}

\begin{figure*}[b]
    \centering
    \vspace{-5mm}
  	\includegraphics[width=\columnwidth]{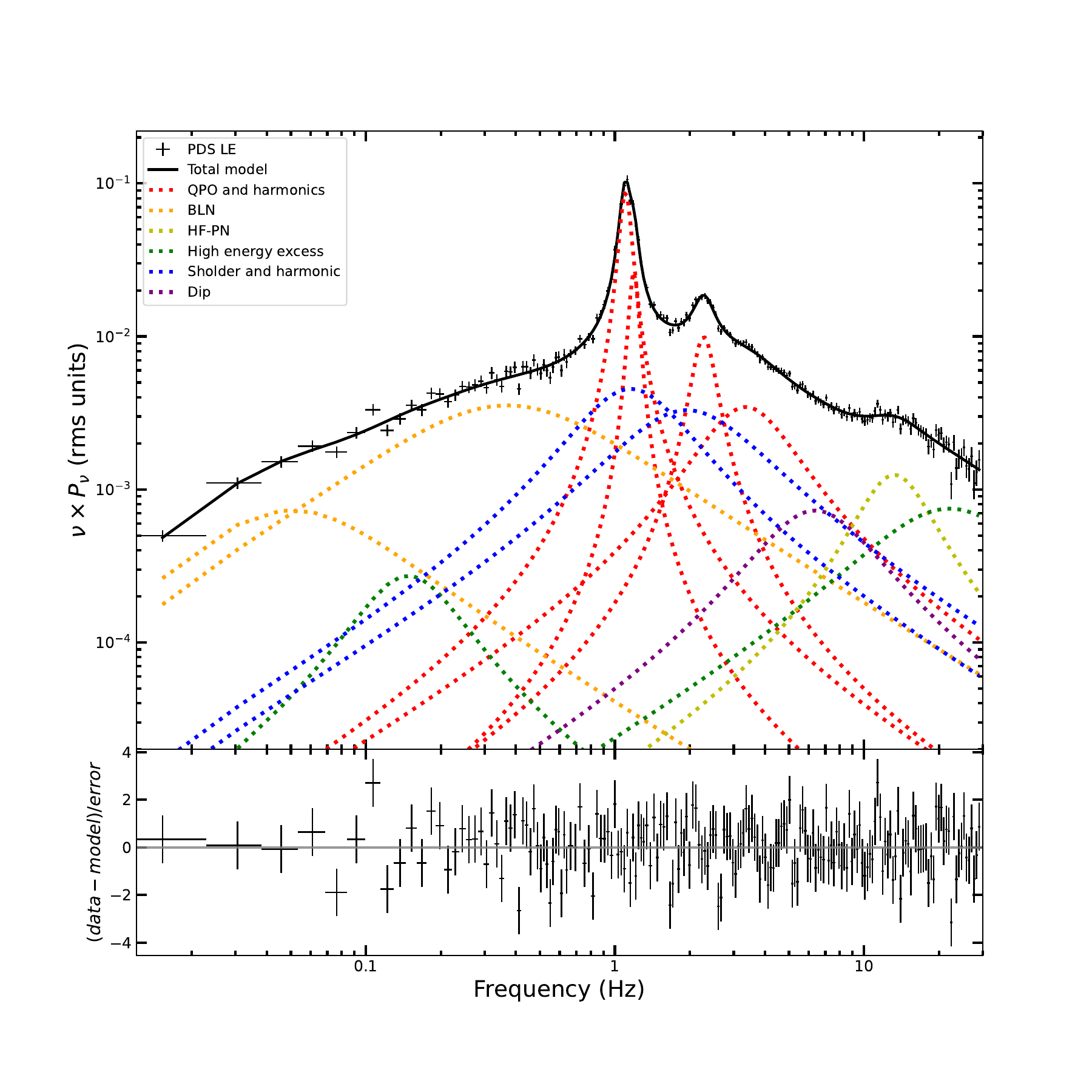}\vspace{-5mm}
    \includegraphics[width=\columnwidth]{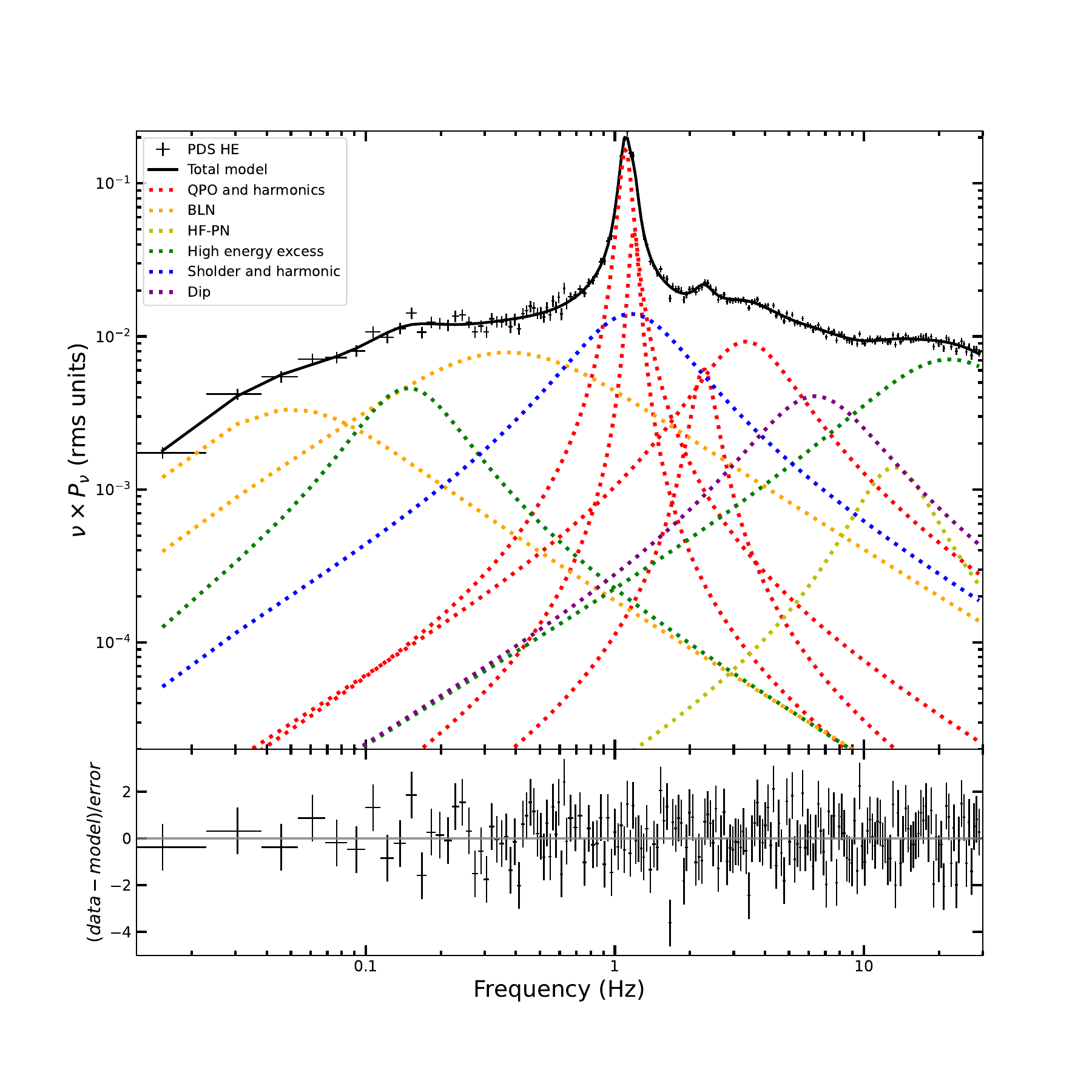}\vspace{-5mm}
	\includegraphics[width=\columnwidth]{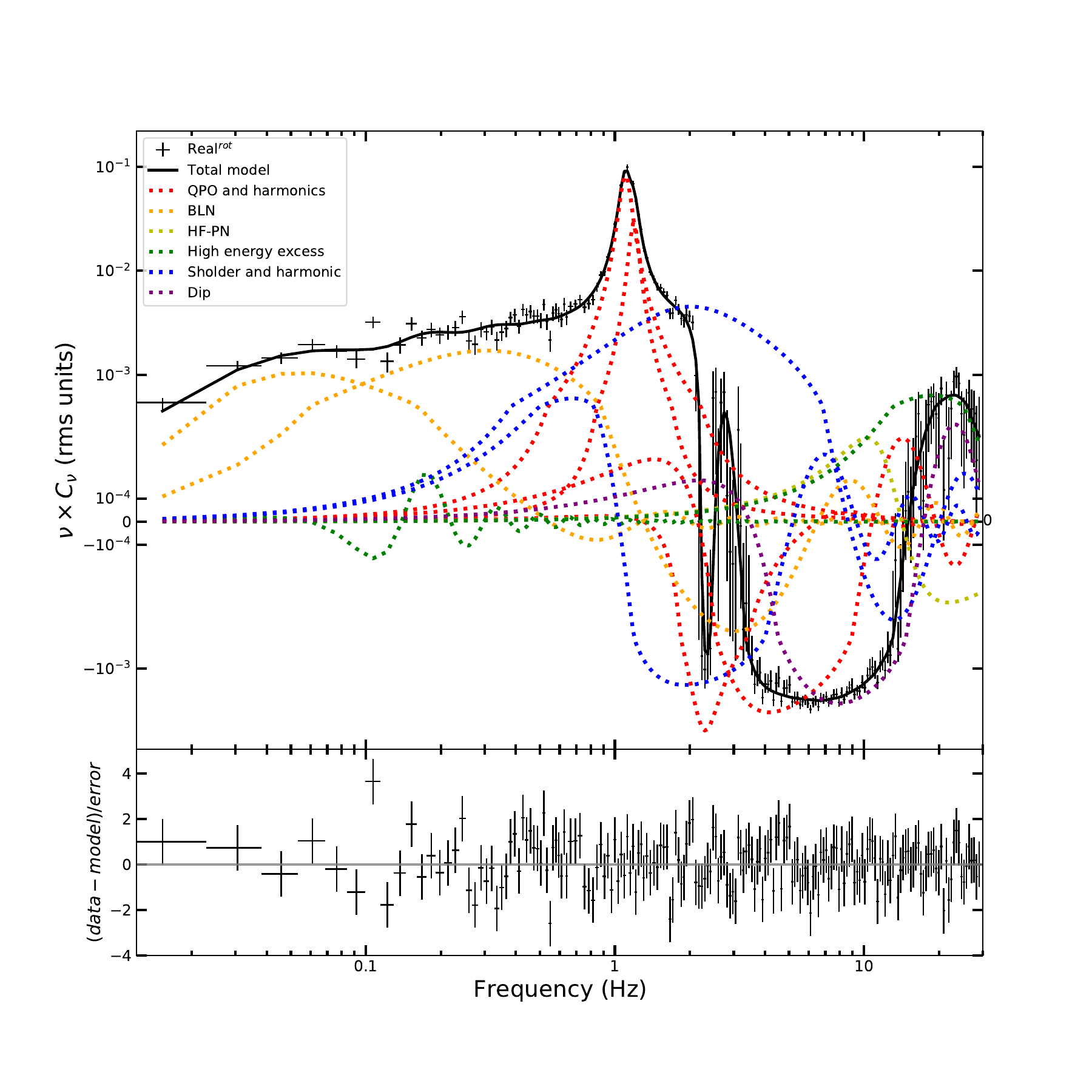}
 	\includegraphics[width=\columnwidth]{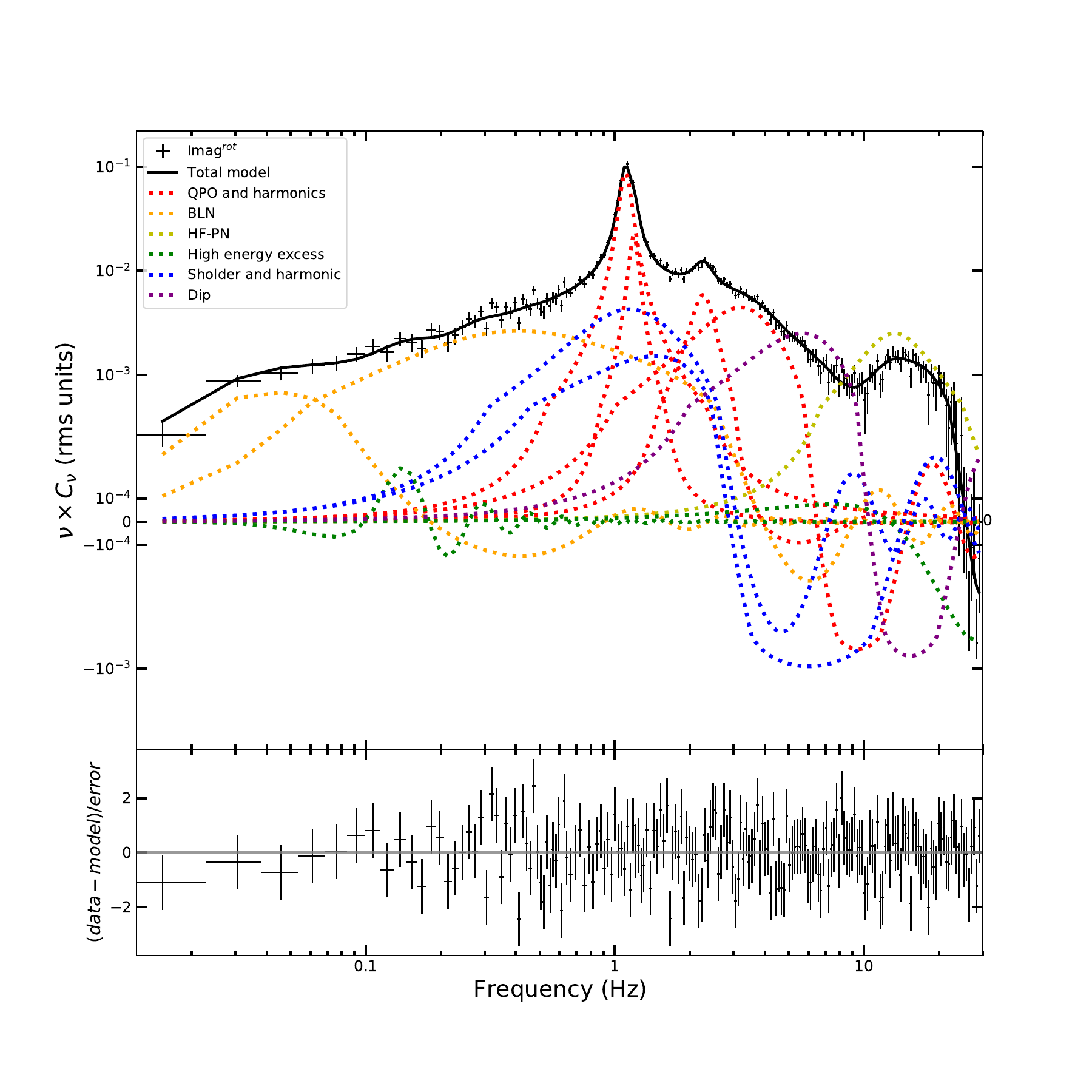}
    \vspace{-5mm}
    \caption{Same as Fig.~\ref{fig:plot_plcon_time}, but assuming the constant time-lag model.}
    \label{fig:plot_tlag_time}
\end{figure*}

In Fig.~\ref{fig:plot_tlag_time} we plot the fit results of the same data shown in Section~\ref{sec:Joint-fit of power and cross spectra} with the constant time-lag model. The model consists of 12 Lorentzians and the fit gives  $\chi^2 = $ 784 for 684 dof. With the constant time-lag model, the phase lag of each component is $2\pi\nu t_c$, where $t_c$ is the constant time lag. Therefore, the real and imaginary parts of the CS oscillate as a function of the frequency with a period of $1/t_{c}$ Hz~\citep{2024MNRAS.527.9405M}. This makes the fit results of the real and imaginary parts of the CS messy at high frequencies, as seen in Fig.~\ref{fig:plot_tlag_time}.

\clearpage
\section{Energy dependence of the coherence function}
\label{sec:coh}

\begin{figure*}[b]
    \centering
  	\includegraphics[width=1.5\columnwidth]{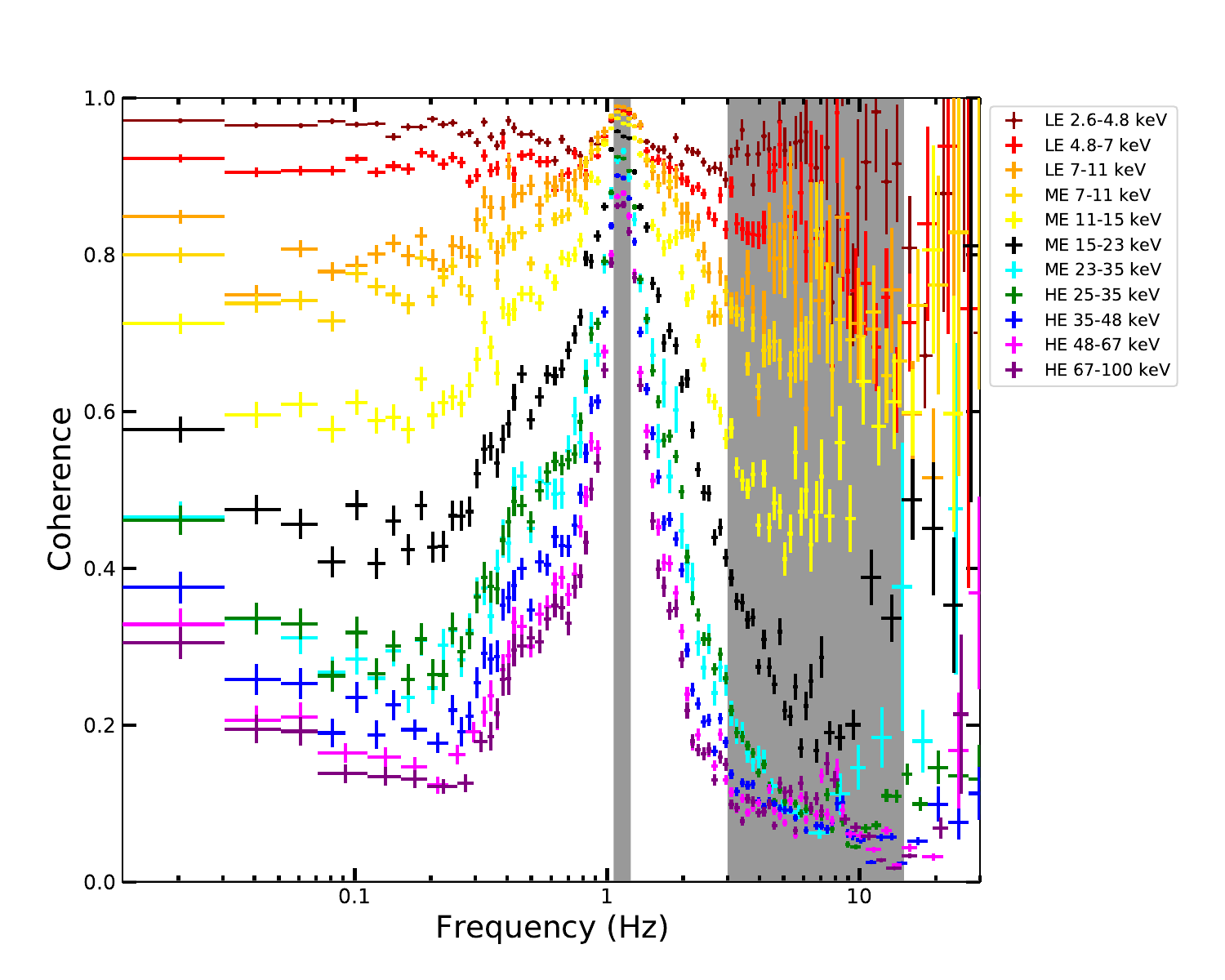}
    \vspace{-5mm}
    \caption{The energy dependence of the coherence function of Swift~J1727.8$-$1613 for Group~\#9 in Table~\ref{tab:group}. The reference energy band is LE 1.0$-$2.6 keV. The colors represent the same energy bands as in Fig~\ref{fig:ene_dependent}.}
    \label{fig:coherence}
\end{figure*}

In Fig.~\ref{fig:coherence} we present the energy dependence of the coherence function of Group~\#9. The energy bands selected are the same as those discussed in Section~\ref{sec:Energy dependent of the dip}, with the reference band being the LE 1.0$-$2.6 keV band. The coherence is close to 1 near the peak of the QPO and shows little variation with energy, suggesting that the QPO dominates the timing variability across all energy bands. As the energy increases, the coherence decreases from above 0.8 to below 0.2 at frequencies lower and higher than the QPO frequency. This decrease is most pronounced in the $\sim$10$-$25 keV energy band. Above $\sim$25~keV, the minimum coherence at low frequencies occurs around 0.1$-$0.3 Hz, where the HE 28$-$200 keV PDS displays a significant excess not present in the LE 2$-$10 keV PDS (see Fig.~\ref{fig:plot_13lor}). The excess is fitted by the low-frequency high-energy excess Lorentzian. At high frequencies, the coherence is very low above  $\sim$25~keV, as we see in Fig.~\ref{fig:plot_13lor}. This indicates either the presence of multiple types of Lorentzians with varying contributions across the low and high energy bands, or that the high-frequency process is incoherent~\citep{1996MNRAS.280..227N, 2000MeScT..11.1825B}.


\end{document}